\documentclass[12pt]{article}

\usepackage{natbib}
\usepackage[nottoc,notlof,notlot]{tocbibind} 

\bibpunct{(}{)}{;}{a}{}{;}
\usepackage{breakcites}
\usepackage[breaklinks=true]{hyperref}
\hypersetup{
    colorlinks=true,
    linkcolor=blue,
    citecolor=blue,
    urlcolor=blue
}

\usepackage{graphicx}
\graphicspath{ {images/} }

\usepackage{xcolor}

\usepackage{amsmath}
\usepackage{bm}
\usepackage{bbm}
\usepackage{amsfonts}%
\usepackage{amssymb}%
\usepackage{amsthm}
\usepackage{geometry} % to change the page dimensions
\usepackage{graphicx} % support the \includegraphics command and options
\usepackage{framed}
\usepackage{caption}
\usepackage{subcaption}

\usepackage{booktabs} % for much better looking tables
\usepackage{bigstrut} % for Excel2Latex
\usepackage{array} % for better arrays (eg matrices) in maths
\usepackage{paralist} % very flexible & customisable lists (eg. enumerate/itemize, etc.)
\usepackage{verbatim} % adds environment for commenting out blocks of text & for better verbatim
\usepackage{lipsum}
\usepackage{setspace}
\usepackage[singlelinecheck=false]{caption}
\usepackage{tabularx}

\usepackage{multirow}
\usepackage{lscape}
\usepackage[linesnumbered,ruled,vlined]{algorithm2e}

\usepackage{tikz}

\usepackage{thmtools} % for running example
\renewcommand{\thmcontinues}[1]{\hyperref[ex:leading]{continued}} % drop page number

\usepackage{diagbox}
\usepackage{rotating}

\title{\setstretch{1} Linear Regression with Centrality Measures}

\author{\parbox{\linewidth}{\centering 
	\normalsize \setstretch{1}
	Yong Cai
	\footnote{I am grateful to Ivan Canay, Eric Auerbach and Joel Horowitz for their guidance on this project. I have also benefited from the comments and suggestions of Santiago Camara, Eduardo Campillo-Betancourt, Piotr Dworczak, Joachim Freyberger, Grant Goehring, Deborah Kim, Jack Porter and Andrei Zeleneev.} 
	\\ 
	Department of Economics \\ 
	University of Wisconsin-Madison \\ 
	\url{yong.cai@wisc.edu}
\endgraf}} 
 \date{ \parbox{\linewidth}{
 \centering \normalsize \setstretch{1}
 \hfill \\ 
  \today
\endgraf}} % Activate to display a given date or no date (if empty),
\theoremstyle{definition}
\newtheorem{definition}{Definition}
\newtheorem{theorem}{Theorem}

\newtheorem{lemma}{Lemma}
\newtheorem{corollary}{Corollary}
\newtheorem{assumption}{Assumption}
\newtheorem{example}{Example}
\newtheorem{remark}{Remark}

\newtheorem{assumpE}{Assumption}

\newtheoremstyle{conjectureStyle}{}{}{}{}{\color{red} \bfseries}{.}{ }{}
\theoremstyle{conjectureStyle}

\newcommand{\prob}{\mathbb{P}}
\newcommand{\cond}{\, \lvert \,}
\newcommand{\expect}{\mathbb{E}}
\newcommand{\cov}{\text{Cov}}

\newcommand{\var}{\text{Var}}

\newcommand{\norm}{\text{N}}
\newcommand{\indep}{\perp\!\!\!\perp}
\newcommand{\pto}{\overset{p}{\to}}
\newcommand{\dto}{\overset{d}{\to}}

\begin{document}
\maketitle
%\centerline{\large PRELIMINARY AND INCOMPLETE}
\begin{abstract} \setstretch{1}\noindent
This paper studies the properties of linear regression on centrality measures when network data is sparse and observed with error. We make three contributions in this setting. First, we show that OLS estimators can become inconsistent under sparsity and characterize the threshold at which this occurs, finding that regression on eigenvector centrality is less robust to sparsity than on degree and diffusion. Second, we derive the asymptotic distributions of the OLS estimators in regimes where they remain consistent. We show that when the target coefficients are non-zero, the estimators exhibit asymptotic bias that can be large relative to their variance, rendering conventional confidence intervals and $t$-tests invalid. Third, we propose bias correction and inference procedures for OLS with sparse, noisy networks. Simulations confirm that our methods perform well in such settings. We demonstrate the empirical relevance of our results in a stylized study of the relationship between consumption smoothing and informal insurance in Nyakatoke, Tanzania.

\end{abstract}



\newpage

\section{Introduction}

Linear regression with centrality measures is a common approach for studying the relationship between network position and economic outcomes. In such analyses, researchers regress an outcome of interest on centrality measures, which are node-level statistics that capture an agent’s importance in a network. For example, \cite{hochberg2007whom} studies the network of venture capital firms and finds that better-connected firms successfully exit a greater proportion of their investments. \cite{cruz2017politician} examines social networks in the Philippines and shows that more central families are disproportionately represented in political office. Similarly, \cite{banerjee2013diffusion} studies the diffusion of microfinance in India and shows that seeding information to more central agents leads to greater program participation.

Across such papers, a variety of centrality measures are used to capture different aspects of network positions. For example, the degree centrality of an agent reflects the number or intensity of their direct links, while eigenvector centrality is designed so that the influence of agents is proportional to that of their connections. The correlation between an outcome variable and a particular centrality measure may therefore reveal the types of interactions that drive a given economic phenomenon: an outcome that is well-predicted solely by degree is likely to be determined in a highly local manner, whereas outcomes that correlate more strongly with eigenvector centrality may reflect interactions that propagate through longer network paths. As such, when researchers estimate these correlations and test their statistical significance, they frequently do so with the goal of drawing conclusions about the economic significance of various centrality measures and the implied mechanisms for outcome determination. Such an exercise is credible provided that standard OLS asymptotic theory provides good approximations in finite samples.

However, network data have two features that may threaten the statistical validity of OLS. First, networks may be sparse, with many more agents than links per agent. This could happen because interactions are observed with low frequency, or because the interactions in question are rare. \cite{chandrasekhar2016formation} argues that many economic networks are sparse, providing evidence from commonly used social network datasets (e.g. AddHealth; Karnataka Villages (\citealt{banerjee2013diffusion}); Harvard social network (\citealt{leider2009directed})). Sparsity poses a challenge to estimation and inference: intuitively, a largely empty network contains little information about the parameters of interest.

Second, the observed network may differ from the true network of interest. Centrality measures are often calculated on data which are obtained by survey or constructed using some proxy for interaction between agents, though subsequent analysis would frequently treat the true network as known. Ignoring observation error may lead to estimates that perform poorly. A growing literature works with networks that are assumed to be observed with error, and increasingly, in combination with sparsity. This is important since the two issues are mutually reinforcing: sparser networks contain weaker signals, which are in turn more difficult to pick out from noisy measurements. The upshot is that OLS estimators computed on sparse, noisy networks may have particularly poor properties. Asymptotic theory that ignores these features will provide similarly poor approximations to their finite sample behavior. Estimation and inference procedures based on these theories may, in turn, lead to invalid conclusions about the economic significance of centrality measures. 

To understand the effect of observation error and sparsity, this paper adopts the asymptotic framework of \cite{bickel2009nonparametric}, featuring a single growing network. Here, agents are endowed with latent types and a graphon (\underline{graph} functi\underline{on}) maps pairs of latent types into link intensities. This is the true network that is unobserved. Instead, researchers observe a single draw from the graphon, in which each link is an independent Bernoulli random variable with success probability given by the link intensity. The difference between the two networks is our observation error. To reflect sparsity, the Bickel–Chen model introduces a scale parameter for the graphon that can shrink it towards zero as sample size increases. As such, the expected number of links per agent can grow much more slowly than $n$ as it tends to infinity.

We study OLS estimation with centrality measures in the Bickel-Chen model, focusing on degree, diffusion and eigenvector centralities. We make the following contributions: (1) We characterize the amount of sparsity at which OLS estimators become inconsistent under observation error, finding that this threshold varies depending on the centrality measure used. Specifically, regression on eigenvector centrality is less robust to sparsity than regressions on degree and diffusion. This suggests that researchers should be cautious about comparing regressions on different centrality measures, since they may differ in statistical properties in addition to economic significance. (2) We develop distributional theory for OLS estimators under observation error and sparsity. We restrict ourselves to sparsity regimes under which OLS is consistent and find that it has an asymptotic bias that can be large when the target coefficient is non-zero. We obtain sharp rates for the asymptotic bias and show that they can be large compared to the variance of the OLS estimators. This implies that the usual confidence intervals do not have the desired coverage probabilities and the $t$-test is invalid for hypothesis of the form $H_0: \beta = b$ when $b \neq 0$. (3) We propose bias-corrected estimators for degree and diffusion that are competitive with the no observation error benchmark. For eigenvector, we propose to conduct inference by widening confidence intervals using an asymptotic bound for the bias. The procedure is conservative when the network is sparse, but exact once the network is moderately dense. 

Because our statistical model captures important features of real-world data, we expect our theory and methods to be more appropriate for settings with observation error and sparsity. We provide simulation evidence supporting this view.  Finally, we demonstrate the empirical relevance of our results in an application inspired by \cite{deweerdt2006risk}, in which we conduct a stylized study of consumption smoothing and social insurance in Nyakatoke, Tanzania. 

\subsection*{Related Literature}

This work is most closely related to papers that study linear regression with centrality measures. \cite{thirkettle2019identification} constructs proposes bounds for linear regression parameters with degree and diffusion centrality under partial sampling of nodes and edges. Several papers consider linear regression with eigenvector centrality. \cite{le2020linear} studies regression on multiple eigenvectors also in the Bickel-Chen framework. Their conditions are denser than necessary for consistency. We obtain sharp thresholds. \cite{cai2021network} focuses on rank-1 networks that are observed with an additive, Gaussian error. In contrast, the Bickel-Chen model features Bernoulli noise and networks with arbitrary rank. However, we will assume finite rank for some results (more details in Remark \ref{remark--compare_caietal}). \cite{cheng2021tackling} considers inference on linear functionals of eigenvectors. The uncertainty that drives their asymptotic results is dominated by the regression error in our setting. 

Our paper also speaks to the literature on the estimation of centrality measures in the presence of noise. Early work provided simulation evidence that centrality measures on noisy networks become less accurate as sparsity increases \citep{costenbader2003stability, borgatti2006robustness}. \cite{segarra2015stability} theoretically studies the stability of network statistics under perturbations, finding that degree and eigenvector centralities are stable, while betweenness is not. \cite{avella2020centrality} and \cite{dasaratha2020distributions} consider settings similar to ours and provide concentration results for degree and eigenvector centralities among others, but not for diffusion centrality. Additionally, they accommodate less sparsity than us, in part because we are not concerned with estimation of centrality measures, only their subsequent use in linear regression.

There is also a large literature involving other types of observation error in network data. In a setting with partial sampling of nodes and edges, \cite{chandrasekhar2016econometrics} characterizes the bias in linear regression on network-level statistics in a setting with many networks. \cite{thirkettle2019identification} provides bounds for centrality statistics and node-level regression parameters in a similar setting. For peer effects regressions, \cite{griffith2022name} examines the effects of selective censoring, while \cite{lewbel2025estimating} considers for link misclassification. \cite{lewbel2021social} examines more general forms of error under the assumption that they vanish asymptotically. A separate body of work assumes that the network is completely unobserved \citep{manresa2016estimating, rose2016identification,lewbel2023social,depaula2025identifying}. With the exception of \cite{thirkettle2019identification}, the aforementioned papers do not consider centrality regressions.

Finally, this work is also connected to papers that emphasizes the statistical challenges arising from sparse networks, whether with network formation models (e.g. \citealt{dePaula2018identifying, jochmans2018semiparametric, graham2020sparse,menzel2022strategic, leung2025normal}) or models of two-sided interactions (e.g. \citealt{jochmans2019fixed,verdier2020estimation, crippa2025identification}). We derive the implications of sparsity for centrality regressions. 

The rest of this paper is organized as follows. Section \ref{section--setup} describes the set-up of our paper. Section \ref{section--theoreticalresults} presents the theoretical results. Simulation results are contained in Section \ref{section--simulations}. In Section \ref{section--application}, we apply our results to the social insurance network in Nyakatoke, Tanzania. Section \ref{section--conclusion} concludes. Proofs are contained in the appendix.

\subsubsection*{Notation}

When $X$ is a square matrix, $\lambda_j(X)$ denotes the $j^\text{th}$ eigenvalue of $X$ while $v_j(X)$ denotes the corresponding eigenvector.  When $f \in L^2([0,1]^2)$ is a symmetric real function,  $\lambda_j(f)$ denotes the $j^{\text{th}}$ eigenvalue of the corresponding Hilbert-Schmidt integral operator, $T_f(g) = \int f(x,y) g(y) dy$, while $\phi_j$ is the corresponding eigenfunction. For deterministic, monotone sequences $x_n$ and $y_n$, we write $x_n \gg z_n$ if $x_n/z_n \to \infty$ and $x_n \ll z_n$ if $x_n/z_n \to 0$. $x_n \approx z_n$ indicates that $x_n/z_n \to k$, where $0 < k <\infty$. We write $x_n \gtrsim z_n$ to mean $\neg (x_n \ll z_n)$ and similarly for $x_n \lesssim z_n$. Let $\iota_n$ be the $n \times 1$ vector of $1$'s. For two $m\times n$ matrices $X$ and $Z$, let $X \circ Z$ denote their entrywise (Hadamard) product.  Finally, $[n]$ denotes the set of integers from $1$ to $n$. 
 
\section{Set-Up}\label{section--setup}
This section outlines the framework for our analysis. Section \ref{subsection--setup-econometric_model} states the linear regression model. Section \ref{section--setup-dgp} describes the data generating process for the true and observed networks. Section \ref{section--setup--sparse_asymp} motivates the sparse network asymptotics of \cite{bickel2009nonparametric}. Section \ref{section--setup-centrality} defines the centrality measures.

\subsection{Regression Model}\label{subsection--setup-econometric_model}

Consider the regression:
\begin{equation*}
	Y_i = C^{(d)}_i \beta^{(d)} + W_i'\gamma^{(d)} + \varepsilon_i^{(d)}  %\quad , \quad E[\varepsilon_i | C^{(d)}_i] = 0~.
\end{equation*}
where $Y_i$ is the outcome of interest and $C^{(d)}_i$ is a network centrality measure of type $d$. $W_i \in \mathbb{R}^L$ is a vector of other covariates. As will become clear in the next section, $W_i$ is allowed to be correlated with the true network but not with the observation errors. We assume that researchers observe $\{(Y_i, W_i)\}_{i=1}^n$ and either the adjacency matrix $A$ or a noisy version $\hat{A}$. $A$ is an $n \times n$ matrix whose $(i,j)^\text{th}$ entry, $A_{ij}$, records the link intensity between agents $i$ and $j$. $\hat{A}$ is some estimate of $A$.  $C^{(d)}_i$ is not directly observed but it can be computed exactly using $A$, or estimated using $\hat{A}$. The parameter of interest is $\beta^{(d)}$, which we interpret as the slope coefficient in the linear conditional expectation function of $Y_i$ on $C_i^{(d)}$ (see Assumption \ref{assumption--varep_condmoments}).

We assume that the data-generating process yields $\{(\varepsilon_i, W_i, U_i)\}_{i=1}^n$ which are independent and identically distributed. $\varepsilon_i$ is the regression error and $U_i$ is an unobserved latent type that will be used to construct the network. Let $\mathbf{Y}, \mathbf{C}^{(d)}, \bm{\varepsilon}$ and $\mathbf{U}$ be vectors which stack the corresponding variables. Similarly for the matrix $\mathbf{W}$. 

In the following sections, we describe (i) the data-generating process for $A$ and $\hat{A}$ via the $U_i$'s, (ii) the Bickel-Chen framework for sparse network asymptotics, and (iii) the use of $A$ and $\hat{A}$ in computing/estimating centrality statistics for OLS estimation. Throughout our discussion, we motivate the econometric framework through the example of consumption smoothing via informal insurance:

\begin{example}  \label{ex:leading}
	Suppose we are interested in the relationship between informal insurance and consumption smoothing. This is a question that has been studied by \cite{deweerdt2006risk, udry1994risk, kinnan2012kinship} and \cite{bourles2021altruism} among many others. Here, we might posit that agents which are more central in the informal insurance network can better smooth consumption. To test this hypothesis, consider the regression where $Y_i$ is standard deviation in $i$'s consumption expenditure and $C^{(d)}_i$ is centrality in the informal insurance network. $\beta^{(d)}$ is then the reduction in consumption dispersion associated with being more central. In the informal insurance network, we might think of $A_{ij}$ as the probability that $i$ lends money to $j$ or vice versa in the event of an adverse income shock. However, $A$ may be hard to elicit via surveys. Instead, we observe the matrix of reported friendships, or a matrix of actual loans which can be taken as noisy measures of $A$.
\end{example}

\subsection{Data-Generating Process for $A$ and $\hat{A}$}\label{section--setup-dgp}

Let $A$ be an $n \times n$ symmetric adjacency matrix. We assume that the relationship between two agents in a network is solely determined by their unobserved latent types $U_i$ through the graphon $f$:
\begin{assumption}[Graphon]\label{assumption--graphon}
 Suppose $U_i \sim U[0,1]$. Let $f:[0,1]^2 \mapsto [0,1]$ and $p_n \in(0, 1]$. For $j > i$, define $A_{ij} = p_n f(U_i, U_j)$. We set $A_{ji} = A_{ij}$ for $j < i$ and normalize $A_{ii} = 0$ for all $i \in [n]$. 
\end{assumption}

In this model, any two agents have a relationship that is between $0$ and $1$, which we interpret as the underlying intensity of interaction between agents. It reflects factors such as the duration of friendship, the frequency of communication, or similarity in characteristics, and is the quantity that is relevant for the outcome variable. The parameter $p_n$ governs the overall density of the network and will be allowed to converge to zero at various rates. As we explain in the next section, it serves as a theoretical device for studying the behavior of OLS estimators when networks are sparse. Finally, we restrict attention to symmetric adjacency matrices because eigenvector centrality is not generally well-defined otherwise. Note that $f$ is not assumed to be known. 

\begin{example}[continues=ex:leading]
	Suppose that $U_i \in [0,1]$ indexes the riskiness of a villager's income as a result of the crops they choose to cultivate. Assumption \ref{assumption--graphon} posits that the relationship between two villagers depends only on their respective income risks. For example, if $f(U_i, U_j) = 1-\left(U_i - U_j\right)^2$, then farmers with similar income risks have higher link intensities between them. $U_i$ can also incorporate other observed or unobserved farmer characteristics, such as place of residence, farming skills or gregariousness. %Together with $f$, the graphon is a rich model of linking behavior. 
\end{example}

When $A$ is observed, we say that there is no observation error. However, researchers do not observe the latent interaction intensity directly in many empirical settings. Instead, they observe some proxy for $A$. We model the observed network, $\hat{A}$, as a matrix of Bernoulli random variables whose success probabilities are given by the latent intensities:
%When $A$ is not observed, we assume that we have access to the noisy version, $\hat{A}$, generated as follows. 
\begin{assumption}[Observed Network]\label{assumption--measurement_model}
	The adjacency matrix with observation error is the $n \times n$ symmetric matrix $\hat{A}$, where for $j > i$,
	\begin{equation*}
		\hat{A}_{ij} \, \lvert \, \mathbf{W, U} \overset{\text{i.i.d.}}{\sim} \text{Bernoulli}\left(A_{ij}\right)~.
	\end{equation*}
	Set $\hat{A}_{ji} = \hat{A}_{ij}$ for $j < i$. $\hat{A}_{ii} = 0$ since $A_{ii} = 0$. 	Let $\xi_{ij} = \hat{A}_{ij} - A_{ij}$ denote the observation error. Suppose also that $\xi_{ij} \perp \!\!\!\perp \varepsilon^{(d)}_k \; \lvert \; \mathbf{W, U}$ for $k \in [n]$ and $d \in \{1,T,\infty\}$. We write $\boldsymbol{\xi} := \hat{A} - A$. 
\end{assumption}
Conditional on $\mathbf{W}$ and $\mathbf{U}$, observation errors are additive, have mean $0$, and are independent across agent pairs. Since we assume conditional independence between $\varepsilon_i$ and $\xi_{ij}$, our model of observation error is akin to classical measurement error. However, additive errors on $\hat{A}_{ij}$ can become non-linear errors on the centrality measures. The covariates $W_i$ are uncorrelated with $\xi_{jk}$ for all $i,j,k$. However, they are allowed to be correlated with $U_i$ and hence the network. 

\begin{example}[continues=ex:leading]
	Suppose we want $A_{ij}$ to be the probability of loans between $i$ and $j$. Instead, we observe whether or not a loan actually occurred. Then, $\hat{A}_{ij}  \sim$ Bernoulli($A_{ij}$). Assumption \ref{assumption--measurement_model} requires that conditional on the loan intensity, the actual decision of $j$ to lend to $i$ is independent of the decision of $k$ to lend to $i$. Such a condition might be satisfied if borrowing is private, so that friends of $i$ cannot coordinate their lending decisions. This might also be the case if loan amounts are small relative to the income shortfall, so that any agent's decision to lend to $i$ does not affect the decisions of their other friends. 
\end{example}

\subsection{Sparse Network Asymptotics}\label{section--setup--sparse_asymp}

To better capture the behavior of estimators when agents in networks have few relationships with one another, we study their properties under sparse network asymptotics. Following \cite{bickel2009nonparametric}, this entails letting $p_n \to 0$ as  $n \to \infty$. 

A vector or matrix is typically said to be sparse if many of its entries are zero. In our setting, we say that $A$ and $\hat{A}$ are sparse if their row sums -- that is, total interactions of agents, actual or observed -- are small. Because the entries of $\hat{A}$ are restricted to be binary, this is the same as having many entries which are $0$. Without such a restriction, the row sums of $A$ could be small even if no entry takes value $0$, as long as non-zero entries are small. In our setting, sparsity should therefore be understood as low interaction intensities in the true network, $A$, but which gives rise to observed networks, $\hat{A}$, that are sparse in the conventional sense.

To see how $p_n \to 0$ gives rise to sparsity, suppose $p_n \to c > 0$. Then $\sum_{j} \expect[A_{ij}] = O(n)$. The network is said to be dense since each agent is linked to a non-vanishing fraction of the sample. In practice, however, researchers may observe sparse networks, in which each agent has few or weak relationships. This is captured by allowing $p_n $ to approach $0$ as $n$ increases. For example, if we set $p_n = k/n$ for some $k > 0$, then $\sum_{j} \expect[A_{ij}] = O(1)$. Each agent then has a bounded number of relationships in expectation even as $n \to \infty$. We say that the network is sparse whenever $p_n \to 0$. The rate at which this occurs corresponds to the amount of sparsity in the data. We will not impose any particular rate on $p_n$, but instead compare centrality measures across sparsity regimes to shed light on the differences between them.

Since network data is often sparse, the Bickel-Chen framework leads to asymptotic approximations that capture more features of observed data. Theory developed in such a framework should therefore better reflect the statistical properties of OLS estimators applied to sparse data. Within econometrics, related approaches have been used to study network formation models (e.g. \citealt{dePaula2018identifying, jochmans2018semiparametric, graham2020sparse,menzel2022strategic, leung2025normal}) and models of two-sided matching (e.g. \citealt{jochmans2019fixed,verdier2020estimation, crippa2025identification}) among others.

The Bickel-Chen model has become a standard data-generating process for network data in nonparametric statistics and machine learning (\citealt{graham2020network}; also see discussion in \citealt{de2017econometrics}). While papers working in this framework are sometimes agnostic about the interpretation of $A$, we interpret it as the matrix of true relationships. By ``true", we mean that these intensities capture the underlying interaction that is relevant for determining the outcome variable $Y_i$. This perspective is consistent with \cite{avella2020centrality}, which argues that centrality measures should be defined with respect to the latent network rather than its noisy realization. Related work studying regressions with eigenvector centrality, such as \cite{le2020linear} and \cite{cai2021network}, also adopts a similar interpretation. More generally, this view is consistent with a broad class of models in which outcomes depend on the network only through the latent network types of agents. In a stochastic block model (see Example \ref{example--low_rank_sbm}), for instance, it implies that outcomes depend solely on block membership, in line with the grouped fixed-effects framework of \cite{bonhomme2015grouped}. This interpretation also fits models in which network data is used to infer unobserved heterogeneity, as in \cite{auerbach2022identification} and \cite{xu2025networks}.

In principle, there are multiple ways of modeling observation error and sparsity. A key motivation for using the Bickel-Chen framework is analytical tractability. Because $\hat{A}$ is a sparse inhomogeneous Erdos-Renyi graph conditional on $\mathbf{W}$ and $\mathbf{U}$, we are able to benefit from a large literature on random graph theory. 

Nonetheless, the model is a reasonable description of network data. In particular, it is consistent with the highly influential weak ties theory of social networks \citep{granovetter1973strength}. This theory posits that low intensity links, which constitute most of any given person's relationships, are the key drivers of many important social and economic outcomes.\footnote{In tracing the network of job referrals, \cite{granovetter1973strength} finds that 83\% of recent job changers in a Boston suburb found their new jobs through friends whom they saw fewer than twice a week, and who were only ``marginally included in the current network of contacts". The author further notes: ``It is remarkable that people receive crucial information from individuals whose very existence they have forgotten." A series of empirical work has found evidence in favor of the weak ties theory across diverse applications such as innovation (\citealt{reagans2001networks}), economic development (\citealt{eagle2010network}) and job referrals (\citealt{rajkumar2022causal}).} Such a view aligns with our model, in which the economically meaningful interactions are numerous and weak. Researchers observe sparse reported connections, which may correspond to strong ties, but are useful only to the extent that they are informative about the weak ties. 

Although our proposed model may not describe every type of economic interaction on networks, it is a plausible model that researchers may not want to rule out ex ante. In that case, the issues that we highlight in the following sections remain relevant.

\subsection{Centrality Statistics and OLS Estimation}\label{section--setup-centrality} 

We next define centrality statistics and the OLS estimators that are based on them. Centrality measures are agent-level measures of importance in a network. Many centrality measures exist, each capturing a different aspect of network position. However, they are all functions of $A$ and can be computed exactly when $A$ is observed. We focus on three popular measures: degree, diffusion and eigenvector centralities. While they are most intuitive when $A$ is binary, centrality measures should be understood as functions of general weighted (symmetric) adjacency matrices. Our definitions are standard up to scaling (see e.g. \citealt{jackson2010social}; \citealt{bloch2021centrality}). 

\begin{definition}[Degree Centrality]
	Degree centrality computed on the $n \times n$ adjacency matrix $A$ is the $n \times 1$ vector:
	\begin{equation*}
		\mathbf{C}^{(1)} = a_n^{(1)}\cdot A\iota_n \quad , \quad a_n^{(1)} = \frac{a_*^{(1)}}{np_n}~.
	\end{equation*}
\end{definition}
Suppose for now that $a_n^{(1)} = 1$. Then agent $i$'s degree centrality is the sum of row $i$ in $A$. If $A$ is binary, degree centrality is simply the number of agents with whom $i$ has a relationship. Under the Bickel-Chen model, it is possible that $\var[(A\iota_n)_i] \to \infty$. The scaling parameter, $a_n^{(1)}$, that we take to be $a_*^{(1)}/np_n$, is introduced to stabilize the limit of $\text{Var}[C^{(1)}_i]$. This ensures that the variance of $\varepsilon_i$ relative to $Y_i$ remains non-trivial as $n \to \infty$. Although $p_n$ is unknown, we can achieve the desired rate if we divide by the mean expected degree $\overline{\text{Deg}} := \frac{1}{n} \sum_{i=1}^n \mathbb{E}\left[\left(A\iota_n\right)\right]_i$ or the standard deviation of unscaled degree, $s.d.(A\iota_n)$. In the latter case, we can interpret $\beta^{(1)}$ as the effect of increasing the number of friends by one standard deviation. %This approach is common in applied work (see e.g. \citealt{banerjee2019using}).

\begin{definition}[Diffusion Centrality]
	For a given $\delta_*$ and $T \in \mathbb{N}$, diffusion centrality computed on the $n \times n$ adjacency matrix $A$ is the $n \times 1$ vector:
	\begin{equation*}
		\mathbf{C}^{(T)} = a_n^{(T)} \cdot \left(\sum_{t = 1}^T \delta_n^{t-1} A^t \right)\iota_n \quad , \quad a_n^{(T)} = \frac{a_*^{(T)}}{np_n} \quad , \quad \delta_n = \frac{\delta_*}{np_n}~.
	\end{equation*}
\end{definition}
Proposed by \cite{banerjee2013diffusion}, diffusion centrality captures the influence of agent $i$ in terms of how many agents they can reach over $T$ periods. Consider again the case of binary $A$. Then the $(i,j)^\text{th}$ entry of $A^t$ is the number of walks from $i$ to $j$ that are of length $t$, which can be thought of as the influence of $i$ on $j$ over $t$ periods of message passing. Without the scaling factor $a_n^{(T)}$, diffusion centrality for agent $i$ is the sum of their influence on all other agents in the network over time up to period $T$, with a decay of $\delta_n$ per period. \cite{bramoulle2018diffusion} provides further discussion on the theoretical foundations of diffusion centrality. 

We treat $T$ as fixed because, as argued by \cite{banerjee2019using}, diffusion centrality tends to perform best in practice when $T$ is small. Accordingly, it is “best thought of as finite and not very large.” The decay parameter $\delta_n$, usually taken to be $<1$, captures the idea that an agent’s influence weakens when their messages take longer to reach others. We choose $\delta_n \propto 1/np_n$ because in our framework, this is the unique rate at which all terms remain non-negligible asymptotically. Outside of this rate, either $A\iota_n$ or $A^T\iota_n$ dominates the other terms. Setting $\delta_n \propto 1/np_n$ therefore leads to a richer asymptotic approximation, but makes analysis more challenging. Our chosen rate is consistent with empirical practice. For example, \cite{banerjee2019using} sets $\delta_n$ to be the inverse of the leading eigenvalue of the adjacency matrix. In our set up, this term would be of order $1/np_n$. Finally, $a_n^{(T)}$ serves the same function as $a_n^{(1)}$ in ensuring that the variance of $\varepsilon_i$ is non-negligible asymptotically. As before, this rate can be achieved, for example, by setting $a_n^{(T)} = 1/\overline{\text{Deg}}$.

\begin{definition}[Eigenvector Centrality]
	Eigenvector centrality computed on the $n \times n$ adjacency matrix $A$ is the $n \times 1$ vector:
	\begin{equation*}
		\mathbf{C}^{(\infty)} = a_n^{(\infty)} \cdot v_1(A) \quad , \quad a_n^{(\infty)} = a_*^{(\infty)} \sqrt{n}
	\end{equation*}
	where $v_1(A)$ is the unit eigenvector corresponding to the eigenvalue of $A$ with the largest absolute value (leading eigenvector).
\end{definition}
Eigenvector centrality is based on the idea that an individual's influence should be proportional to the influence of their friends. That is, for some $k > 0$, we seek the following property:
\begin{equation}\label{equation--eigenvector_centrality_intuition}
	kC^{(\infty)}_i = \sum_{j \neq i} A_{ij}C^{(\infty)}_{j} \quad \mbox{for all } i \in [n]~.
\end{equation}
The eigenvectors of $A$ solve the above equations, with $k$ being the corresponding eigenvalue. By the Perron-Frobenius Theorem, the leading eigenvector is the unique eigenvector that can be chosen so that every entry is non-negative, motivating its use as a centrality measure. 

The leading eigenvector is well-defined only if the leading eigenvalue of $A$ has multiplicity 1, that is, if $\lambda_1(A) \neq \lambda_2(A)$. To ensure that this occurs with high probability, we will make the following assumption when analyzing eigenvector centrality:
\begin{assumpE}\label{assumption--eigenvector_identification}
	Suppose $\lambda_1(f) \neq \lambda_2(f)$. 
\end{assumpE}

Eigenvectors are defined only up to scale: if $\mathbf{C}$ satisfies Equation \ref{equation--eigenvector_centrality_intuition}, so will $a^{(\infty)}_n \cdot \mathbf{C}$ for any $a^{(\infty)}_n \in \mathbb{R}$. We set $a_n^{(\infty)} \propto \sqrt{n}$ to stabilize the variance of $C_i^{(\infty)}$ as $n \to \infty$. This is in line with \cite{cai2021network}. \cite{le2020linear} does not fix the length of the eigenvector, but their goal is to recover the projection $\mathbf{C}^{(d)}\beta^{(d)}$ and not $\beta^{(d)}$ itself.

This paper focuses on the above three centrality measures, which are closely related (\citealt{bloch2021centrality}). When $T = 1$, $\mathbf{C}^{(1)} \propto \mathbf{C}^{(T)}$. Furthermore, as shown by \cite{banerjee2019using}, if $\delta_n \geq 1/\lambda_1(A)$, then $\lim_{T \to \infty} \mathbf{C}^{(T)} \propto \mathbf{C}^{(\infty)}$, where the limit is taken with $n$ fixed. We can therefore understand the centrality measures as sums of $A^t$ up to different $T$'s, motivating our notational choice. Notably, we do not discuss Katz-Bonacich centrality which is the limit of diffusion as $T \to \infty$ when $\delta_n < 1/\lambda_1(A)$. Its properties are challenging to analyze in sparse regimes. 

\begin{example}[continues=ex:leading]
	With risk sharing and social insurance, we can interpret 
	\begin{itemize}
		\item $C^{(1)}_i$ as the probability-weighted number of friends who will lend to or borrow from $i$.
		\item $C^{(T)}_i$ as the probability-weighted number of friends who will lend to or borrow from $i$ directly or through their friends. $T$ is the maximum length of the borrowing chain. For example if $T$ is $2$, $i$ can borrow from friends of friends but not friends of friends of friends. $\delta_n$ is the increased difficulty of borrowing from a person that is one step further, e.g. of borrowing from friends of friends relative to borrowing from a friend directly. 
		\item $C^{(\infty)}_i$ as requiring the borrowing ability of $i$ to be proportional to the borrowing ability of their friends. Implicitly, this means agents might form borrowing chains that are arbitrarily long.
	\end{itemize}
\end{example}

When $A$ is observed,we have access to the following estimators. 
\begin{definition}[OLS Estimators without Observation Error]
	Suppose $A$ is observed. For $d \in \{1, T, \infty\}$, let $(\tilde{\beta}^{(d)}, \tilde{\gamma}^{(d)})$ be the OLS estimators from regressing $Y_i$ on $(C_i^{(d)}, W_i)$. 
\end{definition}

When networks are observed with errors, we assume that network centralities are estimated using $\hat{A}$ in place of $A$:
\begin{definition}[Centralities with Observation Error]
	Suppose $\hat{A}$ is observed but not $A$. Define:
	\begin{align*}
		\hat{\mathbf{C}}^{(1)}  = a_n^{(1)}\cdot \hat{A}\iota_n \quad , \quad 
		\hat{\mathbf{C}}^{(T)}  = a_n^{(T)} \cdot \left(\sum_{t = 1}^T \delta_n^{t-1} \hat{A}^t \right)\iota_n \quad , \quad
		\hat{\mathbf{C}}^{(\infty)}  = a_n^{(\infty)} \cdot v_1(\hat{A})~.
	\end{align*}
\end{definition}
We assume that $a_n^{(d)}$ and $\delta_n$ are known even when $\hat{A}$ is unknown. Estimation of these constants will not affect our result if their rates of convergence is faster than $1/\sqrt{n}$. By Theorem 1 in \cite{bickel2011method}, this is true for a large class of statistics that they call network moments, including mean degree. However, statistics such as standard deviation may require bias correction.% of the form described Section \ref{section--bias-correction--analytical}. 

In the case with observation error, we consider plug-in estimators based on noisy centrality measures:
\begin{definition}[OLS Estimators with Observation Error]
	Suppose $A$ is observed. For $d \in \{1, T, \infty\}$, let $(\hat{\beta}^{(d)}, \hat{\gamma}^{(d)})$ be the OLS estimators from regressing $Y_i$ on $(\hat{C}_i^{(d)}, W_i)$. 
\end{definition}

For the two procedures, we also define the corresponding regression residuals:
\begin{definition}[Regression Residuals]\label{definition--reg_resids}
For $d \in \{1, T,\infty\}$, define:
\begin{align}
	\tilde{\varepsilon}_{i}^{(d)} & := Y_i - \tilde{\beta}^{(d)}C_i^{(d)} - \tilde{\gamma}'W_i~, \label{equation--reg_resid_tilde}\\
	\hat{\varepsilon}_{i}^{(d)} & := Y_i - \hat{\beta}^{(d)}\hat{C}_i^{(d)} - \hat{\gamma}'W_i~.\label{equation--reg_resid_hat}
\end{align}
\end{definition}

We conclude this section with two assumptions. The first concerns the moments of $\varepsilon_i$ and $W_i$: 
\begin{assumption}\label{assumption--varep_condmoments} 
Suppose for $d \in \{1, T, \infty\}$ that:
\begin{enumerate}
	\item[(a)]$E\left[\varepsilon^{(d)}_i \,\big|\, W_i, U_i\right] = 0$
	\item[(b)] $0 < \underline{\sigma}^2 \leq E\left[\left(\varepsilon^{(d)}_i\right)^2 \, \Big| \, W_i, U_i\right] \leq \bar{\sigma}^2 < \infty$.
	\item[(c)] $E\left[\left\lvert\varepsilon^{(d)}_i\right\rvert^3 \,\Big| \,  W_i, U_i\right] \leq \bar{\kappa}_3$.
	\item [(d)] $\mathbb{E}[\lVert W_i \rVert^2] < \infty$. 
\end{enumerate}
\end{assumption}
Assumption (a) above implies that $E[ \varepsilon^{(d)}_{i} \, \big\lvert \, W_i, C_i^{(d)}  ] = 0$, justifying linear regression. 
Meanwhile, (c) ensures that $\varepsilon_i^{(d)}$ converges sufficiently quickly and implies the upper bound in (b). The lower bound in (b) prevents degeneracy. Finally (d) requires $W_i$ to have finite second moment, which is standard. 

Our second assumption rules out perfect collinearity between $W_i$ and the centrality measures. We state our assumption
in terms of the following:
\begin{definition}
Let the asymptotic version of each centrality measure be:
\begin{align*}
	C^{*,(1)}(U_i) & = a_*^{(1)}\int f(U_i, u_1) \; du_1 \\
	C^{*,(T)}(U_i) & = a_*^{(T)}\sum_{t=1}^T (\delta^{*})^{t-1} \int f(U_i, u_1)  f(u_1, u_2) \cdots  f(u_{t-1}, u_{t})\; du_1 \cdots du_T \\
	C^{*,(\infty)}(U_i) & = a_*^{(\infty)} \phi_1(U_i)	
\end{align*}
\end{definition}
Here, $\phi_1$ is the leading eigenfunction of the operator $T(g) = \int f(x, y) g(y) \; dy$. It is well known that $C^{(d)}_i \mid U_i \overset{p}{\to} C^{*,(d)}(U_i)$ for $d \in \{1, \infty\}$ (see e.g. \citealt{avella2020centrality}). The case of $C^{*,(T)}(U_i)$ follows straightforwardly from the law of large numbers for U-statistics. We can now state our final  assumption:
\begin{assumption}[Covariates]\label{assumption--no_collinearity}
	For $d \in \{1,T, \infty\}$, let $X_i = (C^{*,(d)} , W_i')'$ and suppose that $\mathbb{E}[X_iX_i']^{-1}$ exists. 
\end{assumption}
Assumption \ref{assumption--no_collinearity} rules out $f(u,v) = 0$ since otherwise $C_i^{*, (d)} = 0$. If $W_i$ includes a constant, it rules out the constant graphon $f(u,v) = k$ since all agents would have identical centrality. Similarly, if $W_i$ includes fixed effects based on some cluster membership $c$, then $f$ cannot be a stochastic block model based on $c$ only.

\section{Consistency}\label{section--theoreticalresults}

This section studies the consistency of OLS estimators under varying degrees of network sparsity. Section \ref{subsection--benchmark} presents the no observation error benchmark. Section \ref{subsection--consistency} characterizes the level of sparsity at which consistency of each $\hat{\beta}^{(d)}$ fails. The upshot is that $\hat{\beta}^{(\infty)}$ is able to accommodate less sparsity than $\hat{\beta}^{(1)}$ and $\hat{\beta}^{(T)}$. %Section \ref{section--rule_of_thumb} provides a rule of thumb for determining consistency of $\hat{\beta}^{(d)}$.

\subsection{No Observation Error Benchmark}\label{subsection--benchmark}

\begin{theorem}\label{theorem--benchmark}
	Suppose Assumptions \ref{assumption--graphon}, \ref{assumption--varep_condmoments} and \ref{assumption--no_collinearity} hold. Then for $d \in \{1, T\}$, we have:
	\begin{equation*}
		\frac{\sqrt{n}\left(\tilde{\beta}^{(d)}- \beta^{(d)}\right)}{\sqrt{V_{11}^{(d)}}} \overset{d}{\to} \text{N}(0,1)~,
	\end{equation*}
	where
	$
		V^{(d)} = \mathbb{E}[X_iX_i']^{-1} \mathbb{E}[X_iX_i'\varepsilon_i^2] \mathbb{E}[X_iX_i']^{-1}
	$.
	Under Assumption \ref{assumption--eigenvector_identification}, the above holds for $d = \infty$ as well. 
\end{theorem}

In the absence of observation error, $\tilde{\beta}^{(d)}$ is consistent for $\beta^{(d)}$ and converges to a standard normal distribution at rate $\sqrt{n}$. The three centrality measures perform equally well in the sense that consistency holds regardless of network sparsity, since $\tilde{\beta}^{(d)}$ is invariant to $p_n$. As usual, the plug-in estimator for $V^{(d)}$ is also consistent. Specifically, let $\tilde{V}^{(d)}$ be the estimator obtained from $V^{(d)}$ by replacing $C_i^{*,(d)}$ with $C_i^{(d)}$ and $\varepsilon_{i}$ with the regression residual $\tilde{\varepsilon}_i$. Then, replacing $V^{(d)}$ with $\tilde{V}^{(d)}$ does not change the result above. 

\subsection{Consistency under Observation Error}\label{subsection--consistency}

In the presence of observation error, consistency of the OLS estimator depends on $p_n$. 

\begin{theorem}[Consistency with Observation Error]\label{theorem--consistency--error}
	Suppose Assumptions \ref{assumption--graphon}, \ref{assumption--measurement_model}, \ref{assumption--varep_condmoments} and \ref{assumption--no_collinearity} hold. Then,
	\begin{enumerate}
		\item[(a)]  For $d \in \{1,T\}$, 
		$\hat{\beta}^{(d)} \overset{p}{\to} \beta^{(d)}$ if and only if $p_n \gg {n^{-1}}$.
		\item[(b)] Suppose also that Assumption \ref{assumption--eigenvector_identification} holds. Then, $\hat{\beta}^{(\infty)} \overset{p}{\to} \beta^{(\infty)}$ if
		\begin{equation}\label{eq--eigen_noise_lower}
			p_n \gg n^{-1} \sqrt{\frac{\log n}{\log \log n}}~.
		\end{equation}
		Suppose $p_n$ satisfies
		\begin{equation} \label{equation--eigenvector_inconsistent_rates}
			n^{-1}\left(\log\log n\right)^4 \ll p_n \lesssim n^{-1}\sqrt{\frac{\log n}{\log \log n}}~.
		\end{equation}	
		Then $\hat{\beta}^{(\infty)}$ is inconsistent for $\beta^{(\infty)}$.
	\end{enumerate}
\end{theorem}
We draw two lessons from the consistency thresholds in Theorem \ref{theorem--consistency--error}. First, whereas $p_n$ did not affect consistency in the absence of observation error, it is relevant for the consistency of $\hat{\beta}^{(d)}$. Sparsity is therefore an issue only under observation error. Likewise, observation error is not an issue when $p_n$ is sufficiently dense. As such, it is the interaction of sparsity and observation error that leads to the poor properties of the plug-in estimators $\hat{\beta}^{(d)}$. 

Second, $\hat{\beta}^{(\infty)}$ requires greater $p_n$ for consistency than $\hat{\beta}^{(1)}$ and $\hat{\beta}^{(T)}$. In other words, $\hat{\beta}^{(\infty)}$ is less robust to sparsity than $\hat{\beta}^{(1)}$ and $\hat{\beta}^{(T)}$.  The threshold for consistency for degree and diffusion is the threshold at which the expected number of reported links diverges to infinity. Since degree and diffusion are mean-type statistics, this is sufficient for them to concentrate to their true values. On the other hand, when $p_n \approx 1/n$, it is well-known that $\sum_{j} \hat{A}_{ij} \approx \text{Poisson}\left(\sum_{j} A_{ij}\right)$, giving rise to classical measurement error, which translates into the usual attenuation bias in the absence of covariates. 

For eigenvector centrality, our inconsistency results include a lower bound on $p_n$. Note, however, that $n^{-1} \sqrt{\frac{\log n}{\log \log n}}$ is a sharp consistency threshold: $\hat{\beta}^{(d)}$ is consistent right above that rate and inconsistency at and below it. To understand how the threshold arises, observe that the eigenvalues of A are of order $np_n$. The matrix $\hat{A}$ has informative eigenvectors which are close to those of $A$, and with similar eigenvalues. However, it also contains junk eigenvectors. Specifically, each high degree node gives rise to an eigenvector with eigenvalue approximately equal to the square root of the degree. It is well known that maximum degree $\hat{d}_\text{max}$ has order:
\begin{equation*}
	\hat{d}_\text{max} \approx \max\left\{ np_n , \frac{\log n}{\log \log n}   \right\}
\end{equation*}
When $\sqrt{\hat{d}_\text{max}} \gg np_n$, the leading value belongs to an uninformative eigenvector with high probability, and we cannot estimate the eigenvectors of $A$ using the eigenvectors of $\hat{A}$. This is the result of \cite{benaych2020spectral} and \cite{alt2021poisson}.

\begin{remark}
	\cite{avella2020centrality} and \cite{dasaratha2020distributions} provide results showing that for $d \in \{1, \infty\}$, $p_n \gg \frac{\log n}{n}$ is sufficient for $\lVert \hat{C}^{(d)} - C^{(d)} \rVert \to 0$ with probability approaching 1. Our focus is on the consistency of OLS estimators. We thresholds that are sharp and strictly below $\frac{\log n}{n}$ for all three statistics.
\end{remark}

\begin{remark}
	Theorem \ref{theorem--consistency--error} does not specify the behavior of $\hat{\beta}^{(\infty)}$ when $p_n \ll n^{-1}\left(\log \log n\right)^4$. Up to this threshold, we know by \cite{alt2021poisson} that OLS is inconsistent only because we have descriptions of both eigenvalues and eigenvectors. To our knowledge, recent developments in random matrix theory do not provide any description of eigenvectors below this threshold. Hence, it is not clear what type of pathology arises when $p_n \ll n^{-1}\left(\log \log n\right)^4$ and how that might affect the behavior of $\hat{\beta}^{(\infty)}$. Since the estimated eigenvalues are noise, we conjecture that the estimated eigenvectors would be as well. If so, we would not expect $\hat{\beta}^{(\infty)}$ to be consistent.
\end{remark}

\section{Asymptotic Distribution}\label{subsection--inference}

This section studies the asymptotic distribution of $\hat{\beta}^{(1)}$, $\hat{\beta}^{(T)}$ and $\hat{\beta}^{(\infty)}$, focusing on the regimes in which they are consistent. Our main result is that observation error leads to asymptotic bias  that can be of larger order than the variance when $\beta^{(d)} \neq 0$. To better characterize the bias of $\hat{\beta}^{(\infty)}$, we introduce the following assumption: 

\begin{assumpE}[Finite Rank]\label{assumption--rank_r}
	Suppose $f$ has rank $R < \infty$:
	\begin{equation}\label{equation--rank_r_assumption}
		f(u, v) = \sum_{r = 1}^R {\lambda}_r\phi_r(u)\phi_r(v)~, 
	\end{equation}	
	where 
	\begin{align*}
		\int_{[0,1]} \phi_r(u)\phi_s(u) du = \begin{cases}
		1 & \mbox{ if } s = r~, \\
		0 & \mbox{ otherwise. }
		\end{cases}
	\end{align*}
	Moreover, suppose $\expect\left[\phi_r(U_i)^4\right] < \infty$ for $r \in [R]$ and that:
	\begin{equation*}
			\Delta_{\min} = \min_{1 \leq r \leq R-1} \left\lvert {\lambda}_r - {\lambda}_{r+1} \right\rvert > 0~.
	\end{equation*}
\end{assumpE}

In Equation \eqref{equation--rank_r_assumption}, we express $f$ in terms of its eigenfunctions $\left\{\phi_r\right\}_{r=1}^R$. Assumption \ref{assumption--rank_r} implies that the true network has low-dimensional structure and is satisfied by many popular network models, such as the stochastic block model (\citealt{holland1983stochastic}, see also Example \ref{example--low_rank_sbm} below) and random dot product graphs (\citealt{young2007random}). This assumption is also commonly found in the network (e.g. \citealt{levin2019bootstrapping}; \citealt{li2020network}) and matrix completion literatures (e.g. \citealt{negahban2012restricted}; \citealt{chatterjee2015matrix}; \citealt{athey2021matrix}). Importantly, existing papers on inference with eigenvectors (\citealt{le2020linear}; \citealt{cai2021network}) also make this assumption. Note that Assumption \ref{assumption--rank_r} implies Assumption \ref{assumption--eigenvector_identification}.

\begin{example}[Stochastic Block Model] \label{example--low_rank_sbm}
	The Stochastic Block Model (SBM) is one of the earliest statistical models of network data. It assumes that individuals fall into groups $g \in \{1, ..., B\}$ and that the true network depends only on group membership. For example, suppose a village can be divided into an upper and a lower class. The SBM posits that the intensity of the interaction between member of the upper class and any member of the lower class is the same. Analogously for that between any two members of the same class, though the intensities can vary. Going back to the general stochastic block model, suppose the proportion of each group is $\pi_g$ and that link probability between the groups $g$ and $g'$ is $p_{g, g'}$. Then the graphon is a step-function on $[0,1]^2$ with $B^2$-steps and rank $B$. It is visualized in Figure \ref{figure--sbm_graphon}. 
\end{example}

\begin{figure}
	\centering
	\tikzset{every picture/.style={line width=0.75pt}} %set default line width to 0.75pt        

\begin{tikzpicture}[x=0.75pt,y=0.75pt,yscale=-1,xscale=1]
	%uncomment if require: \path (0,262); %set diagram left start at 0, and has height of 262
	
	%Straight Lines [id:da09657789498195746] 
	\draw    (50,219.5) -- (50,21.5) ;
	\draw [shift={(50,19.5)}, rotate = 90] [color={rgb, 255:red, 0; green, 0; blue, 0 }  ][line width=0.75]    (10.93,-3.29) .. controls (6.95,-1.4) and (3.31,-0.3) .. (0,0) .. controls (3.31,0.3) and (6.95,1.4) .. (10.93,3.29)   ;
	%Straight Lines [id:da7733432241161489] 
	\draw    (50,219.5) -- (266,219.5) ;
	\draw [shift={(268,219.5)}, rotate = 180] [color={rgb, 255:red, 0; green, 0; blue, 0 }  ][line width=0.75]    (10.93,-3.29) .. controls (6.95,-1.4) and (3.31,-0.3) .. (0,0) .. controls (3.31,0.3) and (6.95,1.4) .. (10.93,3.29)   ;
	%Straight Lines [id:da11964339247289224] 
	\draw  [dash pattern={on 4.5pt off 4.5pt}]  (50.5,49.5) -- (230.75,49.5) ;
	%Straight Lines [id:da878890637424866] 
	\draw  [dash pattern={on 4.5pt off 4.5pt}]  (230.75,49.5) -- (230.75,219.5) ;
	%Straight Lines [id:da7805823147077475] 
	\draw  [dash pattern={on 4.5pt off 4.5pt}]  (80.75,49.5) -- (80.75,219.5) ;
	%Straight Lines [id:da9830619134141065] 
	\draw  [dash pattern={on 4.5pt off 4.5pt}]  (49.5,189.5) -- (228.75,189.5) ;
	%Straight Lines [id:da6227652572342003] 
	\draw  [dash pattern={on 4.5pt off 4.5pt}]  (130.75,49) -- (130.75,219) ;
	%Straight Lines [id:da25450346888286024] 
	\draw  [dash pattern={on 4.5pt off 4.5pt}]  (51,140.5) -- (230.25,140.5) ;
	%Shape: Brace [id:dp37610176307681176] 
	\draw   (50,220) .. controls (50,224.15) and (52.08,226.23) .. (56.23,226.23) -- (56.23,226.23) .. controls (62.16,226.23) and (65.13,228.31) .. (65.13,232.46) .. controls (65.13,228.31) and (68.09,226.23) .. (74.02,226.23)(71.35,226.23) -- (74.02,226.23) .. controls (78.17,226.23) and (80.25,224.15) .. (80.25,220) ;
	%Shape: Brace [id:dp38762810724659635] 
	\draw   (81.5,219.5) .. controls (81.5,224.17) and (83.83,226.5) .. (88.5,226.5) -- (96.13,226.5) .. controls (102.8,226.5) and (106.13,228.83) .. (106.13,233.5) .. controls (106.13,228.83) and (109.46,226.5) .. (116.13,226.5)(113.13,226.5) -- (123.75,226.5) .. controls (128.42,226.5) and (130.75,224.17) .. (130.75,219.5) ;
	%Straight Lines [id:da0748207367158864] 
	\draw  [dash pattern={on 4.5pt off 4.5pt}]  (50.5,80) -- (229.75,80) ;
	%Straight Lines [id:da030840461130633212] 
	\draw  [dash pattern={on 4.5pt off 4.5pt}]  (200.25,49.5) -- (200.25,219.5) ;
	%Shape: Brace [id:dp6823964147103206] 
	\draw   (200,220.5) .. controls (200,224.65) and (202.08,226.73) .. (206.23,226.73) -- (206.23,226.73) .. controls (212.16,226.73) and (215.13,228.81) .. (215.13,232.96) .. controls (215.13,228.81) and (218.09,226.73) .. (224.02,226.73)(221.35,226.73) -- (224.02,226.73) .. controls (228.17,226.73) and (230.25,224.65) .. (230.25,220.5) ;
	%Shape: Brace [id:dp037081563802510265] 
	\draw   (50,189.5) .. controls (45.95,189.5) and (43.93,191.52) .. (43.93,195.57) -- (43.93,195.57) .. controls (43.93,201.36) and (41.9,204.25) .. (37.85,204.25) .. controls (41.9,204.25) and (43.93,207.14) .. (43.93,212.93)(43.93,210.32) -- (43.93,212.93) .. controls (43.93,216.98) and (45.95,219) .. (50,219) ;
	%Shape: Brace [id:dp18777067390046231] 
	\draw   (49.5,50) .. controls (45.45,50) and (43.43,52.02) .. (43.43,56.07) -- (43.43,56.07) .. controls (43.43,61.86) and (41.4,64.75) .. (37.35,64.75) .. controls (41.4,64.75) and (43.43,67.64) .. (43.43,73.43)(43.43,70.82) -- (43.43,73.43) .. controls (43.43,77.48) and (45.45,79.5) .. (49.5,79.5) ;
	%Shape: Brace [id:dp39497205765266274] 
	\draw   (50,141) .. controls (45.33,141) and (43,143.33) .. (43,148) -- (43,155) .. controls (43,161.67) and (40.67,165) .. (36,165) .. controls (40.67,165) and (43,168.33) .. (43,175)(43,172) -- (43,182) .. controls (43,186.67) and (45.33,189) .. (50,189) ;
	%Shape: Circle [id:dp2630824921413869] 
	\draw  [draw opacity=0][fill={rgb, 255:red, 0; green, 0; blue, 0 }  ,fill opacity=1 ] (46.38,219.5) .. controls (46.38,217.5) and (48,215.88) .. (50,215.88) .. controls (52,215.88) and (53.63,217.5) .. (53.63,219.5) .. controls (53.63,221.5) and (52,223.13) .. (50,223.13) .. controls (48,223.13) and (46.38,221.5) .. (46.38,219.5) -- cycle ;
	%Shape: Circle [id:dp5161218843187431] 
	\draw  [draw opacity=0][fill={rgb, 255:red, 0; green, 0; blue, 0 }  ,fill opacity=1 ] (227.13,49.5) .. controls (227.13,47.5) and (228.75,45.88) .. (230.75,45.88) .. controls (232.75,45.88) and (234.38,47.5) .. (234.38,49.5) .. controls (234.38,51.5) and (232.75,53.13) .. (230.75,53.13) .. controls (228.75,53.13) and (227.13,51.5) .. (227.13,49.5) -- cycle ;
	%Shape: Circle [id:dp12372702715793926] 
	\draw  [draw opacity=0][fill={rgb, 255:red, 0; green, 0; blue, 0 }  ,fill opacity=1 ] (227.13,219.5) .. controls (227.13,217.5) and (228.75,215.88) .. (230.75,215.88) .. controls (232.75,215.88) and (234.38,217.5) .. (234.38,219.5) .. controls (234.38,221.5) and (232.75,223.13) .. (230.75,223.13) .. controls (228.75,223.13) and (227.13,221.5) .. (227.13,219.5) -- cycle ;
	%Shape: Circle [id:dp5577541821150872] 
	\draw  [draw opacity=0][fill={rgb, 255:red, 0; green, 0; blue, 0 }  ,fill opacity=1 ] (46.88,49.5) .. controls (46.88,47.5) and (48.5,45.88) .. (50.5,45.88) .. controls (52.5,45.88) and (54.13,47.5) .. (54.13,49.5) .. controls (54.13,51.5) and (52.5,53.13) .. (50.5,53.13) .. controls (48.5,53.13) and (46.88,51.5) .. (46.88,49.5) -- cycle ;
	
	% Text Node
	\draw (48,46.5) node [anchor=south east] [inner sep=0.75pt]  [font=\footnotesize] [align=left] {$\displaystyle ( 1,0)$};
	% Text Node
	\draw (48,222) node [anchor=north east] [inner sep=0.75pt]  [font=\footnotesize] [align=left] {$\displaystyle ( 0,0)$};
	% Text Node
	\draw (233.5,224.5) node [anchor=north west][inner sep=0.75pt]  [font=\footnotesize] [align=left] {$\displaystyle ( 0,1)$};
	% Text Node
	\draw (65.7,241) node  [font=\footnotesize] [align=left] {$\displaystyle \pi _{1}$};
	% Text Node
	\draw (107.2,241.5) node  [font=\footnotesize] [align=left] {$\displaystyle \pi _{2}$};
	% Text Node
	\draw (232.75,47.13) node [anchor=south west] [inner sep=0.75pt]  [font=\footnotesize] [align=left] {$\displaystyle ( 1,1)$};
	% Text Node
	\draw (215.26,241.5) node  [font=\footnotesize] [align=left] {$\displaystyle \pi _{B}$};
	% Text Node
	\draw (20.7,164) node  [font=\footnotesize] [align=left] {$\displaystyle \pi _{2}$};
	% Text Node
	\draw (20.7,204.5) node  [font=\footnotesize] [align=left] {$\displaystyle \pi _{1}$};
	% Text Node
	\draw (20.26,61.5) node  [font=\footnotesize] [align=left] {$\displaystyle \pi _{B}$};
	% Text Node
	\draw (65.87,203.5) node  [font=\footnotesize] [align=left] {$\displaystyle p_{1,1}$};
	% Text Node
	\draw (107.37,204) node  [font=\footnotesize] [align=left] {$\displaystyle p_{1,2}$};
	% Text Node
	\draw (66.37,164.5) node  [font=\footnotesize] [align=left] {$\displaystyle p_{1,2}$};
	% Text Node
	\draw (106.87,165) node  [font=\footnotesize] [align=left] {$\displaystyle p_{2,2}$};
	% Text Node
	\draw (216.98,64) node  [font=\footnotesize] [align=left] {$\displaystyle p_{B,B}$};
	% Text Node
	\draw (66.92,64.5) node  [font=\footnotesize] [align=left] {$\displaystyle p_{1,B}$};
	% Text Node
	\draw (216.42,204.5) node  [font=\footnotesize] [align=left] {$\displaystyle p_{1,B}$};
	% Text Node
	\draw (105.92,64.5) node  [font=\footnotesize] [align=left] {$\displaystyle p_{2,B}$};
	% Text Node
	\draw (216.92,164.5) node  [font=\footnotesize] [align=left] {$\displaystyle p_{2,B}$};
	% Text Node
	\draw (165.59,204.5) node   [align=left] {$\displaystyle \cdots $};
	% Text Node
	\draw (166.09,64.5) node   [align=left] {$\displaystyle \cdots $};
	% Text Node
	\draw (165.59,164.5) node   [align=left] {$\displaystyle \cdots $};
	% Text Node
	\draw (66,110.59) node  [rotate=-90] [align=left] {$\displaystyle \cdots $};
	% Text Node
	\draw (216.5,110.09) node  [rotate=-90] [align=left] {$\displaystyle \cdots $};
	% Text Node
	\draw (105.5,110.09) node  [rotate=-90] [align=left] {$\displaystyle \cdots $};

\end{tikzpicture}
	\vspace{-10mm}
	\caption{The graphon $f$ of a stochastic block model with $B$ blocks. $f$ is a step-function with $B^2$ steps and is of rank $B$.}\label{figure--sbm_graphon}
\end{figure}	 

We can now describe the asymptotic distribution of the estimators:
\begin{theorem}\label{theorem--distribution_obs_error}
	Suppose the consistency conditions given in Theorem \ref{theorem--consistency--error} hold. Then, for $\hat{\beta}^{(d)}$, $d \in \{1, T, \infty\}$, we have that:
	%Assumptions \ref{assumption--graphon}, \ref{assumption--measurement_model}, \ref{assumption--varep_condmoments} and \ref{assumption--no_collinearity} hold. 
	\begin{equation*}
		\frac{\sqrt{n}\left(\hat{\beta}^{(d)}- \beta^{(d)}\cdot\left(1-B^{(d)}\right)\right)}{\sqrt{V_{11}^{(d)}}} \overset{d}{\to} \text{N}(0,1)~,
	\end{equation*}
	where $V^{(d)}$ is as defined in Theorem \ref{theorem--benchmark}. Moreover,  
	\begin{gather*}
		B^{(1)} = O_p\left(\frac{1}{np_n}\right) \quad \mbox{ , } \quad  B^{(T)} = O_p\left(\frac{1}{np_n}\right)
	\end{gather*}
	and 
	\begin{equation*}
		B^{(\infty)} =O_p\left(\frac{1}{np_n}\right) \quad \mbox{ under Assumption \ref{assumption--rank_r}}~.
	\end{equation*}
\end{theorem}

Under the conditions for consistency, observation error does not affect the asymptotic variance of the OLS estimator. However, it gives rise to an asymptotic bias that is of order $1/np_n$ when $\beta^{(d)} \neq 0$. Equivalently, $\hat{\beta}^{(d)}$ has an ``attenuation" factor of $\frac{1}{1-B^{(d)}}$, where:
\begin{equation*}
	%\hat{\beta}^{(d)} \approx \beta^{(d)} \cdot 
	1-B^{(d)} = \frac{(\hat{\mathbf{C}}^{(d)})'\mathbf{M}_W{\mathbf{C}^{(d)}}}{(\hat{\mathbf{C}}^{(d)})'\mathbf{M}_W{\mathbf{\hat{C}}^{(d)}}} = \frac{\frac{1}{n}(\hat{\mathbf{C}}^{(d)})'\mathbf{M}_W\hat{\mathbf{C}}^{(d)} - (B_1^{(d)} - B_2^{(d)})}{\frac{1}{n}(\hat{\mathbf{C}}^{(d)})'\mathbf{M}_W\hat{\mathbf{C}}^{(d)}}
\end{equation*}
where
\begin{equation}\label{equation--bias-expression}
	\begin{aligned}
		B_1^{(d)} = \frac{1}{n} (\hat{\mathbf{C}}^{(d)})'(\hat{\mathbf{C}}^{(d)} - \mathbf{C}^{(d)}) \quad \mbox{ , } \quad 
		B_2^{(d)} = (\hat{\pi}^{(d)})'  \frac{1}{n} \mathbf{W}'(\hat{\mathbf{C}}^{(d)} - \mathbf{C}^{(d)})
	\end{aligned}
\end{equation}
and $\hat{\pi}^{(d)}$ are the OLS estimates from the regression of $\hat{\mathbf{C}}^{(d)}$ on $\mathbf{W}$.

The above display shows that there are two sources of bias. The first is the attenuation bias term, $B_1^{(d)}$, which is approximately $\var[\hat{C}^{(d)}_i] - \var[C^{(d)}_i] > 0$. In our setup, the variance inflation vanishes asymptotically, but potentially too slowly for inference. The second source of bias, $B_2^{(d)}$, arises only in the presence of covariates. Under observation error, it is possible that $\cov[W_{i}, \hat{C}_i] \neq \cov[W_{i}, C_i]$, and as a result, variation due to $W_i$ can be misattributed to $\hat{C}_i$ and vice versa. This happens when the entry-wise bias of centrality statistics, $\expect[\hat{C}^{(d)}_i - C^{(d)}_i \mid U_i]$, is correlated with $C_i^{(d)}$ and hence $W_i$. $B_2^{(d)}$ can be positive or negative depending on the relationship between $W_i$ and $U_i$. Note that this second channel is absent for degree centrality since $\hat{C}_i^{(1)}$ is unbiased for $C_i^{(1)}$.

Asymptotic bias is exacerbated when the networks are sparse. When $p_n \lesssim 1/\sqrt{n}$, bias is of the same or larger order compared to the variance ($\approx 1/\sqrt{n}$). The plug-in $t$-statistic is therefore incorrectly centered when $\beta^{(d)} \neq 0$, rendering the $t$-test and the associated confidence intervals invalid. When the asymptotic bias dominates the variance, $\hat{\beta}^{(d)}$ does not have a non-degenerate asymptotic distribution. That is, there are no scaling factors $\alpha, \beta$ for which $n^{\alpha}p_n^{\beta}(\hat{\beta}^{(d)}-\beta^{(d)})$ converges to a distribution with non-zero variance. Valid inference therefore requires alternative procedures, which we develop in the next section. We note an important exception to the above discussion: the plug-in $t$-test is valid for the null hypothesis $H_0: {\beta}^{(d)} = 0$. 

Theorem \ref{theorem--distribution_obs_error} provides upper bounds on the rates at which the asymptotic bias goes to $0$. The next result shows that these rates are attained by a homogeneous network with no covariates. The implication is that practitioners should be concerned about asymptotic bias even for relatively dense networks $(p_n \approx 1/\sqrt{n})$. 

\begin{corollary} \label{theorem--bias_magnitude}
	Suppose $f=1$, there are no other covariates and that the consistency conditions in Theorem \ref{theorem--consistency--error} hold. Then,
		\begin{equation*}
			B^{(1)} = \Theta_p\left(\frac{1}{np_n}\right) \quad \mbox{ and } \quad B^{(T)} = \Theta_p\left(\frac{1}{np_n}\right)~.
		\end{equation*}
	Under Assumption \ref{assumption--rank_r},  
	\begin{equation*}
		B^{(\infty)} = \Theta_p\left( \frac{1}{np_n}\right) \quad \mbox{ if } \quad p_n \gg n^{-1} \frac{\log n}{\log \log n}~.
	\end{equation*}
\end{corollary}

Degree and diffusion attain the stated bounds across the full range of consistency. For eigenvector centrality, we require slightly more density. %A naive application of Davis-Kahan inequality yields an upper bound of order $1/\sqrt{np_n}$ for general graphons. Given \ref{assumption--rank_r}, we are able to obtain an approximation of $v_1(\hat{A})$ with a leading term of order $1/np_n$ and a residual that is bounded by $(\lVert \boldsymbol{\xi} \rVert / np_n)^3$. The leading term dominates as long as $np_n \gg \frac{\log n}{\log \log n}$, yielding the rate above. 
It is possible to extend our arguments into slightly sparser regimes but this would not change the conclusion of the present exercise.

\begin{remark}\label{remark--compare_leli}
\cite{le2020linear} considers inference on $\lVert \beta^{(\infty)}C^{(\infty)} \rVert^2 = 0$ when $p_n \gg n^{-1/2}$, under which asymptotic bias is negligible. We establish the consistency thresholds for eigenvector regressions and derive sharp rates on the asymptotic bias. Together with the approach in the next section, our method facilitates inference on $\beta^{(d)}$ up to the consistency threshold $p_n \gg n^{-1}$. 
\end{remark}

\begin{remark}\label{remark--compare_caietal}
\cite{cai2021network} also considers inference on $\beta^{(\infty)}$ and requires $p_n \gg n^{-1}$. Their setting features observation errors which are additive and i.i.d. Gaussian. 
\end{remark}

\section{Bias Correction}\label{section--bias-correction}

This section proposes strategies for addressing the asymptotic bias of $\hat{\beta}^{(d)}$. Section \ref{section--bias-correction--analytical} considers analytical bias correction for degree and diffusion. Section \ref{section--bias_aware} proposes a conservative form of bias-aware inference for eigenvector centrality.

\subsection{Analytical Bias Correction for Degree and Diffusion}\label{section--bias-correction--analytical}

This section develops analytical bias corrections for the degree and diffusion regressions. We first express the bias terms as sums of terms we call mixed products, defined below. We then show that the asymptotically non-negligible components of mixed products can be expressed as counts of particular trees. This yields plug-in estimators for mixed products that can be combined to construct estimators for $B_1^{(d)}$ and $B_2^{(d)}$.

\begin{definition}[Mixed Product of Order $(t,\tau)$]\label{definition--B}
A mixed product of order $(t,\tau)$ is a term of the form
$
	B = \prod_{j=1}^{t} B_j
$
where $B_j \in \left\{A , \boldsymbol{\xi} \right\}$ and $B_j = \boldsymbol{\xi}$ for exactly $\tau$ number of $j$'s. Let $\mathcal{B}(t,\tau)$ denote the set of all mixed products of order $(t,\tau)$. Let $p = (p_1, ..., p_r)'$ record lengths of the contiguous blocks of $A$ and $\boldsymbol{\xi}$ in $B$. Let $q = (q_1, ..., q_r)$ be an indicator that turns on when the corresponding block is one of $\boldsymbol{\xi}$'s. The dependence of $p$ and $q$ on $B$ is suppressed for convenience. 
\end{definition}

\begin{example} 
	Suppose $B = A^2\boldsymbol{\xi}^3 A \boldsymbol{\xi}^2 A$. Then it is a mixed product of order $(9, 5)$, with $p = (2,3,1,2,1)$ and $q = (0,1,0,1,0)$. 
\end{example}

For degree and diffusion, $B_1^{(d)}$ can be written as:
\begin{align}
	B_1^{(d)} & 
	= \frac{1}{n} \cdot a_n^{(d)} \iota_n' \left(\sum_{t=1}^T \delta_n^{t-1}\left(A + \boldsymbol{\xi}\right)^t\right)'\left(\sum_{t=1}^T \delta_n^{t-1}\left(\left(A + \boldsymbol{\xi}\right)^t - A^t\right)\right) \iota_n a_n^{(d)} \nonumber \\
	& = \frac{1}{n} \cdot \left(a_n^{(d)} \right)^2 \sum_{t=2}^{2T} \sum_{\tau \geq 1} \sum_{B \in \mathcal{\tilde{B}}(t,\tau)} \delta_n^{t-2} \iota_n'B\iota_n~. \label{equation--B_1d--wBv}
\end{align}
The final sum is over $\tilde{\mathcal{B}}(t,\tau)$, which is defined to be the multiset obtained by collecting terms of the same order. This set is straightforward to enumerate.

The same reasoning applies to $B_2^{(d)}$:
\begin{align}
	B_2^{(d)} &= \frac{1}{n} \sum_{j} \hat{\pi}_j^{(d)}  \mathbf{W}_{\cdot, j}'\left(\sum_{t=1}^T \delta_n^{t-1}\left(\left(A + \boldsymbol{\xi}\right)^t - A^t\right)\right)a_n^{(d)} \nonumber \\
	& = \frac{1}{n} \cdot a_n^{(d)} \sum_{j} \hat{\pi}_j^{(d)} \sum_{t=1}^T \sum_{\tau \geq 1} \sum_{B \in \mathcal{B}(t,\tau)}  \delta_n^{t-1} \mathbf{W}_{\cdot, j}' B \iota_n~. \label{equation--B_2d--wBv}
\end{align}
Here, $\mathcal{B}(t,\tau)$ is the full set of mixed products as in Definition \ref{definition--B} and $\mathbf{W}_{\cdot, j}$ refers to the $j$-th column of $\mathbf{W}$. We have seen that $B_1^{(d)}$ and $B_2^{(d)}$ are finite sums of terms of the form $w'Bv/n^{t+1}p_n^t$. It is therefore sufficient to find good estimators for such terms.

It turns out that the asymptotically non-negligible components of the mixed product $B$ have a tree structure. Algorithm \ref{alg:construct_tree} describes how the relevant trees can be constructed for a mixed product of order $(t, \tau)$.  Figure \ref{figure--contruct_trees} provides an example. Intuitively, the algorithm replaces contiguous blocks of $A$ with path graphs and contiguous blocks of $\boldsymbol{\xi}$ with appropriate trees from the set $\mathcal{G}(p_s)$. We keep track of the number of each tree using the function $\gamma: \mathcal{G}(p_s) \mapsto \mathbb{N}$. To describe the construction of $\mathcal{G}(p_s)$ and $\gamma(g)$, we introduce the following definitions. 

\begin{algorithm}
\caption{Tree Construction}\label{alg:construct_tree}
 Define $g_0$ to be the tree with 1 node and no edges. Label this node as both the initial and terminal node for $g_0$. Initialize $\mathcal{T}(B) = \{(g_0, 1)\}$

\For{$s=1$ \KwTo $r$}{
	\If{$q_s = 0$ }{
		\ForEach{$(g,k) \in \mathcal{T}(B)$}{
			Remove $(g,k)$ from $\mathcal{T}(B)$. 
			
			Construct $g_\text{new}$ by appending a path of length $p_s$ to the terminal node of $g$. The terminal node of $g_\text{new}$ is the terminal node of the path. 
			
			Add $(g_\text{new}, k)$ to $\mathcal{T}(B)$.
		}
	}
	\Else{
		\ForEach{$(g,k) \in \mathcal{T}(B)$}{
			Remove $(g,k)$ from $\mathcal{T}(B)$. 
			
			\ForEach{$g_\xi \in \mathcal{G}(p_s)$}{
				Construct $g_\text{new}$ by appending the tree $g_\xi$ to the terminal node of $g$. The terminal node of $g_\text{new}$ is the terminal node of $g_\xi$.

				Update $k_\text{new}$ = $k \times \gamma(g_\xi)$.
				
				Add $(g_\text{new}, k_\text{new})$ to $\mathcal{T}(B)$.
			}
		}
	}
}
\Return $\mathcal{T}(B)$.
\end{algorithm}

\begin{definition}\label{definition--induced_graphs}
	For indices $\mathbf{i} = (i_1,...,i_{t+1})$, let $g(\mathbf{i})$ be the labelled (undirected) multigraph obtained by
	the walk $i_1 \to i_2 \to ... \to i_{t+1}$. To be clear, this is a walk on the complete graph with the node set $[n]$ and no self-loops. We label the nodes based on when they are first reached.  For a given $g(\mathbf{i})$, let $g^{\text{simp}}(\mathbf{i})$ be the simple graph obtained from $g(\mathbf{i})$ by deleting all but one edge from every parallel class. Finally, let $V(\mathbf{i})$ be the set of unique indices and let $ E(\mathbf{i})$ be the set of unique undirected edges. 
	See Figure \ref{figure--g_simp_example} for an example. 
\end{definition}

Note that the edge set of $g^{\text{simp}}(\mathbf{i})$, denoted $E(\mathbf{i})$, is the set of unique edges in $g(\mathbf{i})$. Similarly, $V(\mathbf{i})$ is the set of unique nodes in $g(\mathbf{i})$. Recall that a tree is a connected, acyclic graph. Now, let $\mathcal{I}$ be the set of walks of length $s$ on the complete graph on $[s+1]$, with ter\underline{m}inal node $i_{s+1}$ having label $m$. From $\mathcal{I}$, delete every $\mathbf{i}$ for which $g(\mathbf{i})$ has at least one edge with multiplicity exactly equal to 1. Next, delete every $\mathbf{i}$ for which $g^{\text{simp}}(\mathbf{i})$ is not a tree (i.e. contains cycles), denote this final set $\tilde{\mathcal{I}}$. Let $\mathcal{G}(s,m)$ be the set of homorphism classes of $\{g^{\text{simp}}(\mathbf{i}) : \mathbf{i} \in \tilde{\mathcal{I}}\}$. Node labels are identical for all graphs in the same homomorphism class. We define the representative element of each class to be the graph for which node identity is equal to the node label. Let $\gamma(g)$ record the number of homorphisms for each $g \in \mathcal{G}(s,m)$. This set is straightforward to enumerate for $s$ of moderate sizes. Appendix \ref{appendix--xi_trees} presents $\mathcal{G}(s,m)$ for $s \leq 10$. Finally, we define $\mathcal{G}(p_s) = \bigcup_{m} \mathcal{G}(p_s,m)$.

Using Algorithm \ref{alg:construct_tree}, we obtain $\mathcal{T}(B)$, defined to be the set of pairs $(g, k)$. With some abuse of notation, we will also treat $\mathcal{T}(B)$ as the set of trees $\{g: (g,k) \in \mathcal{T}(B)\}$ and 
 $k_B(g) = k \mbox{ for } (g,k) \in \mathcal{T}(B)$ as the function that keeps track of the multiplicity of $g$. Let $E(g)$ denote the edge set of $g$. Recall that $n(g)$ is the number of nodes in $g$. Then, the asymptotically non-negligible components of $w'Bv$ are
 \begin{equation*}
 	{Q}(B,w,v) = \sum_{g \in \mathcal{T}(B)} k_B(g) \cdot \sum_{i_1, ..., i_{n(g)}} w_{i_1} \left(\prod_{jk \in E(g)} {A}_{i_j, i_k} \right)v_{i_m}~.
 \end{equation*}
Here, $m = m(g)$ is the label on the terminal node of $g$. Note that the index of summation should not repeat i.e. $i_j \neq i_k$ for all $j \neq k$. See the caption of Figure \ref{figure--contruct_trees} for an example. 

\begin{figure}
	\centering
	\include{figures/construct_trees}
	\vspace{-10mm}
	\caption{\small This figure illustrates the process for constructing the trees for $B = A\boldsymbol{\xi}^4A^2$. Here, $p = (1,4,2)$, $q = (0,1,0)$. Since $\gamma(g) = 1$ for all $g \in \mathcal{G}(4)$, each tree occurs exactly once ($k_B(g)=1$). $\mathcal{G}(4)$ is obtained from Table \ref{tab:xi_trees_1} in Appendix \ref{appendix--xi_trees}. Across iterations, it is important to keep track of the initial and terminal nodes, labelled 1 and $m$ respectively. The labels on the other nodes are interchangeable. Using the constructed trees, the non-negligible terms of $w'Bv$ are:
	\vspace{-1mm}
	\begin{align*}
		{Q}(B,w, v) & = \sum_{i_1, ..., i_5} w_{i_1} {A}_{i_1,i_2}{A}_{i_2,i_3}{A}_{i_2,i_4}{A}_{i_4,i_5} v_{i_5} \\
		& \qquad + \sum_{i_1, ..., i_6} w_{i_1} {A}_{i_1,i_2}{A}_{i_2,i_3}{A}_{i_2,i_4}{A}_{i_2,i_5} {A}_{i_5,i_6} v_{i_6} \\
		& \qquad + \sum_{i_1, ..., i_6} w_{i_1} {A}_{i_1,i_2}{A}_{i_2,i_3}{A}_{i_3,i_4}{A}_{i_2,i_5} {A}_{i_5,i_6} v_{i_6}~.
	\end{align*}
	%\vspace{-2mm}
	Note that the indices should not repeat i.e. $i_j \neq i_k$ for all $j \neq k$. %Appendix \ref{appendix--speed_up_B} describes a faster procedure based on sampling indices. 
	}\label{figure--contruct_trees}
\end{figure}

\begin{figure}
	\centering
	\tikzset{every picture/.style={line width=0.75pt}} %set default line width to 0.75pt        

\begin{tikzpicture}[x=0.75pt,y=0.75pt,yscale=-1,xscale=1]
%uncomment if require: \path (0,372); %set diagram left start at 0, and has height of 372

%Straight Lines [id:da5596851414839269] 
\draw [color={rgb, 255:red, 37; green, 49; blue, 217 }  ,draw opacity=1 ][line width=1.5]    (40,80) -- (130,80) ;
\draw [shift={(91.8,80)}, rotate = 180] [fill={rgb, 255:red, 37; green, 49; blue, 217 }  ,fill opacity=1 ][line width=0.08]  [draw opacity=0] (11.61,-5.58) -- (0,0) -- (11.61,5.58) -- cycle    ;
%Straight Lines [id:da5512057861086972] 
\draw [color={rgb, 255:red, 37; green, 49; blue, 217 }  ,draw opacity=1 ][line width=1.5]    (130,80) -- (130,165) ;
\draw [shift={(130,129.3)}, rotate = 270] [fill={rgb, 255:red, 37; green, 49; blue, 217 }  ,fill opacity=1 ][line width=0.08]  [draw opacity=0] (11.61,-5.58) -- (0,0) -- (11.61,5.58) -- cycle    ;
%Curve Lines [id:da7195505357062547] 
\draw [color={rgb, 255:red, 37; green, 49; blue, 217 }  ,draw opacity=1 ][line width=1.5]    (410,80) .. controls (410.5,79.75) and (500.5,79.25) .. (500,80) ;
%Curve Lines [id:da2424945078763847] 
\draw [color={rgb, 255:red, 37; green, 49; blue, 217 }  ,draw opacity=1 ][line width=1.5]    (130,80) .. controls (130.5,39.25) and (220.5,39.75) .. (220,80) ;
\draw [shift={(182.1,49.95)}, rotate = 183] [fill={rgb, 255:red, 37; green, 49; blue, 217 }  ,fill opacity=1 ][line width=0.08]  [draw opacity=0] (11.61,-5.58) -- (0,0) -- (11.61,5.58) -- cycle    ;
%Curve Lines [id:da14686325663534716] 
\draw [color={rgb, 255:red, 37; green, 49; blue, 217 }  ,draw opacity=1 ][line width=1.5]    (220,80) .. controls (220.5,121.25) and (130.5,120.25) .. (130,80) ;
\draw [shift={(167.98,110.15)}, rotate = 3.7] [fill={rgb, 255:red, 37; green, 49; blue, 217 }  ,fill opacity=1 ][line width=0.08]  [draw opacity=0] (11.61,-5.58) -- (0,0) -- (11.61,5.58) -- cycle    ;
%Shape: Ellipse [id:dp3349383053491227] 
\draw  [fill={rgb, 255:red, 0; green, 0; blue, 0 }  ,fill opacity=1 ] (35,80) .. controls (35,77.24) and (37.24,75) .. (40,75) .. controls (42.76,75) and (45,77.24) .. (45,80) .. controls (45,82.76) and (42.76,85) .. (40,85) .. controls (37.24,85) and (35,82.76) .. (35,80) -- cycle ;
%Shape: Ellipse [id:dp5691841684773952] 
\draw  [fill={rgb, 255:red, 0; green, 0; blue, 0 }  ,fill opacity=1 ] (125,80) .. controls (125,77.24) and (127.24,75) .. (130,75) .. controls (132.76,75) and (135,77.24) .. (135,80) .. controls (135,82.76) and (132.76,85) .. (130,85) .. controls (127.24,85) and (125,82.76) .. (125,80) -- cycle ;
%Shape: Ellipse [id:dp1058592189241696] 
\draw  [fill={rgb, 255:red, 0; green, 0; blue, 0 }  ,fill opacity=1 ] (215,80) .. controls (215,77.24) and (217.24,75) .. (220,75) .. controls (222.76,75) and (225,77.24) .. (225,80) .. controls (225,82.76) and (222.76,85) .. (220,85) .. controls (217.24,85) and (215,82.76) .. (215,80) -- cycle ;
%Shape: Ellipse [id:dp4350789607916453] 
\draw  [fill={rgb, 255:red, 0; green, 0; blue, 0 }  ,fill opacity=1 ] (125,165) .. controls (125,162.24) and (127.24,160) .. (130,160) .. controls (132.76,160) and (135,162.24) .. (135,165) .. controls (135,167.76) and (132.76,170) .. (130,170) .. controls (127.24,170) and (125,167.76) .. (125,165) -- cycle ;
%Curve Lines [id:da8147824877335997] 
\draw [color={rgb, 255:red, 37; green, 49; blue, 217 }  ,draw opacity=1 ][line width=1.5]    (320,80) .. controls (320.5,79.75) and (410.5,79.25) .. (410,80) ;
%Curve Lines [id:da2752314137771378] 
\draw [color={rgb, 255:red, 37; green, 49; blue, 217 }  ,draw opacity=1 ][line width=1.5]    (410,80) .. controls (410,79.75) and (411,164.75) .. (410,165) ;
%Shape: Ellipse [id:dp5052636176551695] 
\draw  [fill={rgb, 255:red, 0; green, 0; blue, 0 }  ,fill opacity=1 ] (315,80) .. controls (315,77.24) and (317.24,75) .. (320,75) .. controls (322.76,75) and (325,77.24) .. (325,80) .. controls (325,82.76) and (322.76,85) .. (320,85) .. controls (317.24,85) and (315,82.76) .. (315,80) -- cycle ;
%Shape: Ellipse [id:dp7827034635266414] 
\draw  [fill={rgb, 255:red, 0; green, 0; blue, 0 }  ,fill opacity=1 ] (405,80) .. controls (405,77.24) and (407.24,75) .. (410,75) .. controls (412.76,75) and (415,77.24) .. (415,80) .. controls (415,82.76) and (412.76,85) .. (410,85) .. controls (407.24,85) and (405,82.76) .. (405,80) -- cycle ;
%Shape: Ellipse [id:dp7278043570130968] 
\draw  [fill={rgb, 255:red, 0; green, 0; blue, 0 }  ,fill opacity=1 ] (495,80) .. controls (495,77.24) and (497.24,75) .. (500,75) .. controls (502.76,75) and (505,77.24) .. (505,80) .. controls (505,82.76) and (502.76,85) .. (500,85) .. controls (497.24,85) and (495,82.76) .. (495,80) -- cycle ;
%Shape: Ellipse [id:dp3614657955809881] 
\draw  [fill={rgb, 255:red, 0; green, 0; blue, 0 }  ,fill opacity=1 ] (405,165) .. controls (405,162.24) and (407.24,160) .. (410,160) .. controls (412.76,160) and (415,162.24) .. (415,165) .. controls (415,167.76) and (412.76,170) .. (410,170) .. controls (407.24,170) and (405,167.76) .. (405,165) -- cycle ;

% Text Node
\draw (235,80) node   [align=left] {3};
% Text Node
\draw (146,167) node   [align=left] {4};
% Text Node
\draw (25,80) node   [align=left] {1};
% Text Node
\draw (145,80) node   [align=left] {2};
% Text Node
\draw (130,210) node   [align=left] {$\displaystyle g(\mathbf{i})$};
% Text Node
\draw (515,80) node   [align=left] {3};
% Text Node
\draw (425,164) node   [align=left] {4};
% Text Node
\draw (305,80) node   [align=left] {1};
% Text Node
\draw (425,90) node   [align=left] {2};
% Text Node
\draw (410,210) node   [align=left] {$\displaystyle g^{\text{simp}}(\mathbf{i})$};

\end{tikzpicture}
	\vspace{-10mm}
	\caption{$g(\mathbf{i})$ and $g^{\text{simp}}(\mathbf{i})$ for $\mathbf{i} = (5,7,4,7,10)$. The nodes are labelled based on when they are first reached by the walk $5 \to 7 \to 4 \to 7 \to 10$. Label 1 corresponds to node 5, label 2 corresponds to node 7, and so on. $V(\mathbf{i}) = \{4,5,7,10\}$ and $E(\mathbf{i}) = \{ (4,7), (5,7), (7,10) \}$. Note that $g(\mathbf{i})$ is undirected. The edges are labelled only to show the direction of the initial walk.
	}\label{figure--g_simp_example}
\end{figure}

 This motivates the plug-in estimator, $\check{Q}(B,w,v)$, which is obtained from $Q(B,w,v)$ by replacing $A_{i_j,i_k}$ with $\hat{A}_{i_j, i_k}$. When $B$ is a mixed product for which $t$ is large, computation of $\check{Q}(B,w,v)$ can be expensive. Appendix \ref{appendix--speed_up_B} discusses the computation of $\check{Q}(B,w,v)$ on a random subset of indices. Finally, using $\check{Q}(B,w,v)$, we can construct the following estimators: 
\begin{align*}
	\check{B}_1^{(d)} & = \frac{1}{n} \cdot \left(a_n^{(d)} \right)^2 \sum_{t=2}^{2T} \sum_{\tau \geq 1} \sum_{B \in \mathcal{\tilde{B}}(t,\tau)} \delta_n^{t-2} \check{Q}(B,\iota_n,\iota_n) \\
	\check{B}_2^{(d)} & = \frac{1}{n} \cdot a_n^{(d)} \sum_{j} \hat{\pi}_j^{(d)} \sum_{t=1}^T \sum_{\tau \geq 1} \sum_{B \in \mathcal{B}(t,\tau)}  \delta_n^{t-1} \check{Q}(B,\mathbf{W}_{\cdot, j}, \iota_n)
\end{align*}
In turn, we define:
\begin{equation*}
	1-\check{B}^{(d)} = \frac{\frac{1}{n}(\hat{\mathbf{C}}^{(d)})'\mathbf{M}_W\hat{\mathbf{C}}^{(d)} - (\check{B}_1^{(d)} - \check{B}_2^{(d)})}{\frac{1}{n}(\hat{\mathbf{C}}^{(d)})'\mathbf{M}_W\hat{\mathbf{C}}^{(d)}}
\end{equation*}
and our debiased-OLS estimator to be $\check{\beta}^{(d)} := \frac{\hat{\beta}^{(d)}}{1-\check{B}^{(d)}}$. Then,
\begin{theorem}\label{theorem--distribution_degree_diff_bias_corrected}
	Suppose Assumptions \ref{assumption--graphon}, \ref{assumption--measurement_model}, \ref{assumption--varep_condmoments} and \ref{assumption--no_collinearity} hold and that $p_n \gg n^{-1}$. Then for $d \in \{1,T\}$,
	\begin{equation*}
		\frac{\sqrt{n}\left(\check{\beta}^{(d)}- \beta^{(d)}\right)}{\sqrt{V_{11}^{(d)}}/(1-\check{B}^{(d)})} \overset{d}{\to} \text{N}(0,1)~,
	\end{equation*}
	where $V^{(d)}$ is as defined in Theorem \ref{theorem--benchmark}.
\end{theorem}
Our debiased estimators for degree and diffusion converge to $\beta^{(d)}$ at the same rate as $\tilde{\beta}^{(d)}$. $\check{\beta}^{(d)}$ and $\tilde{\beta}^{(d)}$ have the same asymptotic variance since $\check{B}^{(d)} = O_p(1/np_n)$. 
Using $\check{\beta}^{(d)}$ in place of $\hat{\beta}^{(d)}$ therefore makes the plug-in $t$-test and confidence intervals valid. Although the adjustment in the denominator is not technically necessary, our simulations suggest that it can lead to substantial differences in size control and coverage probability in finite sample, particularly when the network is sparse. As such, we recommend rescaling the variance estimator by the same factor. 

Theorem \ref{theorem--distribution_degree_diff_bias_corrected} holds for any sequence $p_n \gg 1/n$. This requires $\check{B}^{(d)}_1$ and $\check{B}^{(d)}_2$ to have errors that are of strictly smaller order than $1/\sqrt{n}$, even when $p_n$ is close to $n^{-1}$. To achieve this, we need a precise characterization of the asymptotic bias as well as sufficiently accurate estimators, necessitating the tree characterization. In particular, a first order bias correction will not lead to valid inference, since that reduces the asymptotic bias only to $1/n^2p_n^2$, which would dominate $1/\sqrt{n}$ if $p_n \ll n^{-3/4}$. On the other hand, if we are willing to assume that $(np_n)^{-k} \ll 1/\sqrt{n}$ for some $k$, then we would only need to perform bias correction for lower order terms. 

The above approach does not apply to eigenvector centrality. Although the asymptotic bias of $\hat{\beta}^{(\infty)}$ also involves mixed products, they are pre-multiplied by terms of the form:
\begin{equation}\label{equation--eigenvector_nuisance_term}
	\frac{\lambda_r({A})}{\lambda_1(\hat{A})} {v}_1(\hat{A})'v_r(A) - v_1(A)'v_r(A)~.
\end{equation}
We show that this term is $O_p(1/np_n)$ when $f$ is has finite rank, which allows us to form improved bounds in the next section (see discussion after Corollary \ref{corollary--bias_aware_bounds}). However, our characterization is not precise enough for analytical bias correction. 

\subsection{Bias-Aware Type Inference}\label{section--bias_aware}

This section considers inference based on asymptotically conservative bounds for the biases $B_1^{(d)}$ and $B_2^{(d)}$. Let the (unnormalized) average degree and its plug-in estimator be:
\begin{equation*}
	\overline{\text{Deg}} = \frac{1}{n} \sum_{i=1}^n \sum_{j\neq i} A_{ij} \quad \mbox{ and } \quad 
	\widehat{\text{Deg}}   = \frac{1}{n} \sum_{i=1}^n \sum_{j\neq i} \hat{A}_{ij}~.
\end{equation*}
It follows from Theorem 1 in \cite{bickel2011method} that 
\begin{equation*}
	\widehat{\text{Deg}} = \overline{\text{Deg}} + O_p\left(\frac{np_n}{\sqrt{n}}\right) = \Theta_p(np_n)
\end{equation*}
Under the conditions in Theorem \ref{theorem--distribution_obs_error}, we have, for $\eta > 0$, that
$\left(\widehat{\text{Deg}} \right)^{-1+\eta} > |B^{(d)}|$ w.p.a. 1 as $n \to \infty$. This motivates the following $(1-\alpha)$ confidence intervals for $\beta^{(d)}$:
\begin{equation*}
	\text{BA-CI}^{(d)}(1-\alpha, \eta) = \left[ \underline{\beta}^{(d)}- z_{1-\alpha/2}\sqrt{\frac{\hat{V}_{11}^{(d)}}{n}} \; , \;  \overline{\beta}^{(d)}+ z_{1-\alpha/2}\sqrt{\frac{\hat{V}_{11}^{(d)}}{n}}  \right]
\end{equation*}
where 
\begin{equation*}\small
	\overline{\beta}^{(d)} = \max\left\{  \frac{\hat{\beta}^{(d)}}{1- \widehat{\text{Deg}}^{-1+\eta}} ,  \frac{\hat{\beta}^{(d)}}{1+\widehat{\text{Deg}}^{-1+\eta}}  \right\}  \mbox{ ,  \,} 	\underline{\beta}^{(d)} = \min\left\{  \frac{\hat{\beta}^{(d)}}{1- \widehat{\text{Deg}}^{-1+\eta}} ,  \frac{\hat{\beta}^{(d)}}{1+\widehat{\text{Deg}}^{-1+\eta}}  \right\}~.
\end{equation*}
The preceding discussion yields:
\begin{corollary}\label{corollary--bias_aware_bounds}
	Suppose the conditions of Theorem \ref{theorem--distribution_obs_error} hold. Then for $d \in \{1, T\}$ and $\eta >0$, we have that
	\begin{equation*}
		\liminf_{n \to \infty} \prob\left( \beta^{(d)} \in \text{BA-CI}^{(d)}(1-\alpha, \eta) \right) \geq 1-\alpha~.
	\end{equation*}
	If \ref{assumption--rank_r} holds, then the above applies for $d = \infty$ as well. 
\end{corollary}

Relative to the plug-in confidence intervals, we essentially replace the point estimate $\hat{\beta}^{(d)}$ with the interval $\left[\underline{\beta}^{(d)}, \overline{\beta}^{(d)}\right]$, chosen to cover the true parameter even in the presence of bias. Our approach is in the spirit of bias-aware inference (see e.g. \citealt{armstrong2018optimal, armstrong2020simple}). These confidence intervals are adaptive to sparsity in that they do not require the users to specify $p_n$. They are conservative when $(np_n)^{1-\eta} \ll \sqrt{n}$, with asymptotic coverage probability of 1. However, they are exact when $(np_n)^{1-\eta} \gg \sqrt{n}$. To bound the bias of eigenvector centrality using the same rate as degree and diffusion, we need the nuisance term in \eqref{equation--eigenvector_nuisance_term} to be $O_p(1/np_n)$. We show that this is the case under Assumption \ref{assumption--rank_r}. Otherwise, the Davis-Kahan inequality only implies a rate that is $O_p(1/\sqrt{np_n})$, leading to much more conservative bounds. 

Our bounds involve a tuning parameter $\eta$. Simulations in the next section suggest that it is reasonable to set it to 0.1, and we defer a data-driven choice to future work. Finally, we emphasize that while bias correction may be challenging, it is not necessary for valid tests of the null hypothesis $H_0: \beta^{(d)} = 0$.

\section{Simulations}\label{section--simulations}

In this section, we present simulation evidence to support our theory. We generate the network from a 2-block stochastic block model with connection matrix:
\begin{equation*}
	B = \begin{pmatrix}
	1 & 0.25 \\
	0.25 & 0
	\end{pmatrix}
\end{equation*}
We can think of block 1 as the extroverts and group 2 as the introverts. Let $c(i)$ indicate the block membership of $i$. Then $c(i) = 1$ if $U_i \leq 0.5$ and $c(i) = 2$ otherwise. As such, blocks are of equal sizes and $f(U_i, U_j) = B_{c(i), c(j)}$. Note that because of $p_n$ (described below), the actual connection probabilities between any two groups are always strictly below $1$. Next, let
\begin{equation*}
	Y_i = C_i^{(d)} \beta +  W_i \gamma + \varepsilon^{(d)}_i~,
\end{equation*}
where $C^{(d)}_i$ are centrality measures calculated on $A$. We will focus on $d \in \{1,2,3,\infty\}$. That is, degree, diffusion with $T \in \{2,3\}$, and eigenvector centrality. We set $\gamma = 1$ and $\beta = 1$. The covariate is $W_i = \sqrt{12} \cdot U_i$, where $U_i \sim \text{Uniform}[0,1]$ is the latent type. It is scaled to have variance $1$. For comparability, we also set $\delta_* = 1$ and choose $a_*^{(d)}$ so that $\var[C_i^{*,(d)}] =1$. Finally, $\varepsilon^{(d)}_{i} \overset{\text{i.i.d.}}{\sim} \text{N}(0,1)$, where $\varepsilon^{(d)}_i \perp\!\!\!\perp U_i$ and $\varepsilon^{(d)}_i \perp\!\!\!\perp \hat{A}_{jk}$ for all $i,j,k \in [n]$. We vary $n$ and $p_n$ across simulation designs, setting $p_n = k \cdot r(n)$ with $k := 2/\int f du dv$ so that mean degree is 2 in the sparsest regime we consider $(r(n) = n^{-1})$. For brevity, the tables report only the rate component of $p_n$, suppressing the multiplicative constant $k$.   Simulations are based on 2,500 draws for each set of parameter values. 

\subsection{Root Mean-Squared Error}

Table \ref{tab:mse} presents the root mean-squared error (RMSE) of the OLS estimator when $A$ is known ($\tilde{\beta}$), when $\hat{A}$ is used as a plug-in ($\hat{\beta}$) and of our bias-corrected estimator ($\check{\beta}$). As $n$ increases from 250 to 4000, the RMSE of $\tilde{\beta}$ converges quickly to $0$. This holds for all centrality measures and the rate of convergence does not depend on $p_n$, consistent with Theorem \ref{theorem--benchmark}. 

In contrast, the RMSE of $\hat{\beta}$ depends on $p_n$. We first observe that there is a sharp degradation in the RMSE when going from the first three regimes of $p_n$, which are moderately sparse, to the last three regimes, which are extremely sparse. For degree and diffusion, RMSE decreases in $n$ for all regimes except $p_n \propto n^{-1}$, where it converges to a constant. For eigenvector, RMSE decreases in $n$ when $p_n \gg n^{-1}\sqrt{\frac{\log n}{\log \log n}}$, the threshold for consistency, but diverges when $p_n \propto n^{-1}$. At the threshold, the RMSE of eigenvector is slowly decreasing. Based on Theorem \ref{theorem--consistency--error}, we would expect the RMSE to stabilize at a value away from $0$, though this might require larger samples to become apparent. Overall, the behavior of $\hat{\beta}$ is in line with threshold provided in Theorem \ref{theorem--consistency--error}. Comparing RMSE across centrality measures also underscores the fact that eigenvector is less robust to sparsity than degree and diffusion centralities. 

Finally, the column labelled $\check{\beta}$ present the RMSE of our bias corrected estimators for degree and diffusion. 
We speed up computation by computing the bias estimator on a randomly sampled subset of indices, as described in Appendix \ref{appendix--speed_up_B}. To reduce variance, we average over 20 random partitions. As the table shows, bias correction is effective. Even in the moderately sparse regime of $p_n \propto n^{-2/3}$, $\hat{\beta}$ has RMSE that is 1.5 times that of $\check{\beta}^{(d)}$, even when $n = 250$. This ratio is greater than 3 by the time $n = 4000$ and can exceed 10 in sparser regimes. 

\subsection{Confidence Intervals and Coverage Probabilities}

Table \ref{tab:coverage} presents the coverage probabilities of 95\% confidence intervals (CIs) under regimes in which  $\tilde{\beta}$, $\hat{\beta}$ and $\check{\beta}$ are consistent. The column headings indicate the estimator on which the CIs are based. BA refers to the bias-aware CIs described in Section \ref{section--bias_aware}. 

As expected, CIs based on $\tilde{\beta}$ achieve close to 95\% coverage across centrality measures and sparsity regimes, in line with Theorem \ref{theorem--benchmark}.

Across centrality measures, CIs based on $\hat{\beta}$ achieve the nominal coverage probability when $p_n \propto n^{-1/3}$. At $p_n \propto n^{-1/2}$, the coverage probability appears to converge to a value below the nominal level as 
$n$ increases. At $p_n \propto n^{-2/3}$ and below, coverage probability converges to $0$. This behavior is consistent with an asymptotic bias that is $\Theta_p(1/np_n)$, as stated in Corollary \ref{theorem--bias_magnitude}.

We address the asymptotic bias using CIs based on $\check{\beta}$ for degree and diffusion. When $p_n$ is $n^{-2/3}$ or above, our proposed CIs have close to nominal coverage. At lower $p_n$'s, coverage probability can be as low as 70\%. This is because $np_n$ ranges from 4 to 6 in these regimes and our asymptotic results are based on $np_n \to \infty$. 
Nonetheless, they perform substantially better than the plug-in CIs based on $\hat{\beta}$. For eigenvector centrality, we propose to construct CIs by bounding $B^{(d)}$ using the mean degree. Their coverage probabilities when $\eta = 0.1$ are presented in the column denoted BA. These CIs are conservative across the board, but the conservativeness is slowly vanishing in $n$ for $p_n \propto n^{-1/3}$. 

\begin{figure}
\centering
\makebox[\textwidth][c]{
\includegraphics[width=1.1\linewidth]{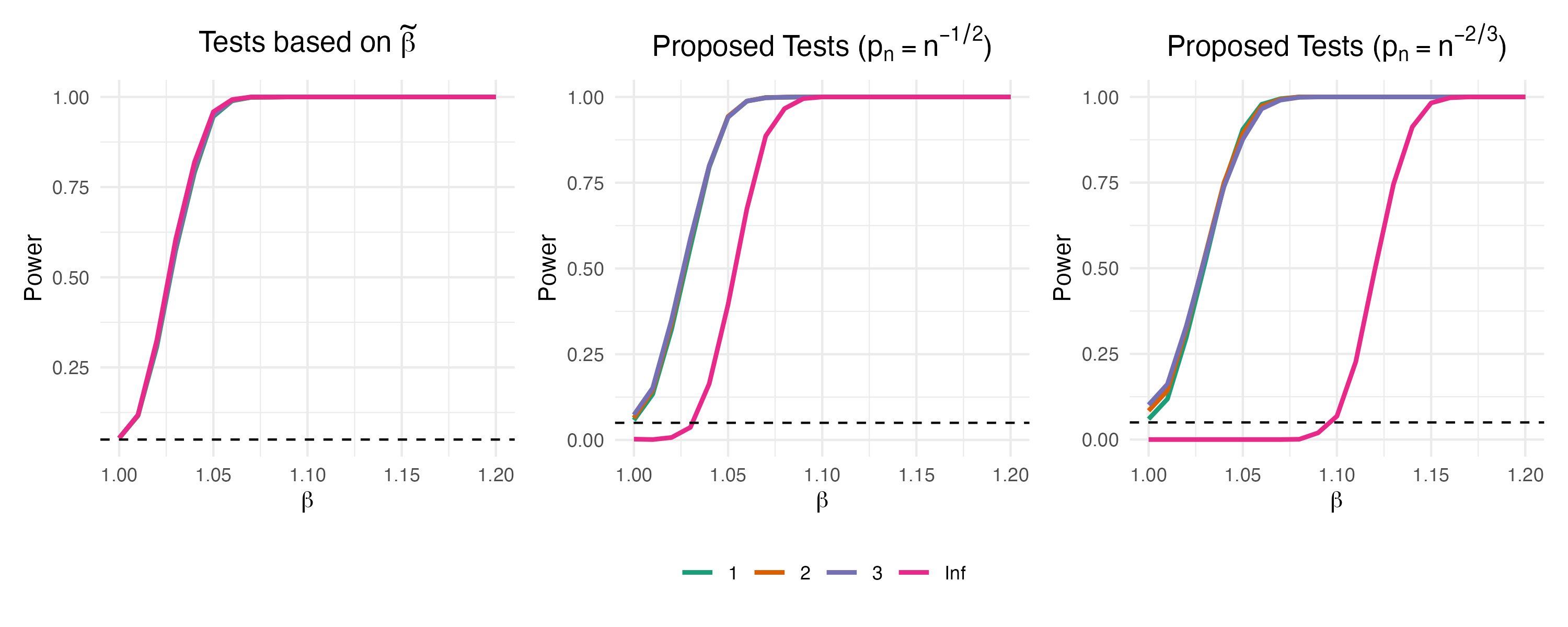}
}
\caption{Power for the two-sided test of $H_0: \beta = 1$ when $n = 2000$. Tests based on $\tilde{\beta}$ presents power of the $t$-test in the absence of observation error, whose power does not depend on $p_n$. For our proposed tests, degree ($d=1$) and diffusion ($d = 2,3$) use the $t$-test with $\check{\beta}$, while eigenvector ($d=\infty$/Inf) uses the test based on the bias-aware CIs.}
\label{fig:powercurves}
\end{figure}

Figure \ref{fig:powercurves} presents power curves for two-sided tests of the hypothesis $H_0: \beta = 1$ when $n=2000$. The first panel presents power of the $t$-tests based on $\tilde{\beta}^{(d)}$. Its power is close to 1 when $\beta = 1.05$. By construction, the power of these tests do not depend on $p_n$. The remaining panels presents the power of our procedures, which are $t$-tests based on $\check{\beta}$ for degree and diffusion, and tests based on the bias-aware confidence intervals for eigenvector. The second panel focuses on the moderately sparse case when $p_n = n^{-1/2}$. We see that tests based on $\check{\beta}$ are competitive with those based on $\tilde{\beta}$. The bias-aware procedure is conservative, but has power 1 once $\beta = 1.1$. In the third panel, $p_n = n^{-2/3}$. Tests based on $\check{\beta}$ continue to do well, even though there is some size distortion when $\beta = 1$. The bias-aware procedure becomes more conservative, since the bounds have to accommodate a bias term that is potentially much bigger. However, its behavior is reasonable, achieving a power of 1 around $\beta = 1.15$. Finally, note that we do not plot the performance of tests based on $\hat{\beta}$ since as Table \ref{tab:coverage} shows, their Type I error can be as high as 20\% when $p_n = n^{-1/2}$ and 90\% when $p_n = n^{-2/3}$.

Overall, our simulations confirm the behavior of $\tilde{\beta}^{(d)}$ and $\hat{\beta}^{(d)}$ as described in Theorem \ref{theorem--benchmark} and \ref{theorem--consistency--error}. Our proposed estimators and confidence intervals perform substantially better than those based on $\hat{\beta}^{(d)}$ and can even be competitive with those based on $\tilde{\beta}^{(d)}$. 

\begin{sidewaystable}
\makebox[\textwidth][c]{
  \centering \footnotesize
    \begin{tabular}{|cc|c|c|ccc|ccc|ccc|cc|}
    \hline
    \multicolumn{2}{|c|}{\multirow{2}[4]{*}{$p_n$}} & \multirow{2}[4]{*}{$n$} & \textbf{Mean} & \multicolumn{3}{c|}{\textbf{Degree ($d=1$)}} & \multicolumn{3}{c|}{\textbf{Diffusion ($d= 2$)}} & \multicolumn{3}{c|}{\textbf{Diffusion ($d = 3$)}} & \multicolumn{2}{c|}{\textbf{Eigen ($d=\infty$)}} \bigstrut\\
\cline{5-15}    \multicolumn{2}{|c|}{} &       & \textbf{Deg} & $\tilde{\beta}$ & $\hat{\beta}$ & $\check{\beta}$ & $\tilde{\beta}$ & $\hat{\beta}$ & $\check{\beta}$ & $\tilde{\beta}$ & $\hat{\beta}$ & $\check{\beta}$ & $\tilde{\beta}$ & $\hat{\beta}$ \bigstrut\\
    \hline
    \multirow{5}[2]{*}{$n^{-1/3}$} & 0.847 & 250   & 79.13 & 0.041 & 0.042 & 0.043 & 0.041 & 0.041 & 0.043 & 0.041 & 0.042 & 0.044 & 0.039 & 0.039 \bigstrut[t]\\
          & 0.672 & 500   & 126.00 & 0.028 & 0.028 & 0.028 & 0.027 & 0.028 & 0.028 & 0.027 & 0.029 & 0.029 & 0.027 & 0.027 \\
          & 0.533 & 1000  & 199.60 & 0.020 & 0.020 & 0.020 & 0.020 & 0.020 & 0.020 & 0.020 & 0.020 & 0.020 & 0.019 & 0.019 \\
          & 0.423 & 2000  & 317.14 & 0.014 & 0.014 & 0.014 & 0.014 & 0.014 & 0.014 & 0.014 & 0.014 & 0.014 & 0.014 & 0.014 \\
          & 0.336 & 4000  & 503.95 & 0.010 & 0.010 & 0.010 & 0.010 & 0.010 & 0.010 & 0.010 & 0.010 & 0.010 & 0.010 & 0.010 \bigstrut[b]\\
    \hline
    \multirow{5}[2]{*}{$n^{-1/2}$} & 0.337 & 250   & 31.41 & 0.041 & 0.047 & 0.045 & 0.041 & 0.051 & 0.047 & 0.041 & 0.055 & 0.050 & 0.039 & 0.042 \bigstrut[t]\\
          & 0.239 & 500   & 44.70 & 0.029 & 0.034 & 0.031 & 0.029 & 0.037 & 0.032 & 0.028 & 0.039 & 0.033 & 0.028 & 0.031 \\
          & 0.169 & 1000  & 63.25 & 0.020 & 0.024 & 0.021 & 0.020 & 0.027 & 0.022 & 0.020 & 0.029 & 0.022 & 0.020 & 0.022 \\
          & 0.119 & 2000  & 89.37 & 0.015 & 0.017 & 0.015 & 0.014 & 0.019 & 0.015 & 0.014 & 0.021 & 0.016 & 0.014 & 0.016 \\
          & 0.084 & 4000  & 126.52 & 0.010 & 0.012 & 0.010 & 0.010 & 0.013 & 0.010 & 0.010 & 0.014 & 0.010 & 0.010 & 0.011 \bigstrut[b]\\
    \hline
    \multirow{5}[2]{*}{$n^{-2/3}$} & 0.134 & 250   & 12.51 & 0.041 & 0.075 & 0.051 & 0.040 & 0.091 & 0.058 & 0.040 & 0.105 & 0.063 & 0.039 & 0.066 \bigstrut[t]\\
          & 0.085 & 500   & 15.84 & 0.028 & 0.059 & 0.034 & 0.028 & 0.074 & 0.038 & 0.028 & 0.086 & 0.041 & 0.027 & 0.052 \\
          & 0.053 & 1000  & 19.97 & 0.020 & 0.047 & 0.023 & 0.020 & 0.059 & 0.025 & 0.020 & 0.069 & 0.027 & 0.019 & 0.042 \\
          & 0.034 & 2000  & 25.17 & 0.014 & 0.036 & 0.016 & 0.014 & 0.046 & 0.017 & 0.014 & 0.054 & 0.018 & 0.013 & 0.032 \\
          & 0.021 & 4000  & 31.74 & 0.010 & 0.029 & 0.011 & 0.010 & 0.038 & 0.012 & 0.010 & 0.044 & 0.012 & 0.010 & 0.026 \bigstrut[b]\\
    \hline
    \multirow{5}[2]{*}{$n^{-1}\sqrt{\log n}$} & 0.050 & 250   & 4.69  & 0.040 & 0.162 & 0.064 & 0.040 & 0.207 & 0.078 & 0.040 & 0.246 & 0.089 & 0.038 & 0.165 \bigstrut[t]\\
          & 0.027 & 500   & 4.97  & 0.029 & 0.152 & 0.046 & 0.029 & 0.197 & 0.056 & 0.029 & 0.235 & 0.063 & 0.028 & 0.158 \\
          & 0.014 & 1000  & 5.26  & 0.020 & 0.147 & 0.032 & 0.020 & 0.191 & 0.039 & 0.020 & 0.229 & 0.044 & 0.020 & 0.153 \\
          & 0.007 & 2000  & 5.51  & 0.014 & 0.141 & 0.022 & 0.014 & 0.184 & 0.026 & 0.014 & 0.220 & 0.030 & 0.014 & 0.146 \\
          & 0.004 & 4000  & 5.76  & 0.010 & 0.135 & 0.015 & 0.010 & 0.177 & 0.018 & 0.010 & 0.212 & 0.021 & 0.010 & 0.140 \bigstrut[b]\\
    \hline
    \multirow{5}[2]{*}{$n^{-1}\sqrt{\frac{\log n}{\log \log n}}$} & 0.038 & 250   & 3.58  & 0.041 & 0.202 & 0.074 & 0.041 & 0.258 & 0.093 & 0.041 & 0.307 & 0.105 & 0.039 & 0.219 \bigstrut[t]\\
          & 0.020 & 500   & 3.69  & 0.028 & 0.197 & 0.051 & 0.028 & 0.255 & 0.063 & 0.028 & 0.304 & 0.072 & 0.027 & 0.219 \\
          & 0.010 & 1000  & 3.77  & 0.020 & 0.192 & 0.037 & 0.020 & 0.250 & 0.046 & 0.020 & 0.299 & 0.052 & 0.020 & 0.217 \\
          & 0.005 & 2000  & 3.87  & 0.014 & 0.189 & 0.025 & 0.014 & 0.246 & 0.030 & 0.014 & 0.295 & 0.035 & 0.014 & 0.213 \\
          & 0.003 & 4000  & 3.96  & 0.010 & 0.185 & 0.018 & 0.010 & 0.242 & 0.022 & 0.010 & 0.290 & 0.025 & 0.010 & 0.209 \bigstrut[b]\\
    \hline
    \multirow{5}[2]{*}{$n^{-1}$} & 0.021 & 250   & 1.99  & 0.040 & 0.310 & 0.099 & 0.040 & 0.395 & 0.132 & 0.040 & 0.469 & 0.153 & 0.038 & 0.399 \bigstrut[t]\\
          & 0.011 & 500   & 2.00  & 0.029 & 0.311 & 0.071 & 0.029 & 0.399 & 0.092 & 0.029 & 0.475 & 0.109 & 0.028 & 0.424 \\
          & 0.005 & 1000  & 2.00  & 0.020 & 0.310 & 0.050 & 0.020 & 0.398 & 0.065 & 0.020 & 0.475 & 0.076 & 0.019 & 0.445 \\
          & 0.003 & 2000  & 2.00  & 0.015 & 0.310 & 0.034 & 0.014 & 0.399 & 0.043 & 0.014 & 0.477 & 0.053 & 0.014 & 0.460 \\
          & 0.001 & 4000  & 2.00  & 0.010 & 0.310 & 0.024 & 0.010 & 0.400 & 0.032 & 0.010 & 0.478 & 0.040 & 0.010 & 0.477 \bigstrut[b]\\
    \hline
    \end{tabular}%
	}
   \caption{RMSE when $\beta = 1$ of the OLS estimator based on the true network $A$ ($\tilde{\beta}^{(d)}$), the OLS estimator based on the plug-in $\hat{A}$ ($\hat{\beta}^{(d)}$) and of the bias-corrected estimator ($\check{\beta}^{(d)}$). Mean degree is the mean of realized degrees, i.e. $\widehat{\text{Deg}}$.}
  \label{tab:mse}%
\end{sidewaystable}%

\begin{sidewaystable}
\makebox[\textwidth][c]{
\centering\footnotesize
    \begin{tabular}{|cc|c|c|ccc|ccc|ccc|ccc|}
    \hline
    \multicolumn{2}{|c|}{\multirow{2}[4]{*}{$p_n$}} & \multirow{2}[4]{*}{$n$} & \textbf{Mean} & \multicolumn{3}{c|}{\textbf{Degree ($d=1$)}} & \multicolumn{3}{c|}{\textbf{Diffusion ($d=2$)}} & \multicolumn{3}{c|}{\textbf{Diffusion ($d=3$)}} & \multicolumn{3}{c|}{\textbf{Eigen ($d = \infty$)}} \bigstrut\\
\cline{5-16}    \multicolumn{2}{|c|}{} &       & \textbf{Degree} & $\tilde{\beta}$ & $\hat{\beta}$ & $\check{\beta}$ & $\tilde{\beta}$ & $\hat{\beta}$ & $\check{\beta}$ & $\tilde{\beta}$ & $\hat{\beta}$ & $\check{\beta}$ & $\tilde{\beta}$ & $\hat{\beta}$ & BA \bigstrut\\
    \hline
    \multirow{5}[2]{*}{$n^{-1/3}$} & 0.847 & 250   & 79.13 & 0.940 & 0.939 & 0.942 & 0.940 & 0.936 & 0.938 & 0.940 & 0.935 & 0.934 & 0.940 & 0.942 & 0.984 \bigstrut[t]\\
          & 0.672 & 500   & 126.00 & 0.961 & 0.956 & 0.959 & 0.961 & 0.954 & 0.956 & 0.961 & 0.952 & 0.953 & 0.961 & 0.957 & 0.987 \\
          & 0.533 & 1000  & 199.60 & 0.951 & 0.950 & 0.950 & 0.951 & 0.947 & 0.946 & 0.951 & 0.944 & 0.944 & 0.951 & 0.950 & 0.984 \\
          & 0.423 & 2000  & 317.14 & 0.947 & 0.945 & 0.946 & 0.947 & 0.943 & 0.945 & 0.947 & 0.940 & 0.943 & 0.947 & 0.947 & 0.980 \\
          & 0.336 & 4000  & 503.95 & 0.950 & 0.944 & 0.951 & 0.950 & 0.942 & 0.950 & 0.949 & 0.939 & 0.951 & 0.950 & 0.947 & 0.977 \bigstrut[b]\\
    \hline
    \multirow{5}[2]{*}{$n^{-1/2}$} & 0.337 & 250   & 31.41 & 0.948 & 0.910 & 0.940 & 0.948 & 0.879 & 0.923 & 0.949 & 0.851 & 0.910 & 0.949 & 0.935 & 0.994 \bigstrut[t]\\
          & 0.239 & 500   & 44.70 & 0.945 & 0.909 & 0.945 & 0.945 & 0.872 & 0.933 & 0.945 & 0.832 & 0.920 & 0.946 & 0.933 & 0.996 \\
          & 0.169 & 1000  & 63.25 & 0.947 & 0.903 & 0.943 & 0.947 & 0.863 & 0.935 & 0.947 & 0.819 & 0.927 & 0.948 & 0.919 & 0.994 \\
          & 0.119 & 2000  & 89.37 & 0.946 & 0.890 & 0.942 & 0.946 & 0.852 & 0.936 & 0.946 & 0.817 & 0.927 & 0.946 & 0.918 & 0.997 \\
          & 0.084 & 4000  & 126.52 & 0.949 & 0.903 & 0.950 & 0.949 & 0.854 & 0.948 & 0.950 & 0.810 & 0.943 & 0.951 & 0.920 & 0.996 \bigstrut[b]\\
    \hline
    \multirow{5}[2]{*}{$n^{-2/3}$} & 0.134 & 250   & 12.51 & 0.950 & 0.711 & 0.925 & 0.950 & 0.548 & 0.884 & 0.950 & 0.428 & 0.856 & 0.950 & 0.782 & 1.000 \bigstrut[t]\\
          & 0.085 & 500   & 15.84 & 0.953 & 0.608 & 0.934 & 0.952 & 0.403 & 0.902 & 0.952 & 0.282 & 0.877 & 0.952 & 0.700 & 0.998 \\
          & 0.053 & 1000  & 19.97 & 0.954 & 0.502 & 0.943 & 0.955 & 0.283 & 0.920 & 0.955 & 0.155 & 0.897 & 0.955 & 0.591 & 1.000 \\
          & 0.034 & 2000  & 25.17 & 0.957 & 0.401 & 0.940 & 0.956 & 0.174 & 0.916 & 0.956 & 0.086 & 0.899 & 0.956 & 0.495 & 1.000 \\
          & 0.021 & 4000  & 31.74 & 0.949 & 0.248 & 0.939 & 0.949 & 0.076 & 0.923 & 0.949 & 0.026 & 0.909 & 0.950 & 0.341 & 1.000 \bigstrut[b]\\
    \hline
    \multirow{5}[2]{*}{$n^{-1}\sqrt{\log n}$} & 0.050 & 250   & 4.69  & 0.952 & 0.094 & 0.911 & 0.952 & 0.027 & 0.841 & 0.952 & 0.009 & 0.784 & 0.952 & 0.076 & 1.000 \bigstrut[t]\\
          & 0.027 & 500   & 4.97  & 0.939 & 0.015 & 0.902 & 0.939 & 0.002 & 0.821 & 0.939 & 0.000 & 0.775 & 0.940 & 0.003 & 1.000 \\
          & 0.014 & 1000  & 5.26  & 0.948 & 0.000 & 0.895 & 0.948 & 0.000 & 0.830 & 0.948 & 0.000 & 0.776 & 0.948 & 0.000 & 1.000 \\
          & 0.007 & 2000  & 5.51  & 0.944 & 0.000 & 0.902 & 0.944 & 0.000 & 0.832 & 0.944 & 0.000 & 0.782 & 0.943 & 0.000 & 1.000 \\
          & 0.004 & 4000  & 5.76  & 0.950 & 0.000 & 0.904 & 0.950 & 0.000 & 0.837 & 0.950 & 0.000 & 0.782 & 0.950 & 0.000 & 1.000 \bigstrut[b]\\
    \hline
    \multirow{5}[2]{*}{$n^{-1}\sqrt{\frac{\log n}{\log \log n}}$} & 0.038 & 250   & 3.58  & 0.944 & 0.025 & 0.873 & 0.944 & 0.007 & 0.787 & 0.944 & 0.002 & 0.739 &       &       &  \bigstrut[t]\\
          & 0.020 & 500   & 3.69  & 0.952 & 0.000 & 0.889 & 0.952 & 0.000 & 0.790 & 0.953 & 0.000 & 0.742 &       &       &  \\
          & 0.010 & 1000  & 3.77  & 0.950 & 0.000 & 0.880 & 0.950 & 0.000 & 0.787 & 0.950 & 0.000 & 0.739 &       &       &  \\
          & 0.005 & 2000  & 3.87  & 0.954 & 0.000 & 0.889 & 0.954 & 0.000 & 0.806 & 0.954 & 0.000 & 0.752 &       &       &  \\
          & 0.003 & 4000  & 3.96  & 0.944 & 0.000 & 0.881 & 0.944 & 0.000 & 0.790 & 0.944 & 0.000 & 0.731 &       &       &  \bigstrut[b]\\
    \hline
    \end{tabular}%
}
\caption{Coverage of 95\% confidence intervals when $\beta = 1$. CIs constructed using either the OLS estimator based on the true network $A$ ($\tilde{\beta}^{(d)}$), the OLS estimator based on the plug-in $\hat{A}$ ($\hat{\beta}^{(d)}$) or the bias-corrected estimator ($\check{\beta}^{(d)}$). For eigenvector, BA refers to the bias-aware CI in Section \ref{section--bias_aware}. We restrict attention to sequences of $p_n$'s for which the estimators are consistent. Mean degree is the mean of realized degrees, i.e. $\widehat{\text{Deg}}$.}
\label{tab:coverage}%

\end{sidewaystable}%

\section{Empirical Demonstration}\label{section--application}

In this section, we demonstrate the utility of our theoretical results via an application inspired by \cite{deweerdt2006risk}.\footnote{The data is obtained from Joachim De Weerdt's website: \url{https://www.uantwerpen.be/en/staff/joachim-deweerdt/public-data-sets/nyakatoke-network/}.} When access to formal credit markets is limited, social insurance is an important mechanism for smoothing consumption. %(\citealt{rosenzweig1988risk,udry1994risk, fafchamps2003risk, kinnan2012kinship}, among many others).
 \cite{deweerdt2006risk} studies Nyakatoke, a village with 119 households in rural Tanzania, and finds that social insurance helps households to smooth consumption following health shocks. The data they use comprises five rounds of panel data on household consumption, illness, and other covariates, collected from February to December 2000. The authors also had access to social network data collected during the first round of the survey, in which households were asked to identify those whom they depend on or who depend on them for help. The authors then regress a household's change in consumption following illness on the mean consumption of its network neighbors, finding evidence of positive co-movements. 

Another way to study the role of social insurance in consumption smoothing is to regress the standard deviation of consumption expenditure on network centrality measures. Specifically, consider the regression:
\begin{equation*}
	Y_i =  C^{(d)}_i \beta+  W_i'\gamma + \varepsilon_i^{(d)}~,
\end{equation*}
where $Y_i$ is the standard deviation in log food expenditure over the five survey rounds. $C^{(d)}_i$ denotes degree ($d=1$), diffusion with 2 periods ($d=2$), diffusion with 3 periods ($d=3$), or eigenvector centrality ($d=\infty$). The vector of covariates $W_i$ includes a constant and the number of major illnesses the household experienced. Because wealthier households may have smoother consumption and higher centralities, we also include mean log food expenditure as a control variable. 

Centrality regression could be preferable to the authors' specification. This might be the case if we do not know which covariates best capture the channels through which social insurance operates. For example, it might be a household's stock of savings that co-moves with the decision to lend to their friends, rather than their own consumption. However, in either case, we might expect households that are more central in an appropriate sense to benefit from more risk-sharing. Centrality regressions may also capture more complex patterns of assistance. For example, there might be a second-order form of social insurance in which friends receive support after lending to a household with illness. Such activity could be captured by an appropriate centrality measure, but might be less tractable to model explicitly. 

\begin{figure}
\centering
\includegraphics[width=1\linewidth]{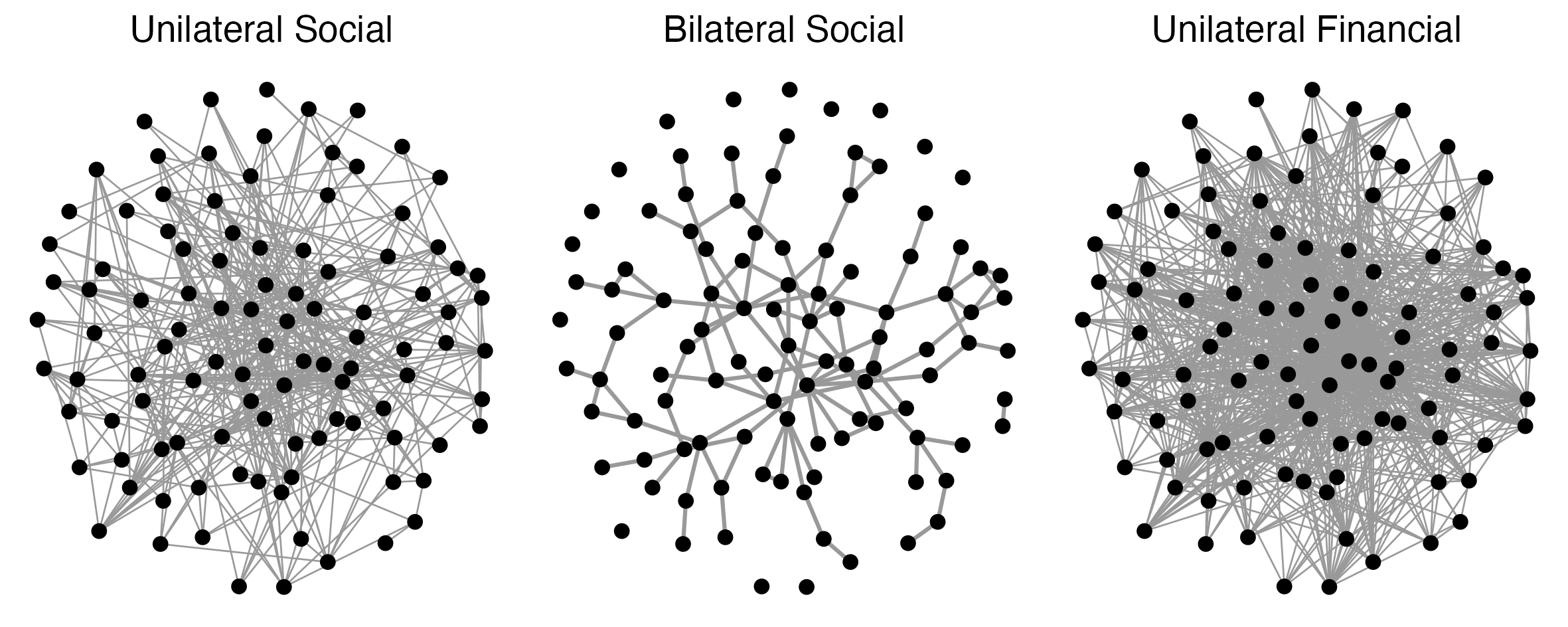}
\caption{Various networks in Nyakatoke.}
\label{fig:nyakatoke_networkscombined}
\end{figure}

The above regression requires information on the network of social insurance. We might think of this as the matrix $A$, where each entry $A_{ij}$ records the propensity or probability that $i$ lends money to $j$, or vice versa, over the survey period. Following the authors, we consider the use of the following proxy networks:
\begin{itemize}
	\item[] \textbf{Unilateral Social (US):}. $\hat{A}_{ij} = 1$ if either $i$ or $j$ names the other household as a party that they could depend on or who depends on them for help.
	\item[] \textbf{Bilateral Social (BS):} $\hat{A}_{ij} = 1$ only if both $i$ and $j$ name the other household as a party that they could depend on or who depends on them for help.
	\item[] \textbf{Unilateral Financial (UF):} $\hat{A}_{ij} = 1$ if either $i$ or $j$ has lent money to the other over the survey period.
\end{itemize}
The networks are plotted in Figure \ref{fig:nyakatoke_networkscombined} and the degree distributions are described in Table \ref{tab:nyakatoke_deg_dist}. UF is the densest network, with a mean degree of 16, followed by US with a mean degree of 8. By construction, BS is extremely sparse, with a mean degree of 2.

% Table generated by Excel2LaTeX from sheet 'deg_distribution'
\begin{table}[h]
  \centering
    \begin{tabular}{lccccc}\hline\hline
     ($n = 119$)     & Mean  & Median & Min   & Max & Isolates\\
    \midrule
    Unilateral Social & 8.02  & 7     & 1     & 31  & 0 \\
    Bilateral Social & 2.30  & 2     & 0     & 10 & 14\\
    Unilateral Financial & 16.53 & 14    & 3     & 79 & 0\\
    \hline\hline
    \end{tabular}%
  \caption{Descriptive statistics for unnormalized degrees in the various Nyakatoke networks.}
  \label{tab:nyakatoke_deg_dist}%
\end{table}%

Regression results are presented in Table \ref{tab:regression_null}. In this exercise, we set $a^{(d)}_n$ and $\delta_n$ equal to the inverse of the mean degree. This scaling achieves the stated rates in Section \ref{section--setup-centrality} without introducing further asymptotic bias. The column labelled $\hat{\beta}$ presents the OLS estimators based on $\hat{A}$, while $p$-values are for two-sided tests of the hypotheses $H_0: \beta^{(d)} = 0$. Across the board, ${\beta}^{(d)}$ is estimated to be weakly negative, suggesting that more central households exhibit lower consumption volatility, in line with improved access to informal insurance. In terms of statistical significance, degree and diffusion centralities yield estimates qualitatively similar to those based on eigenvector centrality in the denser networks US and UF. For BS, however, we see that degree and diffusion are substantially more predictive than eigenvector centrality. Although this could be taken to mean that the latter is not economically meaningful, our results suggest that the observed difference could simply be due to the eigenvector centrality's relatively worse statistical properties. 

Table \ref{tab:regression_null} also presents our bias corrected estimators for degree and diffusion, as well as the estimated asymptotic bias, $\check{\beta} - \hat{\beta}$. For UF, the bias is between -0.003 to -0.029, so that bias correction makes a substantial difference to the estimates, particularly for diffusion centrality. US is sparser and consequently, the bias is of larger absolute value, ranging from 0.003 to -0.103. For BS, bias ranges between -0.02 to 0.042. However, with a mean degree of 2.3, BS is so sparse that we do not expect our estimators to be reliable. These estimates should therefore not be overinterpreted. Finally, the last column presents bias-corrected CIs for degree and diffusion, and bias-aware type CIs for eigenvector. As expected, these CIs are generally wider than the CIs based on $\hat{\beta}$. 

In sum, centrality regressions on the Nyakatoke networks exhibit behavior in line with the predictions of our theory. Apart from the sparsest network, our methods appear to work well in this setting. 

% Table generated by Excel2LaTeX from sheet 'reg_null'
\begin{table}[t]
  \centering \footnotesize
    \begin{tabular}{cccccccc}\hline\hline
    Network & d     & $\hat{\beta}$ & $\check{\beta}$ & Bias  & $p$-val & $\hat{\beta}$-CI & $\check{\beta}$-CI / BA-CI \bigstrut[b]\\
    \hline
          & 1     & -0.011 & -0.007 & 0.003 & 0.347 & [-0.033, 0.012] & [-0.022, 0.008] \bigstrut[t]\\
    Unilateral & 2     & -0.010 & -0.030 & -0.020 & 0.369 & [-0.032, 0.012] & [-0.097, 0.036] \\
    Social & 3     & -0.010 & -0.113 & -0.103 & 0.379 & [-0.032, 0.012] & [-0.363, 0.138] \\
          & $\infty$ & -0.009 &       &       & 0.423 & [-0.031, 0.013] & [-0.036, 0.018] \bigstrut[b]\\
    \hline
          & 1     & -0.025 & -0.027 & -0.002 & 0.013 & [-0.044, -0.005] & [-0.048, -0.006] \bigstrut[t]\\
    Bilateral & 2     & -0.021 & 0.020 & 0.042 & 0.036 & [-0.041, -0.001] & [0.001, 0.039] \\
    Social & 3     & -0.018 & 0.004 & 0.022 & 0.072 & [-0.039, 0.002] & [-0.000, 0.007] \\
          & $\infty$ & -0.009 &       &       & 0.414 & [-0.030, 0.012] & [-0.056, 0.034] \bigstrut[b]\\
    \hline
          & 1     & -0.023 & -0.026 & -0.003 & 0.021 & [-0.042, -0.003] & [-0.048, -0.004] \bigstrut[t]\\
    Unilateral & 2     & -0.021 & -0.041 & -0.020 & 0.041 & [-0.041, -0.001] & [-0.080, -0.002] \\
    Finance & 3     & -0.021 & -0.049 & -0.029 & 0.046 & [-0.041, -0.000] & [-0.098, -0.001] \\
          & $\infty$ & -0.019 &       &       & 0.077 & [-0.040, 0.002] & [-0.044, 0.005] \\
    \hline\hline
    \end{tabular}%
    \caption{Regression results for various networks. Mean degrees are in brackets. $\hat{\beta}$ is the plug-in estimator. $\check{\beta}$ is our bias corrected estimator. Bias is the estimated asymptotic bias, $\check{\beta} - \hat{\beta}$. $p$-val is for the two-sided test that $H_0: \beta = 0$. $\hat{\beta}$-CI are 95\% CIs based on the plug-in estimator. The final column presents $\check{\beta}$-CI for degree and diffusion and bias aware CI with $\eta = 0.1$ for eigenvector.}
  \label{tab:regression_null}%
\end{table}%

\section{Conclusion}\label{section--conclusion}
This paper studies the properties of linear regression on degree, diffusion and eigenvector centrality when networks are sparse and observed with error. We show that these issues threaten the consistency of OLS estimators and characterize the amount of sparsity at which inconsistency occurs. In doing so, we find that eigenvector centrality is less robust to sparsity than the others and that the statistical properties of the corresponding regression are sensitive to the scaling. 

Additionally, we show that an asymptotic bias arises whenever the true slope parameter is not zero and that the bias can be of larger order than the variance, invalidating plug-in $t$-tests and confidence intervals even when the estimators are consistent. For degree and diffusion, we obtain a precise characterization of the asymptotic bias by relating them to counts of particular trees, and we provide algorithms for enumerating such trees. This allows us to construct bias corrected estimator with substantially better performance compared to the plug-in estimator. For eigenvector, we propose to widen the confidence intervals using a bound of the asymptotic bias. Although the procedure is conservative when networks are sparse, they become exact once the network is moderately dense. 

Our results suggest that applied researchers should view their estimates and confidence intervals with caution when applying OLS to sparse, noisy networks. Specifically, comparing the statistical significance of eigenvector centrality with degree or diffusion may yield misleading conclusions since they differ not only in economic significance but also in their statistical properties. Provided that the networks are not too sparse, the usual $t$-test is valid for the null hypothesis that the slope parameter is $0$. However, alternative inference procedures will be necessary for non-zero null hypotheses and for constructing valid confidence intervals. Additionally, there may be scope for improving estimation through the use of bias-corrected estimators. Estimation and inference under bounded degrees remain an open question, though as \cite{le2017concentration} and \cite{graham2020sparse} show, parametric models may point to a way forward.

\begin{spacing}{0.5}
\bibliographystyle{chicago}
\bibliography{eigenReg}

\begin{thebibliography}{}

\bibitem[\protect\citeauthoryear{Alt, Ducatez, and Knowles}{Alt
  et~al.}{2021}]{alt2021poisson}
Alt, J., R.~Ducatez, and A.~Knowles (2021).
\newblock {Poisson statistics and localization at the spectral edge of sparse
  Erd{\H{o}}s--R{\'e}nyi graphs}.
\newblock {\em arXiv preprint arXiv:2106.12519\/}.

\bibitem[\protect\citeauthoryear{Armstrong and Koles{\'a}r}{Armstrong and
  Koles{\'a}r}{2018}]{armstrong2018optimal}
Armstrong, T.~B. and M.~Koles{\'a}r (2018).
\newblock Optimal inference in a class of regression models.
\newblock {\em Econometrica\/}~{\em 86\/}(2), 655--683.

\bibitem[\protect\citeauthoryear{Armstrong and Koles{\'a}r}{Armstrong and
  Koles{\'a}r}{2020}]{armstrong2020simple}
Armstrong, T.~B. and M.~Koles{\'a}r (2020).
\newblock Simple and honest confidence intervals in nonparametric regression.
\newblock {\em Quantitative Economics\/}~{\em 11\/}(1), 1--39.

\bibitem[\protect\citeauthoryear{Athey, Bayati, Doudchenko, Imbens, and
  Khosravi}{Athey et~al.}{2021}]{athey2021matrix}
Athey, S., M.~Bayati, N.~Doudchenko, G.~Imbens, and K.~Khosravi (2021).
\newblock Matrix completion methods for causal panel data models.
\newblock {\em Journal of the American Statistical Association\/}~{\em
  116\/}(536), 1716--1730.

\bibitem[\protect\citeauthoryear{Auerbach}{Auerbach}{2022}]{auerbach2022identification}
Auerbach, E. (2022).
\newblock {Identification and estimation of a partially linear regression model
  using network data}.
\newblock {\em Econometrica\/}~{\em 90\/}(1), 347--365.

\bibitem[\protect\citeauthoryear{Avella-Medina, Parise, Schaub, and
  Segarra}{Avella-Medina et~al.}{2020}]{avella2020centrality}
Avella-Medina, M., F.~Parise, M.~T. Schaub, and S.~Segarra (2020).
\newblock Centrality measures for graphons: Accounting for uncertainty in
  networks.
\newblock {\em IEEE Transactions on Network Science and Engineering\/}~{\em
  7\/}(1), 520--537.

\bibitem[\protect\citeauthoryear{Banerjee, Chandrasekhar, Duflo, and
  Jackson}{Banerjee et~al.}{2013}]{banerjee2013diffusion}
Banerjee, A., A.~G. Chandrasekhar, E.~Duflo, and M.~O. Jackson (2013).
\newblock {The diffusion of microfinance}.
\newblock {\em Science\/}~{\em 341\/}(6144), 1236498.

\bibitem[\protect\citeauthoryear{Banerjee, Chandrasekhar, Duflo, and
  Jackson}{Banerjee et~al.}{2019}]{banerjee2019using}
Banerjee, A., A.~G. Chandrasekhar, E.~Duflo, and M.~O. Jackson (2019).
\newblock {Using gossips to spread information: Theory and evidence from two
  randomized controlled trials}.
\newblock {\em The Review of Economic Studies\/}~{\em 86\/}(6), 2453--2490.

\bibitem[\protect\citeauthoryear{Benaych-Georges, Bordenave, and
  Knowles}{Benaych-Georges et~al.}{2020}]{benaych2020spectral}
Benaych-Georges, F., C.~Bordenave, and A.~Knowles (2020).
\newblock Spectral radii of sparse random matrices.
\newblock In {\em Annales de l'Institut Henri Poincar{\'e}, Probabilit{\'e}s et
  Statistiques}, Volume~56, pp.\  2141--2161. Institut Henri Poincar{\'e}.

\bibitem[\protect\citeauthoryear{Bickel and Chen}{Bickel and
  Chen}{2009}]{bickel2009nonparametric}
Bickel, P.~J. and A.~Chen (2009).
\newblock {A nonparametric view of network models and Newman--Girvan and other
  modularities}.
\newblock {\em Proceedings of the National Academy of Sciences\/}~{\em
  106\/}(50), 21068--21073.

\bibitem[\protect\citeauthoryear{Bickel, Chen, and Levina}{Bickel
  et~al.}{2011}]{bickel2011method}
Bickel, P.~J., A.~Chen, and E.~Levina (2011).
\newblock {The method of moments and degree distributions for network models}.
\newblock {\em The Annals of Statistics\/}~{\em 39\/}(5), 2280--2301.

\bibitem[\protect\citeauthoryear{Bloch, Jackson, and Tebaldi}{Bloch
  et~al.}{2021}]{bloch2021centrality}
Bloch, F., M.~O. Jackson, and P.~Tebaldi (2021).
\newblock Centrality measures in networks.
\newblock {\em arXiv preprint arXiv:1608.05845\/}.

\bibitem[\protect\citeauthoryear{Blom}{Blom}{1976}]{blom1976some}
Blom, G. (1976).
\newblock Some properties of incomplete u-statistics.
\newblock {\em Biometrika\/}, 573--580.

\bibitem[\protect\citeauthoryear{Bonhomme and Manresa}{Bonhomme and
  Manresa}{2015}]{bonhomme2015grouped}
Bonhomme, S. and E.~Manresa (2015).
\newblock Grouped patterns of heterogeneity in panel data.
\newblock {\em Econometrica\/}~{\em 83\/}(3), 1147--1184.

\bibitem[\protect\citeauthoryear{Borgatti, Carley, and Krackhardt}{Borgatti
  et~al.}{2006}]{borgatti2006robustness}
Borgatti, S.~P., K.~M. Carley, and D.~Krackhardt (2006).
\newblock {On the robustness of centrality measures under conditions of
  imperfect data}.
\newblock {\em Social networks\/}~{\em 28\/}(2), 124--136.

\bibitem[\protect\citeauthoryear{Bourl{\`e}s, Bramoull{\'e}, and
  Perez-Richet}{Bourl{\`e}s et~al.}{2021}]{bourles2021altruism}
Bourl{\`e}s, R., Y.~Bramoull{\'e}, and E.~Perez-Richet (2021).
\newblock Altruism and risk sharing in networks.
\newblock {\em Journal of the European Economic Association\/}~{\em 19\/}(3),
  1488--1521.

\bibitem[\protect\citeauthoryear{Bramoull{\'e} and Genicot}{Bramoull{\'e} and
  Genicot}{2018}]{bramoulle2018diffusion}
Bramoull{\'e}, Y. and G.~Genicot (2018).
\newblock Diffusion centrality: Foundations and extensions.

\bibitem[\protect\citeauthoryear{Cai, Yang, Zhu, Shen, and Zhao}{Cai
  et~al.}{2021}]{cai2021network}
Cai, J., D.~Yang, W.~Zhu, H.~Shen, and L.~Zhao (2021).
\newblock {Network regression and supervised centrality estimation}.
\newblock {\em Available at SSRN 3963523\/}.

\bibitem[\protect\citeauthoryear{Chandrasekhar}{Chandrasekhar}{2016}]{chandrasekhar2016formation}
Chandrasekhar, A. (2016).
\newblock Econometrics of network formation.
\newblock {\em The Oxford handbook of the economics of networks\/}, 303--357.

\bibitem[\protect\citeauthoryear{Chandrasekhar and Lewis}{Chandrasekhar and
  Lewis}{2016}]{chandrasekhar2016econometrics}
Chandrasekhar, A. and R.~Lewis (2016).
\newblock {Econometrics of sampled networks}.
\newblock {\em Working Paper\/}.

\bibitem[\protect\citeauthoryear{Chatterjee}{Chatterjee}{2015}]{chatterjee2015matrix}
Chatterjee, S. (2015).
\newblock Matrix estimation by universal singular value thresholding.
\newblock {\em The Annals of Statistics\/}~{\em 43\/}(1), 177--214.

\bibitem[\protect\citeauthoryear{Chen and Kato}{Chen and
  Kato}{2019}]{chen2019randomized}
Chen, X. and K.~Kato (2019).
\newblock Randomized incomplete u-statistics in high dimensions.

\bibitem[\protect\citeauthoryear{Cheng, Wei, and Chen}{Cheng
  et~al.}{2021}]{cheng2021tackling}
Cheng, C., Y.~Wei, and Y.~Chen (2021).
\newblock Tackling small eigen-gaps: Fine-grained eigenvector estimation and
  inference under heteroscedastic noise.
\newblock {\em IEEE Transactions on Information Theory\/}~{\em 67\/}(11),
  7380--7419.

\bibitem[\protect\citeauthoryear{Costenbader and Valente}{Costenbader and
  Valente}{2003}]{costenbader2003stability}
Costenbader, E. and T.~W. Valente (2003).
\newblock The stability of centrality measures when networks are sampled.
\newblock {\em Social networks\/}~{\em 25\/}(4), 283--307.

\bibitem[\protect\citeauthoryear{Crippa}{Crippa}{2025}]{crippa2025identification}
Crippa, F. (2025).
\newblock Identification, estimation, and inference in two-sided interaction
  models.
\newblock {\em arXiv preprint arXiv:2510.22884\/}.

\bibitem[\protect\citeauthoryear{Cruz, Labonne, and Querubin}{Cruz
  et~al.}{2017}]{cruz2017politician}
Cruz, C., J.~Labonne, and P.~Querubin (2017).
\newblock Politician family networks and electoral outcomes: Evidence from the
  philippines.
\newblock {\em American Economic Review\/}~{\em 107\/}(10), 3006--37.

\bibitem[\protect\citeauthoryear{Dasaratha}{Dasaratha}{2020}]{dasaratha2020distributions}
Dasaratha, K. (2020).
\newblock Distributions of centrality on networks.
\newblock {\em Games and Economic Behavior\/}~{\em 122}, 1--27.

\bibitem[\protect\citeauthoryear{De~Paula}{De~Paula}{2017}]{de2017econometrics}
De~Paula, A. (2017).
\newblock {Econometrics of network models}.
\newblock In {\em Advances in economics and econometrics: Theory and
  applications, eleventh world congress}, pp.\  268--323. Cambridge University
  Press Cambridge.

\bibitem[\protect\citeauthoryear{De~Paula, Rasul, and Souza}{De~Paula
  et~al.}{2025}]{depaula2025identifying}
De~Paula, A., I.~Rasul, and P.~C. Souza (2025).
\newblock Identifying network ties from panel data: Theory and an application
  to tax competition.
\newblock {\em Review of Economic Studies\/}~{\em 92\/}(4), 2691--2729.

\bibitem[\protect\citeauthoryear{De~Paula, Richards-Shubik, and Tamer}{De~Paula
  et~al.}{2018}]{dePaula2018identifying}
De~Paula, {\'A}., S.~Richards-Shubik, and E.~Tamer (2018).
\newblock Identifying preferences in networks with bounded degree.
\newblock {\em Econometrica\/}~{\em 86\/}(1), 263--288.

\bibitem[\protect\citeauthoryear{De~Weerdt and Dercon}{De~Weerdt and
  Dercon}{2006}]{deweerdt2006risk}
De~Weerdt, J. and S.~Dercon (2006).
\newblock Risk-sharing networks and insurance against illness.
\newblock {\em Journal of development Economics\/}~{\em 81\/}(2), 337--356.

\bibitem[\protect\citeauthoryear{Eagle, Macy, and Claxton}{Eagle
  et~al.}{2010}]{eagle2010network}
Eagle, N., M.~Macy, and R.~Claxton (2010).
\newblock Network diversity and economic development.
\newblock {\em Science\/}~{\em 328\/}(5981), 1029--1031.

\bibitem[\protect\citeauthoryear{Graham}{Graham}{2020a}]{graham2020network}
Graham, B.~S. (2020a).
\newblock {Network data}.
\newblock In {\em Handbook of Econometrics}, Volume~7, pp.\  111--218.
  Elsevier.

\bibitem[\protect\citeauthoryear{Graham}{Graham}{2020b}]{graham2020sparse}
Graham, B.~S. (2020b).
\newblock {Sparse network asymptotics for logistic regression}.

\bibitem[\protect\citeauthoryear{Granovetter}{Granovetter}{1973}]{granovetter1973strength}
Granovetter, M.~S. (1973).
\newblock The strength of weak ties.
\newblock {\em American journal of sociology\/}~{\em 78\/}(6), 1360--1380.

\bibitem[\protect\citeauthoryear{Griffith}{Griffith}{2022}]{griffith2022name}
Griffith, A. (2022).
\newblock Name your friends, but only five? the importance of censoring in peer
  effects estimates using social network data.
\newblock {\em Journal of Labor Economics\/}~{\em 40\/}(4), 779--805.

\bibitem[\protect\citeauthoryear{Hochberg, Ljungqvist, and Lu}{Hochberg
  et~al.}{2007}]{hochberg2007whom}
Hochberg, Y.~V., A.~Ljungqvist, and Y.~Lu (2007).
\newblock Whom you know matters: Venture capital networks and investment
  performance.
\newblock {\em The Journal of Finance\/}~{\em 62\/}(1), 251--301.

\bibitem[\protect\citeauthoryear{Hoeffding}{Hoeffding}{1963}]{hoeffding1963probability}
Hoeffding, W. (1963).
\newblock Probability inequalities for sums of bounded random variables.
\newblock {\em Journal of the American statistical association\/}~{\em
  58\/}(301), 13--30.

\bibitem[\protect\citeauthoryear{Holland, Laskey, and Leinhardt}{Holland
  et~al.}{1983}]{holland1983stochastic}
Holland, P.~W., K.~B. Laskey, and S.~Leinhardt (1983).
\newblock Stochastic blockmodels: First steps.
\newblock {\em Social networks\/}~{\em 5\/}(2), 109--137.

\bibitem[\protect\citeauthoryear{Jackson}{Jackson}{2010}]{jackson2010social}
Jackson, M.~O. (2010).
\newblock {\em Social and economic networks}.
\newblock Princeton university press.

\bibitem[\protect\citeauthoryear{Jochmans}{Jochmans}{2018}]{jochmans2018semiparametric}
Jochmans, K. (2018).
\newblock {Semiparametric analysis of network formation}.
\newblock {\em Journal of Business \& Economic Statistics\/}~{\em 36\/}(4),
  705--713.

\bibitem[\protect\citeauthoryear{Jochmans and Weidner}{Jochmans and
  Weidner}{2019}]{jochmans2019fixed}
Jochmans, K. and M.~Weidner (2019).
\newblock Fixed-effect regressions on network data.
\newblock {\em Econometrica\/}~{\em 87\/}(5), 1543--1560.

\bibitem[\protect\citeauthoryear{Kinnan and Townsend}{Kinnan and
  Townsend}{2012}]{kinnan2012kinship}
Kinnan, C. and R.~Townsend (2012).
\newblock Kinship and financial networks, formal financial access, and risk
  reduction.
\newblock {\em American Economic Review\/}~{\em 102\/}(3), 289--93.

\bibitem[\protect\citeauthoryear{Le, Levina, and Vershynin}{Le
  et~al.}{2017}]{le2017concentration}
Le, C.~M., E.~Levina, and R.~Vershynin (2017).
\newblock {Concentration and regularization of random graphs}.
\newblock {\em Random Structures \& Algorithms\/}~{\em 51\/}(3), 538--561.

\bibitem[\protect\citeauthoryear{Le and Li}{Le and Li}{2020}]{le2020linear}
Le, C.~M. and T.~Li (2020).
\newblock {Linear regression and its inference on noisy network-linked data}.
\newblock {\em arXiv preprint arXiv:2007.00803\/}.

\bibitem[\protect\citeauthoryear{Lei and Rinaldo}{Lei and
  Rinaldo}{2015}]{lei2015consistency}
Lei, J. and A.~Rinaldo (2015).
\newblock Consistency of spectral clustering in stochastic block models.
\newblock {\em The Annals of Statistics\/}~{\em 43\/}(1), 215--237.

\bibitem[\protect\citeauthoryear{Leider, M{\"o}bius, Rosenblat, and Do}{Leider
  et~al.}{2009}]{leider2009directed}
Leider, S., M.~M. M{\"o}bius, T.~Rosenblat, and Q.-A. Do (2009).
\newblock Directed altruism and enforced reciprocity in social networks.
\newblock {\em The Quarterly Journal of Economics\/}~{\em 124\/}(4),
  1815--1851.

\bibitem[\protect\citeauthoryear{Leung and Moon}{Leung and
  Moon}{2025}]{leung2025normal}
Leung, M.~P. and H.~R. Moon (2025).
\newblock Normal approximation in large network models.
\newblock {\em arXiv preprint arXiv:1904.11060\/}.

\bibitem[\protect\citeauthoryear{Levin and Levina}{Levin and
  Levina}{2019}]{levin2019bootstrapping}
Levin, K. and E.~Levina (2019).
\newblock Bootstrapping networks with latent space structure.
\newblock {\em arXiv preprint arXiv:1907.10821\/}.

\bibitem[\protect\citeauthoryear{Lewbel, Qu, and Tang}{Lewbel
  et~al.}{2023}]{lewbel2023social}
Lewbel, A., X.~Qu, and X.~Tang (2023).
\newblock Social networks with unobserved links.
\newblock {\em Journal of Political Economy\/}~{\em 131\/}(4), 898--946.

\bibitem[\protect\citeauthoryear{Lewbel, Qu, and Tang}{Lewbel
  et~al.}{2025}]{lewbel2025estimating}
Lewbel, A., X.~Qu, and X.~Tang (2025).
\newblock Estimating social network models with link misclassification.
\newblock {\em arXiv preprint arXiv:2509.07343\/}.

\bibitem[\protect\citeauthoryear{Lewbel, Qu, Tang, et~al.}{Lewbel
  et~al.}{2021}]{lewbel2021social}
Lewbel, A., X.~Qu, X.~Tang, et~al. (2021).
\newblock {\em Social Networks with Mismeasured Links}.
\newblock Boston College.

\bibitem[\protect\citeauthoryear{Li, Levina, and Zhu}{Li
  et~al.}{2020}]{li2020network}
Li, T., E.~Levina, and J.~Zhu (2020).
\newblock Network cross-validation by edge sampling.
\newblock {\em Biometrika\/}~{\em 107\/}(2), 257--276.

\bibitem[\protect\citeauthoryear{Lov{\'a}sz}{Lov{\'a}sz}{2012}]{lovasz2012large}
Lov{\'a}sz, L. (2012).
\newblock {\em Large networks and graph limits}, Volume~60.
\newblock American Mathematical Soc.

\bibitem[\protect\citeauthoryear{Manresa}{Manresa}{2016}]{manresa2016estimating}
Manresa, E. (2016).
\newblock Estimating the structure of social interactions using panel data.
\newblock {\em Working Paper\/}.

\bibitem[\protect\citeauthoryear{Menzel}{Menzel}{2022}]{menzel2022strategic}
Menzel, K. (2022).
\newblock Strategic network formation with many agents.

\bibitem[\protect\citeauthoryear{Negahban and Wainwright}{Negahban and
  Wainwright}{2012}]{negahban2012restricted}
Negahban, S. and M.~J. Wainwright (2012).
\newblock Restricted strong convexity and weighted matrix completion: Optimal
  bounds with noise.
\newblock {\em The Journal of Machine Learning Research\/}~{\em 13\/}(1),
  1665--1697.

\bibitem[\protect\citeauthoryear{Rajkumar, Saint-Jacques, Bojinov,
  Brynjolfsson, and Aral}{Rajkumar et~al.}{2022}]{rajkumar2022causal}
Rajkumar, K., G.~Saint-Jacques, I.~Bojinov, E.~Brynjolfsson, and S.~Aral
  (2022).
\newblock A causal test of the strength of weak ties.
\newblock {\em Science\/}~{\em 377\/}(6612), 1304--1310.

\bibitem[\protect\citeauthoryear{Reagans and Zuckerman}{Reagans and
  Zuckerman}{2001}]{reagans2001networks}
Reagans, R. and E.~W. Zuckerman (2001).
\newblock Networks, diversity, and productivity: The social capital of
  corporate r\&d teams.
\newblock {\em Organization science\/}~{\em 12\/}(4), 502--517.

\bibitem[\protect\citeauthoryear{Rose}{Rose}{2016}]{rose2016identification}
Rose, C. (2016).
\newblock Identification of spillover effects using panel data.
\newblock Technical report, Working Paper.

\bibitem[\protect\citeauthoryear{Segarra and Ribeiro}{Segarra and
  Ribeiro}{2015}]{segarra2015stability}
Segarra, S. and A.~Ribeiro (2015).
\newblock Stability and continuity of centrality measures in weighted graphs.
\newblock {\em IEEE Transactions on Signal Processing\/}~{\em 64\/}(3),
  543--555.

\bibitem[\protect\citeauthoryear{Shao, Xia, and Zhang}{Shao
  et~al.}{2025}]{shao2025u}
Shao, M., D.~Xia, and Y.~Zhang (2025).
\newblock U-statistic reduction: Higher-order accurate risk control and
  statistical-computational trade-off.
\newblock {\em Journal of the American Statistical Association\/}, 1--14.

\bibitem[\protect\citeauthoryear{Thirkettle}{Thirkettle}{2019}]{thirkettle2019identification}
Thirkettle, M. (2019).
\newblock Identification and estimation of network statistics with missing link
  data.
\newblock {\em Working Paper\/}.

\bibitem[\protect\citeauthoryear{Udry}{Udry}{1994}]{udry1994risk}
Udry, C. (1994).
\newblock Risk and insurance in a rural credit market: An empirical
  investigation in northern nigeria.
\newblock {\em The Review of Economic Studies\/}~{\em 61\/}(3), 495--526.

\bibitem[\protect\citeauthoryear{Verdier}{Verdier}{2020}]{verdier2020estimation}
Verdier, V. (2020).
\newblock Estimation and inference for linear models with two-way fixed effects
  and sparsely matched data.
\newblock {\em Review of Economics and Statistics\/}~{\em 102\/}(1), 1--16.

\bibitem[\protect\citeauthoryear{Vershynin}{Vershynin}{2018}]{vershynin2018high}
Vershynin, R. (2018).
\newblock {\em High-dimensional probability: An introduction with applications
  in data science}, Volume~47.
\newblock Cambridge university press.

\bibitem[\protect\citeauthoryear{Xu}{Xu}{2025}]{xu2025networks}
Xu, G. (2025).
\newblock Networks as control functions: Nonparametric identification and
  estimation of partial effects.

\bibitem[\protect\citeauthoryear{Young and Scheinerman}{Young and
  Scheinerman}{2007}]{young2007random}
Young, S.~J. and E.~R. Scheinerman (2007).
\newblock Random dot product graph models for social networks.
\newblock In {\em International Workshop on Algorithms and Models for the
  Web-Graph}, pp.\  138--149. Springer.

\end{thebibliography}
\end{spacing}

\appendix

\section*{Appendices}

\section{Enumerated Trees}\label{appendix--xi_trees}

Tables \ref{tab:xi_trees_1} and \ref{tab:xi_trees_2} contains $\mathcal{G}(t,m)$ and $\gamma(g)$ for $t \leq 10$. 
These are used construct estimators for $\boldsymbol{\xi}(t)$. Trees are stated as edge lists. The number of trees grow exponentially, but computation time is fast even for moderate $t$'s. Also see Figure \ref{figure--contruct_trees} for an example of how to use these tables.

% Table generated by Excel2LaTeX from sheet 'tree_table'
\begin{sidewaystable}[htbp]
  \centering 
  \caption{$\mathcal{G}(t,m)$ for $t \leq 9$. Not all $m \in [t]$ can be reached by the required walks. Trees are given as edge lists.}\footnotesize
    \begin{tabular}{clccc|rclccr|rclccr|rclcc}
    t     & Edge List & m     & $\gamma$ &       &       & t     & Edge List & m     & $\gamma$ &       &       & t     & Edge List & m     & $\gamma$ &       &       & t     & Edge List & m     & $\gamma$ \bigstrut[b]\\
\cline{1-4}\cline{7-10}\cline{13-16}\cline{19-22}    2     & 12    & 1     & 1     &       &       & 7     & 12, 14, 23 & 4     & 1     &       &       & 8     & 12, 23, 24, 45 & 1     & 1     &       &       & 9     & 12, 13, 25, 34 & 2     & 1 \bigstrut[t]\\
    3     & 12    & 2     & 1     &       &       & 7     & 12, 23 & 2     & 6     &       &       & 8     & 12, 23, 25, 34 & 1     & 1     &       &       & 9     & 12, 13, 34 & 3     & 10 \\
    4     & 12    & 1     & 1     &       &       & 7     & 12, 23, 24 & 2     & 3     &       &       & 8     & 12, 23, 34 & 3     & 4     &       &       & 9     & 12, 13, 34 & 2     & 5 \\
    4     & 12, 13 & 1     & 1     &       &       & 7     & 12, 23, 34 & 2     & 2     &       &       & 8     & 12, 23, 34 & 1     & 5     &       &       & 9     & 12, 13, 34, 35 & 3     & 3 \\
    4     & 12, 23 & 1     & 1     &       &       & 8     & 12    & 1     & 1     &       &       & 8     & 12, 23, 34, 35 & 1     & 1     &       &       & 9     & 12, 13, 34, 35 & 2     & 1 \\
    5     & 12    & 2     & 1     &       &       & 8     & 12, 13 & 1     & 7     &       &       & 8     & 12, 23, 34, 45 & 1     & 1     &       &       & 9     & 12, 13, 34, 45 & 3     & 2 \\
    5     & 12, 13 & 2     & 1     &       &       & 8     & 12, 13, 14 & 1     & 6     &       &       & 9     & 12    & 2     & 1     &       &       & 9     & 12, 13, 34, 45 & 2     & 1 \\
    5     & 12, 13 & 3     & 1     &       &       & 8     & 12, 13, 14, 15 & 1     & 1     &       &       & 9     & 12, 13 & 2     & 7     &       &       & 9     & 12, 14, 15, 23 & 2     & 1 \\
    5     & 12, 23 & 2     & 2     &       &       & 8     & 12, 13, 14, 45 & 1     & 1     &       &       & 9     & 12, 13 & 3     & 7     &       &       & 9     & 12, 14, 15, 23 & 4     & 1 \\
    6     & 12    & 1     & 1     &       &       & 8     & 12, 13, 15, 34 & 1     & 1     &       &       & 9     & 12, 13, 14 & 2     & 6     &       &       & 9     & 12, 14, 15, 23 & 5     & 1 \\
    6     & 12, 13 & 1     & 3     &       &       & 8     & 12, 13, 24 & 1     & 1     &       &       & 9     & 12, 13, 14 & 3     & 6     &       &       & 9     & 12, 14, 23 & 2     & 6 \\
    6     & 12, 13, 14 & 1     & 1     &       &       & 8     & 12, 13, 24 & 4     & 1     &       &       & 9     & 12, 13, 14 & 4     & 6     &       &       & 9     & 12, 14, 23 & 4     & 5 \\
    6     & 12, 13, 34 & 1     & 1     &       &       & 8     & 12, 13, 34 & 1     & 5     &       &       & 9     & 12, 13, 14, 15 & 2     & 1     &       &       & 9     & 12, 14, 23, 25 & 2     & 1 \\
    6     & 12, 14, 23 & 1     & 1     &       &       & 8     & 12, 13, 34 & 4     & 2     &       &       & 9     & 12, 13, 14, 15 & 3     & 1     &       &       & 9     & 12, 14, 23, 45 & 4     & 2 \\
    6     & 12, 23 & 1     & 3     &       &       & 8     & 12, 13, 34, 35 & 1     & 1     &       &       & 9     & 12, 13, 14, 15 & 4     & 1     &       &       & 9     & 12, 14, 23, 45 & 2     & 1 \\
    6     & 12, 23 & 3     & 2     &       &       & 8     & 12, 13, 34, 45 & 1     & 1     &       &       & 9     & 12, 13, 14, 15 & 5     & 1     &       &       & 9     & 12, 15, 23, 24 & 2     & 1 \\
    6     & 12, 23, 24 & 1     & 1     &       &       & 8     & 12, 14, 15, 23 & 1     & 1     &       &       & 9     & 12, 13, 14, 25 & 2     & 1     &       &       & 9     & 12, 15, 23, 24 & 5     & 1 \\
    6     & 12, 23, 34 & 1     & 1     &       &       & 8     & 12, 14, 23 & 1     & 5     &       &       & 9     & 12, 13, 14, 35 & 3     & 1     &       &       & 9     & 12, 15, 23, 34 & 2     & 1 \\
    7     & 12    & 2     & 1     &       &       & 8     & 12, 14, 23 & 3     & 1     &       &       & 9     & 12, 13, 14, 45 & 4     & 2     &       &       & 9     & 12, 15, 23, 34 & 5     & 1 \\
    7     & 12, 13 & 2     & 3     &       &       & 8     & 12, 14, 23, 45 & 1     & 1     &       &       & 9     & 12, 13, 14, 45 & 2     & 1     &       &       & 9     & 12, 23 & 2     & 14 \\
    7     & 12, 13 & 3     & 3     &       &       & 8     & 12, 15, 23, 24 & 1     & 1     &       &       & 9     & 12, 13, 14, 45 & 3     & 1     &       &       & 9     & 12, 23, 24 & 2     & 18 \\
    7     & 12, 13, 14 & 2     & 1     &       &       & 8     & 12, 15, 23, 34 & 1     & 1     &       &       & 9     & 12, 13, 15, 34 & 2     & 1     &       &       & 9     & 12, 23, 24, 25 & 2     & 4 \\
    7     & 12, 13, 14 & 3     & 1     &       &       & 8     & 12, 23 & 1     & 7     &       &       & 9     & 12, 13, 15, 34 & 3     & 1     &       &       & 9     & 12, 23, 24, 45 & 2     & 3 \\
    7     & 12, 13, 14 & 4     & 1     &       &       & 8     & 12, 23 & 3     & 6     &       &       & 9     & 12, 13, 15, 34 & 5     & 1     &       &       & 9     & 12, 23, 25, 34 & 2     & 3 \\
    7     & 12, 13, 24 & 2     & 1     &       &       & 8     & 12, 23, 24 & 1     & 6     &       &       & 9     & 12, 13, 24 & 2     & 5     &       &       & 9     & 12, 23, 34 & 2     & 11 \\
    7     & 12, 13, 34 & 3     & 2     &       &       & 8     & 12, 23, 24 & 3     & 3     &       &       & 9     & 12, 13, 24 & 3     & 1     &       &       & 9     & 12, 23, 34 & 4     & 4 \\
    7     & 12, 13, 34 & 2     & 1     &       &       & 8     & 12, 23, 24 & 4     & 3     &       &       & 9     & 12, 13, 24, 25 & 2     & 1     &       &       & 9     & 12, 23, 34, 35 & 2     & 2 \\
    7     & 12, 14, 23 & 2     & 1     &       &       & 8     & 12, 23, 24, 25 & 1     & 1     &       &       & 9     & 12, 13, 24, 45 & 2     & 1     &       &       & 9     & 12, 23, 34, 45 & 2     & 2 \\
    \end{tabular}%
  \label{tab:xi_trees_1}%
\end{sidewaystable}%

% Table generated by Excel2LaTeX from sheet 'tree_table'
\begin{sidewaystable}[htbp]
  \centering
  \caption{$\mathcal{G}(t,m)$ for $t = 10$. Not all $m \in [t]$ can be reached by the required walks. Trees are given as edge lists.}\footnotesize
  \begin{tabular}{clccc|cclccc|cclccc|cclcc}
    t     & Edge List & m     & $\gamma$ &       &       & t     & Edge List & m     & $\gamma$ &       &       & t     & Edge List & m     & $\gamma$ &       &       & t     & Edge List & m     & \multicolumn{1}{c}{$\gamma$} \bigstrut[b]\\
\cline{1-4}\cline{7-10}\cline{13-16}\cline{19-22}    10    & 12    & 1     & 1     &       &       & 10    & 12, 13, 24, 45 & 1     & 1     &       &       & 10    & 12, 14, 23, 45 & 5     & 2     &       &       & \multicolumn{1}{r}{10} & 12, 23, 24, 25, 56 & \multicolumn{1}{l}{1} & \multicolumn{1}{c}{1} \bigstrut[t]\\
    10    & 12, 13 & 1     & 15    &       &       & 10    & 12, 13, 25, 34 & 1     & 1     &       &       & 10    & 12, 14, 23, 45 & 3     & 1     &       &       & \multicolumn{1}{r}{10} & 12, 23, 24, 26, 45 & \multicolumn{1}{l}{1} & \multicolumn{1}{c}{1} \\
    10    & 12, 13, 14 & 1     & 25    &       &       & 10    & 12, 13, 25, 34 & 5     & 1     &       &       & 10    & 12, 14, 23, 45, 46 & 1     & 1     &       &       & \multicolumn{1}{r}{10} & 12, 23, 24, 35 & \multicolumn{1}{l}{3} & \multicolumn{1}{c}{3} \\
    10    & 12, 13, 14, 15 & 1     & 10    &       &       & 10    & 12, 13, 34 & 1     & 18    &       &       & 10    & 12, 14, 23, 45, 56 & 1     & 1     &       &       & \multicolumn{1}{r}{10} & 12, 23, 24, 35 & \multicolumn{1}{l}{1} & \multicolumn{1}{c}{1} \\
    10    & 12, 13, 14, 15, 16 & 1     & 1     &       &       & 10    & 12, 13, 34 & 4     & 10    &       &       & 10    & 12, 15, 16, 23, 24 & 1     & 1     &       &       & \multicolumn{1}{r}{10} & 12, 23, 24, 45 & \multicolumn{1}{l}{4} & \multicolumn{1}{c}{6} \\
    10    & 12, 13, 14, 15, 56 & 1     & 1     &       &       & 10    & 12, 13, 34, 35 & 1     & 8     &       &       & 10    & 12, 15, 16, 23, 34 & 1     & 1     &       &       & \multicolumn{1}{r}{10} & 12, 23, 24, 45 & \multicolumn{1}{l}{1} & \multicolumn{1}{c}{8} \\
    10    & 12, 13, 14, 16, 45 & 1     & 1     &       &       & 10    & 12, 13, 34, 35 & 4     & 3     &       &       & 10    & 12, 15, 23, 24 & 1     & 8     &       &       & \multicolumn{1}{r}{10} & 12, 23, 24, 45 & \multicolumn{1}{l}{3} & \multicolumn{1}{c}{3} \\
    10    & 12, 13, 14, 25 & 1     & 1     &       &       & 10    & 12, 13, 34, 35 & 5     & 3     &       &       & 10    & 12, 15, 23, 24 & 3     & 1     &       &       & \multicolumn{1}{r}{10} & 12, 23, 24, 45, 46 & \multicolumn{1}{l}{1} & \multicolumn{1}{c}{1} \\
    10    & 12, 13, 14, 25 & 5     & 1     &       &       & 10    & 12, 13, 34, 35, 36 & 1     & 1     &       &       & 10    & 12, 15, 23, 24 & 4     & 1     &       &       & \multicolumn{1}{r}{10} & 12, 23, 24, 45, 56 & \multicolumn{1}{l}{1} & \multicolumn{1}{c}{1} \\
    10    & 12, 13, 14, 35 & 1     & 1     &       &       & 10    & 12, 13, 34, 35, 56 & 1     & 1     &       &       & 10    & 12, 15, 23, 24, 56 & 1     & 1     &       &       & \multicolumn{1}{r}{10} & 12, 23, 25, 26, 34 & \multicolumn{1}{l}{1} & \multicolumn{1}{c}{1} \\
    10    & 12, 13, 14, 35 & 5     & 1     &       &       & 10    & 12, 13, 34, 36, 45 & 1     & 1     &       &       & 10    & 12, 15, 23, 34 & 1     & 7     &       &       & \multicolumn{1}{r}{10} & 12, 23, 25, 34 & \multicolumn{1}{l}{1} & \multicolumn{1}{c}{8} \\
    10    & 12, 13, 14, 45 & 1     & 8     &       &       & 10    & 12, 13, 34, 45 & 1     & 7     &       &       & 10    & 12, 15, 23, 34 & 3     & 1     &       &       & \multicolumn{1}{r}{10} & 12, 23, 25, 34 & \multicolumn{1}{l}{3} & \multicolumn{1}{c}{3} \\
    10    & 12, 13, 14, 45 & 5     & 2     &       &       & 10    & 12, 13, 34, 45 & 4     & 4     &       &       & 10    & 12, 15, 23, 34, 56 & 1     & 1     &       &       & \multicolumn{1}{r}{10} & 12, 23, 25, 34 & \multicolumn{1}{l}{5} & \multicolumn{1}{c}{3} \\
    10    & 12, 13, 14, 45, 46 & 1     & 1     &       &       & 10    & 12, 13, 34, 45, 46 & 1     & 1     &       &       & 10    & 12, 16, 23, 24, 25 & 1     & 1     &       &       & \multicolumn{1}{r}{10} & 12, 23, 25, 34, 56 & \multicolumn{1}{l}{1} & \multicolumn{1}{c}{1} \\
    10    & 12, 13, 14, 45, 56 & 1     & 1     &       &       & 10    & 12, 13, 34, 45, 56 & 1     & 1     &       &       & 10    & 12, 16, 23, 24, 45 & 1     & 1     &       &       & \multicolumn{1}{r}{10} & 12, 23, 26, 34, 35 & \multicolumn{1}{l}{1} & \multicolumn{1}{c}{1} \\
    10    & 12, 13, 15, 16, 34 & 1     & 1     &       &       & 10    & 12, 14, 15, 16, 23 & 1     & 1     &       &       & 10    & 12, 16, 23, 25, 34 & 1     & 1     &       &       & \multicolumn{1}{r}{10} & 12, 23, 26, 34, 45 & \multicolumn{1}{l}{1} & \multicolumn{1}{c}{1} \\
    10    & 12, 13, 15, 24 & 1     & 1     &       &       & 10    & 12, 14, 15, 23 & 1     & 8     &       &       & 10    & 12, 16, 23, 34, 35 & 1     & 1     &       &       & \multicolumn{1}{r}{10} & 12, 23, 34 & \multicolumn{1}{l}{3} & \multicolumn{1}{c}{21} \\
    10    & 12, 13, 15, 34 & 1     & 8     &       &       & 10    & 12, 14, 15, 23 & 3     & 1     &       &       & 10    & 12, 16, 23, 34, 45 & 1     & 1     &       &       & \multicolumn{1}{r}{10} & 12, 23, 34 & \multicolumn{1}{l}{1} & \multicolumn{1}{c}{18} \\
    10    & 12, 13, 15, 34 & 4     & 1     &       &       & 10    & 12, 14, 15, 23, 56 & 1     & 1     &       &       & 10    & 12, 23 & 1     & 15    &       &       & \multicolumn{1}{r}{10} & 12, 23, 34, 35 & \multicolumn{1}{l}{3} & \multicolumn{1}{c}{6} \\
    10    & 12, 13, 15, 34, 56 & 1     & 1     &       &       & 10    & 12, 14, 16, 23, 45 & 1     & 1     &       &       & 10    & 12, 23 & 3     & 14    &       &       & \multicolumn{1}{r}{10} & 12, 23, 34, 35 & \multicolumn{1}{l}{1} & \multicolumn{1}{c}{8} \\
    10    & 12, 13, 16, 34, 35 & 1     & 1     &       &       & 10    & 12, 14, 23 & 1     & 18    &       &       & 10    & 12, 23, 24 & 1     & 25    &       &       & \multicolumn{1}{r}{10} & 12, 23, 34, 35, 36 & \multicolumn{1}{l}{1} & \multicolumn{1}{c}{1} \\
    10    & 12, 13, 16, 34, 45 & 1     & 1     &       &       & 10    & 12, 14, 23 & 3     & 6     &       &       & 10    & 12, 23, 24 & 3     & 18    &       &       & \multicolumn{1}{r}{10} & 12, 23, 34, 35, 56 & \multicolumn{1}{l}{1} & \multicolumn{1}{c}{1} \\
    10    & 12, 13, 24 & 1     & 6     &       &       & 10    & 12, 14, 23, 25 & 1     & 1     &       &       & 10    & 12, 23, 24 & 4     & 18    &       &       & \multicolumn{1}{r}{10} & 12, 23, 34, 36, 45 & \multicolumn{1}{l}{1} & \multicolumn{1}{c}{1} \\
    10    & 12, 13, 24 & 4     & 5     &       &       & 10    & 12, 14, 23, 25 & 3     & 1     &       &       & 10    & 12, 23, 24, 25 & 1     & 10    &       &       & \multicolumn{1}{r}{10} & 12, 23, 34, 45 & \multicolumn{1}{l}{3} & \multicolumn{1}{c}{4} \\
    10    & 12, 13, 24, 25 & 1     & 1     &       &       & 10    & 12, 14, 23, 25 & 5     & 1     &       &       & 10    & 12, 23, 24, 25 & 3     & 4     &       &       & \multicolumn{1}{r}{10} & 12, 23, 34, 45 & \multicolumn{1}{l}{1} & \multicolumn{1}{c}{7} \\
    10    & 12, 13, 24, 25 & 4     & 1     &       &       & 10    & 12, 14, 23, 35 & 3     & 1     &       &       & 10    & 12, 23, 24, 25 & 4     & 4     &       &       & \multicolumn{1}{r}{10} & 12, 23, 34, 45, 46 & \multicolumn{1}{l}{1} & \multicolumn{1}{c}{1} \\
    10    & 12, 13, 24, 25 & 5     & 1     &       &       & 10    & 12, 14, 23, 45 & 1     & 7     &       &       & 10    & 12, 23, 24, 25 & 5     & 4     &       &       & \multicolumn{1}{r}{10} & 12, 23, 34, 45, 56 & \multicolumn{1}{l}{1} & \multicolumn{1}{c}{1} \\
    10    & 12, 13, 24, 45 & 4     & 2     &       &       &       &       &       &       &       &       & 10    & 12, 23, 24, 25, 26 & 1     & 1     &       &       &       &       &       &  \\
    \end{tabular}%
  \label{tab:xi_trees_2}%
\end{sidewaystable}%

\newpage 

\section{Reducing Computation for $\check{Q}(B,w,v)$}\label{appendix--speed_up_B}

The estimator $\check{Q}(B,w,v)$ has the structure of U-statistics, which can be computationally intensive (see e.g. \citealt{blom1976some,chen2019randomized,shao2025u}). It averages only over index tuples without repeated nodes. Although tuples with repeated nodes are asymptotically rare, their contribution to the bias is non-negligible in sparse regimes, making their removal necessary. We therefore consider a randomized estimator. To estimate the bias of a term whose associated tree has $n(g)$ nodes, first randomly partition the observations into $n(g)$ equally sized subsets $\mathcal{N}_1,\cdots,\mathcal{N}_{n(g)}$, dropping observations as needed. The estimator is then constructed using indices $i_j$ drawn only from $\mathcal{N}_j$:
\begin{equation*}\small
	\check{Q}^r(B,w,v) = \sum_{g \in \mathcal{T}(B)} k_B(g) \cdot n(g)^{n(g)}\cdot \sum_{\substack{i_1 \in \mathcal{N}_1, ...,\\  i_{n(g)} \in \mathcal{N}_{n(g)}}} w_{i_1} \left(\prod_{jk \in E(g)} \hat{A}_{i_j, i_k} \right)v_{i_m}~.
\end{equation*}
The $n(g)^{n(g)}$ factor corrects for the fact that each tree in $\check{Q}^r(B,w,v)$ is estimated using approximately $1/n(g)^{n(g)}$ of the full index set. Since this fraction is constant, $\check{Q}^r(B,w,v)$ has the same rate of convergence as $\check{Q}(B,w,v)$ so that
\begin{equation*}\small
	\sqrt{n}\delta_n^t \cdot (\check{Q}^r(B,w,v) - Q(B,w,v)) = o_p(1)~.
\end{equation*}

Let $\hat{A}_{j\to k}$ denote the submatrix of links between individuals in $\mathcal{N}_j$ and $\mathcal{N}_k$. Then $\check{Q}^r(B,w,v)$ can be written as a sequence of matrix and Hadamard products involving $\hat{A}_{j\to k}$. Specifically, multiply the terminal node by $v$ and each other leaf by the all-ones vector of length $n/n(g)$. Working from the leaves upward, replace each non-branching node with a matrix product and each branching node with a Hadamard product across its branches. As an example, the above algorithm applied to the first term in Figure \ref{figure--contruct_trees} yields:
\begin{align*}\footnotesize
	& \sum_{\substack{i_1 \in \mathcal{N}_1, ..., 
	i_5 \in \mathcal{N}_5}} w_{i_1} \hat{A}_{i_1,i_2}\hat{A}_{i_2,i_3}\hat{A}_{i_2,i_4}\hat{A}_{i_4,i_5} v_{i_5}  =  w_1'\times \hat{A}_{1 \to 2} \times ((\hat{A}_{2 \to 3}\times \iota) \circ (\hat{A}_{2 \to 4} \times \hat{A}_{4 \to 5}\times v_5))~.% \\[1em]
%	& \sum_{\substack{i_1 \in \mathcal{N}_1, ..., 
%		i_6 \in \mathcal{N}_6}} w_{i_1} \hat{A}_{i_1,i_2}\hat{A}_{i_2,i_3}\hat{A}_{i_2,i_4}\hat{A}_{i_2,i_5} \hat{A}_{i_5,i_6} v_{i_6} \\ 	& \qquad \qquad = \quad w_1'\times \hat{A}_{1 \to 2} \times ((\hat{A}_{2 \to 3}\times \iota) \circ (\hat{A}_{2 \to 4}\times \iota)  \circ (\hat{A}_{2 \to 5} \times \hat{A}_{5 \to 6}\times v_6))~, \\[1em]
%	& \sum_{\substack{i_1 \in \mathcal{N}_1, ..., 
%		i_6 \in \mathcal{N}_6}} w_{i_1} \hat{A}_{i_1,i_2}\hat{A}_{i_2,i_3}\hat{A}_{i_3,i_4}\hat{A}_{i_2,i_5} \hat{A}_{i_5,i_6} v_{i_6} \\	& \qquad \qquad = \quad w_1'\times \hat{A}_{1 \to 2} \times ((\hat{A}_{2 \to 3}\times \hat{A}_{3 \to 4} \times \iota) \circ (\hat{A}_{2 \to 5} \times \hat{A}_{5 \to 6}\times v_6))~.
\end{align*}

To de-randomize this estimator, draw multiple such partitions, compute $\check{Q}^r$ on each partition and take the average. 

\clearpage
\newpage

\section{Proofs}\label{appendix--proofs}

We introduce some additional notation and useful lemmas before proving our results. Proofs of the following lemmas are contained in Appendix \ref{appendix--lemma_proofs}.

Define the following: 
\begin{align*}\small
	\mathbf{C}^{(d)} = \begin{pmatrix}
		C_1^{(d)} \\
		\vdots \\
		C_n^{(d)}\\
	\end{pmatrix} 
	\; , \; 	
	\hat{\mathbf{C}}^{(d)} = \begin{pmatrix}
		\hat{C}_1^{(d)} \\
		\vdots \\
		\hat{C}_n^{(d)}\\
	\end{pmatrix} 
	\; , \; 	
	\mathbf{C}^{*,(d)} = \begin{pmatrix}
			C_1^{*,(d)} \\
			\vdots \\
			C_n^{*,(d)}\\
		\end{pmatrix}
		\; , \; 	
	\mathbf{W} = \begin{pmatrix}
			W_1'\\
			\vdots \\
			W_n'\\
		\end{pmatrix}
	\; , \; 	
	\bm{\varepsilon} =  \begin{pmatrix}
				\varepsilon_1\\
				\vdots \\
				\varepsilon_n\\
			\end{pmatrix}
\end{align*}
We omit the dependence on $(d)$ and refer instead to $\mathbf{C}$/$C_i$, $\hat{\mathbf{C}}$/$\hat{C}_i$ or $\mathbf{C}^*$/$C_i^*$ when the centrality measure in question is clear. Also define the projection and residual maker matrices: $\mathbf{P}_W = \mathbf{W}(\mathbf{W}'\mathbf{W})^{-1}\mathbf{W}'$ and $\mathbf{M}_W = \mathbf{I} - \mathbf{P}_W$.

\begin{lemma}\label{lemma--graphon-centrality}
	Suppose Assumption \ref{assumption--graphon} holds. Let $\eta > 0$. For $d\in \{1,T\}$,
	\begin{equation}\label{equation--graphon_centrality_1}
		\max_{i\in [n]} \; \lVert C_i^{(d)} - C_i^{*,(d)} \rVert  \leq n^{-1/2+\eta} \quad \mbox{ w.p.a. 1}~.
	\end{equation}
	Suppose additionally that Assumption \ref{assumption--eigenvector_identification} holds. Then,
	\begin{equation}\label{equation--graphon_centrality_2}
		\mathbb{E}\left[\frac{1}{n} \sum_{i=1}^n \left(C_i^{(\infty)} - C_i^{*,(\infty)} \right)^2\right] \to 0
	\end{equation}
	Note that \eqref{equation--graphon_centrality_1} implies \eqref{equation--graphon_centrality_2} for $d \in \{1,T\}$ since  $|C_i^{(d)} - C_i^{*,(d)}| \leq 2$.
\end{lemma}

\begin{definition}\label{definition--induced_graphs_classes}
	Let $\mathcal{T}(t)$ be the set of indices with length $t$ for which every edge in $g(\mathbf{i})$ has multiplicity at least 2 and $g^\text{simp}(\mathbf{i})$ is a tree. Let $\mathcal{T}^c(t)$ be be the set of indices with length $t$ for which every edge in $g(\mathbf{i})$ has multiplicity at least 2 but $g^\text{simp}(\mathbf{i})$ is not a tree (i.e. contains cycles). 
\end{definition}

\begin{lemma}\label{lemma--order-xit}
	Suppose Assumptions \ref{assumption--graphon} and \ref{assumption--measurement_model} hold and that $np_n = \Omega(1)$. Let $t \geq 2$ and the deterministic vectors $w$ and $v$ be given. Define the matrix $\tilde{\boldsymbol{\xi}}(t)$ such that
	\begin{equation*}
		\left[\tilde{\boldsymbol{\xi}}(t)\right]_{l,m} =  \sum_{\substack{\mathbf{i} \in \mathcal{T}(t) \\ i_1 = l, i_{t+1} = m}}     \left(\prod_{jk \in E(\mathbf{i})} p_nf(U_j, U_k)  \right)
	\end{equation*}
	Then
\begin{align*}
	& \expect\left[w'\boldsymbol{\xi}^t v \mid \mathbf{U}, \mathbf{W}\right] = (1 + O(p_n)) \cdot 	\left(w' \tilde{\boldsymbol{\xi}}(t) v + w'R(\boldsymbol{\xi}^t)v \right)
\end{align*}
where the $(l,m)$-entry of the matrix $\tilde{\boldsymbol{\xi}}(t)$ satisfies:
\begin{equation*}
		\Big\lvert [\tilde{\boldsymbol{\xi}}(t)]_{l,m} \Big\rvert \leq (t+1)^t \cdot n^{\lfloor t/2 \rfloor-1}p_n^{\lfloor t/2 \rfloor}~,
\end{equation*}
and the $(l,m)$-entry of the matrix $R(\boldsymbol{\xi}^t)$ satisfies:
\begin{align*}
	\Big\lvert [R(\boldsymbol{\xi}^t)]_{l,m}\Big\rvert \leq \begin{cases}
		 (t+1)^t \cdot n^{{\lfloor t/2 \rfloor}-2}p_n^{{\lfloor t/2 \rfloor}} & \mbox{ if } t \geq 6 \\
		 0 & \mbox{ if } t < 6
	\end{cases}
\end{align*}
\end{lemma}

Note that $R$ contains contributions from $\mathbf{i}$ for which $g^{\text{simp}}(\mathbf{i})$ contains cycles. Such cycles exist only for walks with at least 6 edges i.e. $t \geq 6$.

\begin{lemma}\label{lemma--order-B}
	Suppose Assumptions \ref{assumption--graphon} and \ref{assumption--measurement_model} hold and that $np_n = \Omega(1)$. Let $B$ be a mixed product of order $(t,\tau)$, where $t \geq \tau \geq 2$. Then,
	\begin{equation*}
	\expect\left[w' B v  \mid \mathbf{U}, \mathbf{W}\right]  =	(1+O(p_n)) \cdot (w'\tilde{B}v + w'R(B)v)
	\end{equation*} 
	where $r$ is the number of blocks in $B$ (i.e. length of $p$) and
	\begin{equation*}
		\tilde{B} =\prod_{s=1}^r  \Big(\mathbbm{1}\left\{q_s = 0\right\} \cdot A^{p_s} + \mathbbm{1}\left\{q_s = 1\right\} \cdot \tilde{\boldsymbol{\xi}}(p_s)  \Big)
	\end{equation*}
	The $(l,m)$-entry of the matrix $\tilde{B}$ satisfies 
	\begin{equation*}
			\left\lvert [\tilde{B}]_{l,m} \right\rvert \leq (t+1)^t \cdot n^{t-2}p_n^{t-1}~,
	\end{equation*} 
	and the $(l,m)$-entry of the matrix $R(\boldsymbol{\xi}^t)$ satisfies 
	\begin{align*}
		\Big\lvert [R(\boldsymbol{\xi}^t)]_{l,m}\Big\rvert \leq \begin{cases}
			 (t+1)^t \cdot n^{t-3}p_n^{t-1} & \mbox{ if } t \geq 3 \\
			 0 & \mbox{ if } t < 3
		\end{cases}~.
	\end{align*}
\end{lemma}
 The lemma shows that the mean of  $w'Bv$ can be approximated by replacing each contiguous block of $\boldsymbol{\xi}^t$ with $\tilde{\boldsymbol{\xi}}(t)$. For example, 
$
	\expect\left[w' \boldsymbol{\xi}^2 A \boldsymbol{\xi}^2 v  \mid \mathbf{U}, \mathbf{W}\right] \approx w'\tilde{\boldsymbol{\xi}}(2)A\tilde{\boldsymbol{\xi}}(2)v
$.
As before, when $\tau = 1$, $\expect\left[w' B v  \mid \mathbf{U}, \mathbf{W}\right] = 0$. Relative to the previous lemma, blocks of mixed order can generate graphs with expectation that are non-zero as long as $t \geq 3$. However, the contributions of these graphs are still asymptotically negligible.

\begin{lemma}\label{lemma--debiased-B}
	Suppose Assumptions \ref{assumption--graphon} and \ref{assumption--measurement_model} hold and that $np_n = \Omega(1)$. When $t \geq 2$,
	\begin{equation*}
	\expect\left[\left(	w'Bv - \expect\left[w' B v  \mid \mathbf{U}, \mathbf{W}\right]  \right)^2  \mid \mathbf{U}, \mathbf{W} \right] \leq (2t+2)^{2t-2} \cdot n^{2t-4} p_n^{2t-1} \cdot \sum_{i_1, i_2, i_3, i_4} |w_{i_1}v_{i_2}w_{i_3} v_{i_4}|
	\end{equation*}
	When $B = \boldsymbol{\xi}$, 
	\begin{equation*}
		\expect\left[\left(	w'\boldsymbol{\xi}v \right)^2  \mid \mathbf{U}, \mathbf{W} \right] \leq 2p_n \sum_{i_1, i_2} w_{i_1}^2 v_{i_2}^2 
	\end{equation*}
	When $B = A$, the remainder is $0$. 
\end{lemma}

\begin{lemma}[Concentration in Spectral Norm]\label{lemma--spectral_conc}
	Suppose Assumptions \ref{assumption--graphon} and \ref{assumption--measurement_model} hold and that
	\begin{equation*}
		np_n \gg \sqrt{\frac{\log n}{\log \log n}}~.
	 \end{equation*}
	Then with probability approaching 1, we have that for $\nu \in (0,1)$,
	\begin{equation*}
		\left\lVert A -\hat{A} \right\rVert \leq k \left(np_n\right)^{(1+\nu)/2} \left(\frac{ \log n}{\log \log n}\right)^{(1-\nu)/4}~,
	\end{equation*}
	where $k$ is a universal constant.
\end{lemma}

\begin{lemma}\label{lemma--eigenvector_of_A}
	Under Assumptions \ref{assumption--graphon} and \ref{assumption--rank_r}, the eigenvectors of $A$ satisfy
	\begin{equation*}
		v_s(A) = \sum_{r=1}^{R} \alpha_{s,r}\frac{\phi_r(\mathbf{U})}{\sqrt{n}}
	\end{equation*}
	where $\prob\left(|\alpha_{s,r}| \leq 1+\varepsilon\right) \to 1$ for all $s,r \in [R]$. Moreover $|\alpha_{s,s}| \pto 1$ and $\alpha_{s,r} \pto 0$ for $r \neq s$ and $\lambda_s(A)/np_n \pto \lambda_s$. By the Perron-Frobenius theorem, we can choose $v_1(A)$ such that $\alpha_{1,1} \pto 1$. 
\end{lemma}

\subsection{Proof of Theorem \ref{theorem--benchmark}}

By the Frisch-Waugh-Lovell theorem, write:
\begin{equation}\label{equation--benchmark-FWL}
	\sqrt{n}\left(\tilde{\beta}- \beta\right) = \frac{\frac{1}{\sqrt{n}}\mathbf{C}' \mathbf{M}_W \bm{\varepsilon}}{\frac{1}{n}\mathbf{C}' \mathbf{M}_W \mathbf{C}}
\end{equation}

Let $\tilde{\pi} = (\mathbf{W}'\mathbf{W})^{-1}\mathbf{W}'\mathbf{C}$. Note that $\tilde{\pi}  \overset{p}{\to} \pi =: \mathbb{E}[W_iW_i']^{-1}\mathbb{E}[W_iC_i^*]$. To see this, write:
\begin{align*}
	\tilde{\pi}  = (\mathbf{W}'\mathbf{W})^{-1}\mathbf{W}'\mathbf{C}^* + (\mathbf{W}'\mathbf{W})^{-1}\mathbf{W}'(\mathbf{C} - \mathbf{C}^*)
\end{align*}
The first term above converges to $\pi$ by standard arguments. The second term is negligible since for all covariate $j$, we have:
\begin{equation*}
	\left(\frac{1}{n} \sum_{i=1}^n W_{i,j}\left(C_i - C_i^*\right) \right)^2 \leq \left(\frac{1}{n}\sum_{i=1}^n W_{i,j}^2 \right)\left(\frac{1}{n}\sum_{i=1}^n\left(C_i- C_i^{*}\right)^2\right) \overset{p}{\to} 0 \quad \mbox{ by Lemma \ref{lemma--graphon-centrality}.}
\end{equation*}

To derive the asymptotic distribution of $\tilde{\beta}$, start with the numerator in \eqref{equation--benchmark-FWL}:
\begin{align*}
	\frac{1}{\sqrt{n}} \mathbf{C}' \mathbf{M}_W \bm{\varepsilon} &= \frac{1}{\sqrt{n}} \sum_{i=1}^n \left(C_i - W_i'\tilde{\pi}\right)\varepsilon_i \\
	& =  \frac{1}{\sqrt{n}} \sum_{i=1}^n \left(C^*_i - W_i'{\pi}\right)\varepsilon_i + \frac{1}{\sqrt{n}} \sum_{i=1}^n \left(C_i - C_i^*\right)\varepsilon_i  - \left(\tilde{\pi} - \pi\right)' \frac{1}{\sqrt{n}} \sum_{i=1}^n W_i\varepsilon_i 
\end{align*}
By the Lindeberg-Levy CLT, the first term is asymptotically normal:
\begin{equation*}
	\frac{1}{\sqrt{n}} \sum_{i=1}^n \left(C^*_i - W_i'{\pi}\right)\varepsilon_i  \dto \norm\left(0, \expect\left[\left(C^*_i - W_i'{\pi}\right)^2\varepsilon_i^2\right]\right)~.
\end{equation*}
The second term is $o_p(1)$ since the Cauchy-Schwarz inequality and Assumption \ref{assumption--varep_condmoments} yield:
\begin{equation*}
	\expect\left[ \left( \frac{1}{\sqrt{n}} \sum_{i=1}^n \left(C_i - C_i^*\right)\varepsilon_i   \right)^2 \; \bigg\lvert\; \mathbf{U}, \mathbf{W}  \right] \leq \bar{\sigma}^2 \cdot \frac{1}{n} \sum_{i=1}^n \left(C_i - C_i^*\right)^2 
\end{equation*}
Taking expectations and applying Lemma \ref{lemma--graphon-centrality} gives us that this term is negligible in the limit. Finally, the last term is $o_p(1)$ since $\tilde{\pi} \pto \pi$. 

An almost identical argument to the above gives us that
\begin{align*}
	\frac{1}{n}\mathbf{C}' \mathbf{M}_W \mathbf{C} & = \frac{1}{n} \sum_{i=1}^n \left(C^*_i - W_i'{\pi}\right)^2 + o_p(1)  \pto \expect\left[\left(C^*_i - W_i'{\pi}\right)^2 \right]~.
\end{align*}
As such, 
\begin{equation*}
	\sqrt{n}\left(\tilde{\beta} - \beta \right)  \dto \norm\left( 0 \; , \; \frac{\expect\left[\left(C^*_i - W_i'{\pi}\right)^2\varepsilon_i^2 \right]}{\expect\left[\left(C^*_i - W_i'{\pi}\right)^2 \right]^2}\right)~.
\end{equation*}

\subsection{Proof of Theorem \ref{theorem--consistency--error}}

We prove only inconsistency here. Consistency follows from the proof of Theorem \ref{theorem--distribution_obs_error}. 

\subsubsection*{Inconsistency of Degree and Diffusion}

Consider a counterexample in which $f = 1$, $a_*^{(1)} = \beta^{(1)}= 1$ and there are no other covariates. Let $\varepsilon_i \indep U_i$ for all $i \in [n]$. We develop the case for degree in detail. The argument for diffusion centrality is essentially the same. Write:
\begin{equation*}
	\hat{\beta}^{(1)} = \frac{\mathbf{\hat{C}}'\mathbf{C}}{\mathbf{\hat{C}}'\mathbf{\hat{C}}} + \frac{\mathbf{\hat{C}}'\bm{\varepsilon}}{\mathbf{\hat{C}}'\mathbf{\hat{C}}}
\end{equation*}
We will show that the first term does not converge in probability to $1$ as long as $p_n = O(1/n)$ and that the second term is asymptotically negligible. 

Start with the numerator of the first term:
\begin{align*}
	\frac{1}{n} \mathbf{\hat{C}}'\mathbf{C} = \frac{1}{n} \mathbf{{C}}'\mathbf{C} + \frac{1}{n} (\mathbf{\hat{C}}-\mathbf{C})'\mathbf{C}~.
\end{align*}
We have already shown that $\frac{1}{n} \mathbf{{C}}'\mathbf{C} \pto \expect\left[\left(C_i^{*}\right)^2\right]$. The remaining term is:
\begin{align}\label{equation--order-Axi}
	\frac{1}{n} \mathbf{C}'(\mathbf{\hat{C}}-\mathbf{C}) & = \frac{\iota' A\boldsymbol{\xi}\iota}{n^3p_n^2}  
	% & =  \frac{1}{n^3p_n^2} \sum_{i_1, i_2, i_3} p_nf(U_{i_1}, U_{i_2}) \xi_{i_2,i_3} \\
	% = O_p\left(\frac{\sqrt{n^4p_n^3}}{n^3p_n^2}\right) 
	= O_p\left(\frac{1}{n\sqrt{p_n}}\right) \quad \mbox{ by Lemma \ref{lemma--debiased-B}} 
\end{align}
Note that even though Lemma \ref{lemma--debiased-B} formally assumes $p_n \gg 1/n$, this condition is used only to simplify the bounds by replacing terms of the form $n^{k+1}p_n^k$ with $n^{t}p_n^{t-1}$. This is not necessary when $t=2$ because there is only one graph associated with a non-zero expectation.

For the denominator,
\begin{equation*}
	\frac{1}{n}\mathbf{\hat{C}}'\mathbf{\hat{C}} = \frac{1}{n}\mathbf{{C}}'\mathbf{{C}} + 2 \cdot  \frac{\iota'A\boldsymbol{\xi}\iota}{n^3p^2_n}   + \frac{\iota \boldsymbol{\xi}^2 \iota}{n^3p_n^2} ~.
\end{equation*}
where the last term is new. Its mean is 
\begin{equation}\label{equation--deg_bias_tight_estimate}
	\expect\left[\frac{\iota \boldsymbol{\xi}^2 \iota}{n^3p_n^2} \right] = \frac{1}{n^3p_n^2} \sum_{i_1, i_2} \expect[\xi_{i_1,i_2}^2] = \frac{n(n-1)p_n}{n^3p_n^2} = \Theta\left(\frac{1}{np_n}\right)~.
\end{equation}
Its variance is
\begin{align*}
	\expect\left[ \left( \frac{\iota \boldsymbol{\xi}^2 \iota}{n^3p_n^2} - \expect\left[ \frac{\iota \boldsymbol{\xi}^2 \iota}{n^3p_n^2} \right]\right)^2\right] = \frac{1}{n^6p_n^4} \sum_{i_1, ..., i_6} \expect\left[\left(\xi_{i_1,i_2}\xi_{i_2,i_3} - \expect[\xi_{i_1,i_2}\xi_{i_2,i_3} ]\right) \left(\xi_{i_4,i_5}\xi_{i_5,i_6} - \expect[\xi_{i_4,i_5}\xi_{i_5,i_6} ]\right)\right]
\end{align*}
As in the proof of Lemma \ref{lemma--debiased-B}, the indices $(i_1, i_2,i_3)$ and $(i_4,i_5,i_6)$ define two walks. The expectation associated with $\mathbf{i} = (i_1, ..., i_6)'$ is non-zero only when these two walks share at least 1 edge. Otherwise, 
\begin{multline*}
	\expect\left[\left(\xi_{i_1,i_2}\xi_{i_2,i_3} - \expect[\xi_{i_1,i_2}\xi_{i_2,i_3} ]\right) \left(\xi_{i_4,i_5}\xi_{i_5,i_6} - \expect[\xi_{i_4,i_5}\xi_{i_5,i_6} ]\right)\right] \\
	= \expect\left[\xi_{i_1,i_2}\xi_{i_2,i_3} - \expect[\xi_{i_1,i_2}\xi_{i_2,i_3} ]\right] \expect\left[\xi_{i_4,i_5}\xi_{i_5,i_6} - \expect[\xi_{i_4,i_5}\xi_{i_5,i_6} ]\right] = 0
\end{multline*}
Moreover, every edge in the graph needs to have multiplicity at least 2. There are two graphs that satisfy the above criteria (up to isomorphism, i.e. reversing start and end points). These are $g_1$ and $g_2$ in Figure \ref{fig:g-variance}. $g_1$ is associated with the expectation $\expect[\xi^2_{i_1,i_2}\xi^2_{i_2,i_3}] = p_n^2(1-p_n)^2$. 
Meanwhile, $g_2$ gives us $\expect[\xi^4_{i_1,i_2}] = p_n(1-p_n)(1-2p_n)(1-3p_n)$. There are $n(n-1)(n-2)$ copies of $g_1$ and $n(n-1)$ copies of $g_2$. As such,
\begin{equation*}
	\expect\left[ \left( \frac{\iota \boldsymbol{\xi}^2 \iota}{n^3p_n^2} - \expect\left[ \frac{\iota \boldsymbol{\xi}^2 \iota}{n^3p_n^2} \right]\right)^2\right]  % = O\left( \frac{n^3p_n^2}{n^6p_n^4} \right) + O\left( \frac{n^2p_n}{n^6p_n^4} \right) 
	% = \Theta\left(\frac{1}{n^3p_n^2}\right) + \Theta\left(\frac{1}{n^4p_n^3}\right)\\
	= \underbrace{\Theta\left(\frac{1}{n^3p_n^2}\right)}_{\text{from }g_1} +  \underbrace{\Theta\left(\frac{1}{n^4p_n^3}\right)}_{\text{from }g_2}~.
\end{equation*}

\begin{figure}
	\centering
	\include{figures/g_variance}
	\vspace{-5mm}
	\caption{Some graphs that can be formed with 2 walks of length 2, colored red and blue respectively. $g_1$ and $g_2$ are the only graphs that make a non-zero contribution to the variance of $\iota'\boldsymbol{\xi}^2\iota$. $g_3$ has zero expectation since the walks do not share any edge. $g_4$ has zero expectation since there are edges with multiplicity $1$.} \label{fig:g-variance}
\end{figure}

Suppose $1/n^2 \ll p_n \leq p/n$. By \eqref{equation--order-Axi}, $ {\iota' A\boldsymbol{\xi}\iota}/{n^3p_n^2}  = o_p(1)$. By Chebychev's inequality, we also have that for $k > 0$ small enough and $n$ large enough, 
\begin{align*}
	\prob\left( \frac{\iota \boldsymbol{\xi}^2 \iota}{n^3p_n^2}  \geq k \right) & \geq \prob\left( \left\vert \frac{\iota \boldsymbol{\xi}^2 \iota}{n^3p_n^2} - \expect\left[ \frac{\iota \boldsymbol{\xi}^2 \iota}{n^3p_n^2}\right] \right\vert  \leq \frac{1}{2}\expect\left[ \frac{\iota \boldsymbol{\xi}^2 \iota}{n^3p_n^2} \right] \right) \\
	& \geq 1 - \frac{4\expect\left[ \left( \frac{\iota \boldsymbol{\xi}^2 \iota}{n^3p_n^2}  - \expect\left[ \frac{\iota \boldsymbol{\xi}^2 \iota}{n^3p_n^2} \right]\right)^2\right]}{\expect\left[ \frac{\iota \boldsymbol{\xi}^2 \iota}{n^3p_n^2} \right]^2 }  = 1-O\left(\frac{1}{n^2p_n}\right) \to 1~.
\end{align*}
Consequently,
\begin{equation*}
	\prob\left( \left\lvert \frac{\mathbf{\hat{C}}'\mathbf{C}}{\mathbf{\hat{C}}'\mathbf{\hat{C}}} \right\rvert \leq \frac{\expect\left[\left(C_i^*\right)^2\right]}{\expect\left[\left(C_i^*\right)^2\right]+k} \right) \to 1~. % \quad \implies \quad  \hat{\beta} \mbox{ is inconsistent}~.
\end{equation*}
In other words, we have attenuation bias w.p.a. 1. The above argument works for any $k > 0$ when $p_n \ll 1/n$ since 
$\expect\left[ \frac{\iota \boldsymbol{\xi}^2 \iota}{n^3p_n^2} \right]  \to \infty$. In this case,  $\mathbf{\hat{C}}'\mathbf{C}/\mathbf{\hat{C}}'\mathbf{\hat{C}}  \pto 0$. 

It remains to check that $\frac{1}{n}\hat{\mathbf{C}}'\bm{\varepsilon}$ is small. Indeed, 
\begin{equation*}
	\expect\left[ \left(\frac{\iota' A\bm{\varepsilon}}{n^2p_n}\right)^2\right] =  \frac{\sum_{j=1}^n (n-1)^2p^2_n \expect[\varepsilon_j^2]}{n^4p_n^2}= \Theta\left(\frac{1}{n}\right)
\end{equation*}
and 
\begin{equation*}
	\expect\left[ \left(\frac{\iota' \boldsymbol{\xi}\bm{\varepsilon}}{n^2p_n}\right)^2\right] =  \frac{\sum_{i,j=1}^n  \expect[\xi_{i,j}^2\varepsilon_j^2]}{n^4p_n^2}= \Theta\left(\frac{1}{n^2p_n}\right)~.
\end{equation*}
As such, when $p_n \gg 1/n^2$, 
\begin{equation*}
	\frac{\iota'A\bm{\varepsilon}}{n^2p_n} = o_p(1) \quad \mbox{ and } \quad \frac{\iota' \boldsymbol{\xi}\bm{\varepsilon}}{n^2p_n} = o_p(1)~.
\end{equation*}
Conclude that $\hat{\beta}$ is inconsistent when $1/n^2 \ll p_n \leq p/n$.

Finally, suppose $p_n \leq p/n^2$. Then, 
\begin{equation*}
	\liminf_{n\to\infty} \prob\left(\hat{A}_{i_1,i_2} = 0 \mbox{ for all } i_1, i_2\right)  \geq 	\liminf_{n\to\infty} \left(1-\frac{p}{n^2}\right)^{n(n-1)/2} \geq e^{-p/2}
\end{equation*}
In other words, $\hat{\mathbf{C}} = \mathbf{0}$ and OLS is undefined with strictly positive probability. Putting the two cases together, we have that $\hat{\beta}$ is inconsistent whenever $p_n = O(1/n)$.

\subsubsection*{Inconsistency of Eigenvector}

Consider again the counterexample in which $f = 1$, $a_*^{(\infty)} = \beta^{(\infty)}  = 1$, $\varepsilon_i \sim \norm(0, \sigma^2)$, $\varepsilon_i \indep U_i$. Since $v_1(\hat{A})'v_1(\hat{A}) = 1$, we can write:
\begin{align*}
	\hat{\beta}^{(\infty)} & = \frac{\mathbf{\hat{C}}'\mathbf{C}}{\mathbf{\hat{C}}'\mathbf{\hat{C}}} + \frac{\mathbf{\hat{C}}'\bm{\varepsilon}}{\mathbf{\hat{C}}'\mathbf{\hat{C}}} =  v_1(\hat{A})'v_1(A) + \frac{v_1(\hat{A})'\bm{\varepsilon}}{\sqrt{n}}~.
\end{align*}
Suppose 
\begin{equation}\label{equation--eigenvalue_sparsity}
	\left(\log \log n\right)^\zeta \ll np_n \lesssim \sqrt{\frac{\log n}{\log \log n}}~.
\end{equation}
By Theorem 1.3 and Remark 1.4 in \cite{alt2021poisson}, there exists a ``stray" eigenvalue $\nu$ with eigenvector $v_\nu$ such that $	\lVert v_1(A)-v_\nu\rVert_2 \pto 0$. Let $\Gamma$ be the event that $\nu < \lambda_1(\hat{A})$. Then, $\lim\inf_{n\to\infty} \prob\left(\Gamma \right) > 0$ under \eqref{equation--eigenvalue_sparsity}. On $\Gamma$, $v_1(\hat{A})'v_\nu = 0$ so that
\begin{equation*}
	\lVert v_1(\hat{A})'v_1(A) \rVert_2^2 = \lVert v_1(\hat{A})'(v_1(A)-v_\nu) \rVert_2^2 \leq \lVert v_1(\hat{A}) \rVert_2^2 \lVert v_1(A)-v_\nu \rVert_2^2 \pto 0
\end{equation*}
By independence of $\bm{\varepsilon}$ and $\mathbf{U}$, $\prob\left(v_1(\hat{A})'\bm{\varepsilon} \leq 0 \mid \Gamma \right) = \frac{1}{2}$. As such, 
$$\liminf_{n \to \infty} \prob\left( \hat{\beta}^{(\infty)} \leq \frac{1}{2}\right) \geq \liminf_{n \to \infty} \prob\left( \Gamma \right)*\frac{1}{2} > 0  $$ 
and $\hat{\beta}^{(\infty)}$ is inconsistent.

\subsection{Proof of Theorem \ref{theorem--distribution_obs_error} and Corollary \ref{theorem--bias_magnitude}}

Write:
\begin{equation*}
	\hat{\beta} = \beta \cdot \frac{\hat{\mathbf{C}}'\mathbf{M}_W\mathbf{C}}{\hat{\mathbf{C}}'\mathbf{M}_W\hat{\mathbf{C}}} + \frac{\hat{\mathbf{C}}'\mathbf{M}_W\bm{\varepsilon}}{\hat{\mathbf{C}}'\mathbf{M}_W\hat{\mathbf{C}}} 
\end{equation*}

\subsubsection*{Asymptotic Distribution of Degree and Diffusion}

Let $\hat{\pi} = (\mathbf{W}'\mathbf{W})^{-1}\mathbf{W}'\mathbf{\hat{C}}$. We first show that $\hat{\pi}  \overset{p}{\to} \pi$ when $p_n \gg 1/n$. Write:
\begin{equation*}
	\hat{\pi} = (\mathbf{W}'\mathbf{W})^{-1}\mathbf{W}'\mathbf{C} + (\mathbf{W}'\mathbf{W})^{-1}\mathbf{W}'(\hat{\mathbf{C}} - \mathbf{C})
\end{equation*}
We have already shown that the first term has probability limit $\pi$. To analyze the second term, write:
\begin{equation}\label{equation--degree_consistency_numerator}
	\frac{1}{n} \mathbf{W}'(\hat{\mathbf{C}} - \mathbf{C}) = 	\frac{1}{n} \mathbf{W}' \left( \frac{a_*^{(T)}}{np_n} \sum_{t=1}^T \left(\frac{\delta_*}{np_n} \right)^{t-1} \sum_{B_t \in  \mathcal{B}_1(t)} B_t \right) \iota_n
\end{equation}
where $\mathcal{B}_1(t)$ is the set of mixed products with order $(t,\tau)$ with $\tau \geq 1$. Let $\mathbf{W}_{\cdot, j}$ denote the $j$-th column of $\mathbf{W}$. Each component of the vector in \eqref{equation--degree_consistency_numerator} sums over terms of the form
\begin{equation*}
	\frac{\mathbf{W}_{\cdot, j}' B_t \iota_n}{n^{t+1}p^t_n}  = \expect\left[  \frac{\mathbf{W}_{\cdot, j}' B_t \iota_n}{n^{t+1}p^t_n} \; \bigg\lvert \; \mathbf{U}, \mathbf{W} \right] + 	\frac{\mathbf{W}_{\cdot, j}' B_t \iota_n}{n^{t+1}p^t_n}  - \expect\left[  \frac{\mathbf{W}_{\cdot, j}' B_t \iota_n}{n^{t+1}p^t_n} \; \bigg\lvert \; \mathbf{U}, \mathbf{W} \right] 
\end{equation*}
By Lemma \ref{lemma--order-B}, 
\begin{align*}
	\expect\left[ \left\lvert  \frac{\mathbf{W}_{\cdot, j}' B_t \iota_n}{n^{t+1}p^t_n}  \right\rvert \; \bigg\lvert \; \mathbf{U}, \mathbf{W} \right]  & \leq  (1+O(p_n)) \cdot \left\lvert \frac{\mathbf{W}_{\cdot, j}'\tilde{B}_t \iota_n}{n^{t+1}p_n}
	\right\rvert  \leq (1+O(p_n))\cdot  \frac{1}{n^3p_n}  \sum_{i_1, i_2} |W_{ij}|
\end{align*}
so that 
\begin{equation*}
	\expect\left[ \left\lvert  \frac{\mathbf{W}_{\cdot, j}' B_t \iota_n}{n^{t+1}p^t_n}\right\rvert \right] \leq (1+O(p_n))\cdot \frac{\expect\left[|W_{ij}|\right]}{np_n} = O\left(\frac{1}{np_n}\right)~.
\end{equation*}
By Lemma \ref{lemma--debiased-B}, 
\begin{align*}
	\expect\left[\left(\frac{\mathbf{W}_{\cdot, j}' B_t \iota_n}{n^{t+1}p^t_n}  - \expect\left[  \frac{\mathbf{W}_{\cdot, j}' B_t \iota_n}{n^{t+1}p^t_n} \; \bigg\lvert \; \mathbf{U}, \mathbf{W} \right] \right)^2  \right] 
	& \leq (2t+2)^{2t-2} \cdot \frac{1}{n^6p_n} \sum_{i_1,i_2,i_3,i_4} \expect\left[W_{i_1,j}W_{i_3,j}\right] \\
	& \leq (2t+2)^{2t-2} \cdot \frac{\expect\left[W_{i,j}^2\right]}{n^2p_n} = O\left(\frac{1}{n^2p^2_n}\right)
\end{align*}
Since $T$ is finite, there is only a finite number of $B_t$ terms. Putting the last two displays together, we conclude that 
\begin{equation}\label{equation--dist_deg_diff_WC-Chat}
	\frac{1}{n} \mathbf{W}'(\hat{\mathbf{C}} - \mathbf{C}) = O_p\left(\frac{1}{np_n}\right)
\end{equation}
so that $\hat{\pi} \pto \pi$.

Next, write:
\begin{align*}
	\hat{\beta} & = \beta \cdot \frac{\hat{\mathbf{C}}'\mathbf{M}_W\mathbf{C}}{\hat{\mathbf{C}}'\mathbf{M}_W\hat{\mathbf{C}}} + \frac{\hat{\mathbf{C}}'\mathbf{M}_W\bm{\varepsilon}}{\hat{\mathbf{C}}'\mathbf{M}_W\hat{\mathbf{C}}} 
	 = \beta\left(1 - \frac{\hat{\mathbf{C}}'\mathbf{M}_W(\mathbf{\hat{C}}-{\mathbf{C}})}{\hat{\mathbf{C}}'\mathbf{M}_W\hat{\mathbf{C}}}\right) + \frac{\hat{\mathbf{C}}'\mathbf{M}_W\bm{\varepsilon}}{\hat{\mathbf{C}}'\mathbf{M}_W\hat{\mathbf{C}}} 
\end{align*}
Now, 
\begin{align}\label{equation--deg_diff_dist_num}
	\frac{1}{n}\hat{\mathbf{C}}'\mathbf{M}_W(\mathbf{\hat{C}}-{\mathbf{C}}) & = \frac{1}{n} \sum_{i=1}^n (\hat{C}_i-C_i)(\hat{C}_i - W_i'\hat{\pi})\nonumber \\
	& = \frac{1}{n} \sum_{i=1}^n (\hat{C}_i-C_i)\hat{C}_i - \hat{\pi}'W_i(\hat{C}_i - C_i)
\end{align}
The second term in \eqref{equation--deg_diff_dist_num} is $O_p(1/np_n)$ by \eqref{equation--dist_deg_diff_WC-Chat} and the fact that $\hat{\pi} \pto \pi$. To bound the first term, write:
\begin{align}\label{equation--chat-c_square}
		& \frac{1}{n} \sum_{i=1}^n (\hat{C}_i-C_i)\hat{C}_i  =  \frac{1}{n} (\hat{\mathbf{C}} - \mathbf{C})' \mathbf{\hat{C}} \nonumber \\			
		& = 	\frac{1}{n} \iota_n' \left( \frac{a_*^{(T)}}{np_n} \sum_{t=1}^T \left(\frac{\delta_*}{np_n} \right)^{t-1} \sum_{B_t \in  \mathcal{B}_1(t)} B_t \right)'\left( \frac{a_*^{(T)}}{np_n} \sum_{t=1}^T \left(\frac{\delta_*}{np_n} \right)^{t-1} \sum_{B_t \in  \mathcal{B}_1(t)} B_t \right) \iota_n
\end{align}
Expanding the above product, we find that it is a sum of the terms that take the form $B_1B_2$, $B_1 \in \mathcal{B}_1(t)$, $B_2 \in \mathcal{B}_0(s)$. Clearly, $B_1B_2$  is a mixed product of order $(t+s, \tau)$ where $s+t \leq 2T$ and $\tau \geq 1$. By Lemma \ref{lemma--order-B}, $B_1B_2$ is of order at most $1/np_n$. Since there is a finite number of such terms, 
$
	\frac{1}{n} \sum_{i=1}^n (\hat{C}_i-C_i)\hat{C}_i = O_p\left(\frac{1}{np_n}\right)
$.
Note that the above bound is attained. For degree centrality, 
\begin{equation*}
	\frac{1}{n}\sum_{i=1}^n (\hat{C}_i - C_i)\hat{C}_i \approx \frac{\iota \boldsymbol{\xi}^2\iota}{n^3p_n^2} \approx \frac{\iota' A \iota}{n^3p_n^2} = \Theta_p\left(\frac{1}{np_n}\right) 
\end{equation*}
Observe that $\iota \boldsymbol{\xi}^2\iota/n^3p_n^2$ is also the leading term in a similar expression for diffusion centrality.

Next, write
\begin{equation}\label{equation--distribution--decomposition-CMC}
	\frac{1}{n} \hat{\mathbf{C}}'\mathbf{M}_W\hat{\mathbf{C}} = \frac{1}{n} {\mathbf{C}}'\mathbf{M}_W{\mathbf{C}} + \frac{1}{n} {\mathbf{C}}'\mathbf{M}_W(\hat{\mathbf{C}} -\mathbf{C}) + \frac{1}{n} (\hat{\mathbf{C}} -\mathbf{C})'\mathbf{M}_W \mathbf{\hat{C}}
\end{equation}
From the proof of Theorem 1, we have that ${\mathbf{C}}'\mathbf{M}_W{\mathbf{C}}/n \pto \expect\left[(C_i^* - W_i'\pi)^2\right] > 0$. The third term is the same as in \eqref{equation--deg_diff_dist_num}, which we have just shown is $O_p(1/np_n)$. The second term is:
\begin{align*}
	& \frac{1}{n} \sum_{i=1}^n (\hat{C}_i-C_i){C}_i  =  \frac{1}{n} (\hat{\mathbf{C}} - \mathbf{C})' \mathbf{\hat{C}} \\			
	&	\qquad = 	\frac{1}{n} \iota_n' \left( \frac{a_*^{(T)}}{np_n} \sum_{t=1}^T \left(\frac{\delta_*}{np_n} \right)^{t-1} \sum_{B_t \in  \mathcal{B}_1(t)} B_t \right)'\left( \frac{a_*^{(T)}}{np_n} \sum_{t=1}^T \left(\frac{\delta_*}{np_n} \right)^{t-1} A^t \right) \iota_n
\end{align*}
Expanding the above product, we find that it is a sum of the terms that take the form $B_1A^s$ where $B_1 \in \mathcal{B}_1(t)$. Clearly, $B_1A^s$  is a mixed product of order $(t+s, \tau)$ where $s+t \leq 2T$ and $\tau \geq 1$. By Lemma \ref{lemma--order-B}, this term is of order at most $1/np_n$. As such, $	\frac{1}{n} \hat{\mathbf{C}}'\mathbf{M}_W\hat{\mathbf{C}}  \pto  \expect\left[(C_i^* - W_i'\pi)^2\right]$ and that the asymptotic bias is
\begin{equation*}
	B^{(d)} := \frac{\hat{\mathbf{C}}'\mathbf{M}_W(\hat{\mathbf{C}}-\mathbf{C}) }{\hat{\mathbf{C}}'\mathbf{M}_W\hat{\mathbf{C}} } = O_p\left(\frac{1}{np_n}\right)~.
\end{equation*}
Finally, consider
\begin{equation*}
	\frac{1}{n}\hat{\mathbf{C}}'\mathbf{M}_W\bm{\varepsilon} = \frac{1}{n} {\mathbf{C}}'\mathbf{M}_W\bm{\varepsilon} + \frac{1}{n}(\hat{\mathbf{C}} -\mathbf{C})'\mathbf{M}_W\bm{\varepsilon}~.
\end{equation*}
We have already shown in the proof of Theorem \ref{theorem--benchmark} that the first term above satisfies
\begin{equation*}
	\frac{1}{\sqrt{n}} \mathbf{C}\mathbf{M}_W\bm{\varepsilon} \dto \norm\left(0, \expect\left[(C_i^*-W_i'\pi)^2\varepsilon_i^2\right]\right)~.
\end{equation*}
It remains to show that the last term is $o_p(1/\sqrt{n})$. Write
\begin{equation}\label{equation--distribution--decomposition-CMvarepsilon}
	\frac{1}{n}(\hat{\mathbf{C}} -\mathbf{C})'\mathbf{M}_W\bm{\varepsilon} = \frac{1}{n} \sum_{i=1}^{n} (\hat{C}_i -C_i)\varepsilon_i + (\hat{\pi} - \tilde{\pi}) \cdot \frac{1}{n} \sum_{i=1}^{n}W_i\varepsilon_i
\end{equation}
First observe that
\begin{equation*}
	(\hat{\pi} - \tilde{\pi}) \cdot \frac{1}{n} \sum_{i=1}^{n}W_i\varepsilon_i = o_p(1) \cdot O_p\left(\frac{1}{\sqrt{n}}\right)
\end{equation*}
as required. Next, write:
\begin{equation}\label{equation--Chat-C_varepsilon}
	\frac{1}{n} \sum_{i=1}^{n} (\hat{C}_i -C_i)\varepsilon_i  = \frac{1}{n} \iota_n' \left( \frac{a_*^{(T)}}{np_n} \sum_{t=1}^T \left(\frac{\delta_*}{np_n} \right)^{t-1} \sum_{B_t \in \mathcal{B}_1(t)} B_t \right) \bm{\varepsilon}
\end{equation}
which comprises terms of the form $\iota_n'B_t\bm{\varepsilon}/n^{t+1}p_n^t$. Since $\expect\left[\varepsilon_i \cond \mathbf{W}, \mathbf{U}\right] = 0$ and $\varepsilon_i \indep \boldsymbol{\xi}$, we have that $\expect[{\iota_n'B_t\bm{\varepsilon}} \; \lvert \; \mathbf{W}, \mathbf{U}]  = 0$. Meanwhile, by independence of $\bm{\varepsilon}$ and $\boldsymbol{\xi}$, {\small
\begin{align*}
		\expect\left[\left(\frac{\iota_n'B_t\bm{\varepsilon}}{n^{t+1}p_n^t}\right)^2 \; \bigg\lvert \; \mathbf{W}, \mathbf{U}, \boldsymbol{\xi} \right] & = \frac{1}{n^{2t+2}p_n^{2t}} \sum_{i_1, ..., i_{2t+2}} b_{i_1, i_2} \cdots b_{i_{t}, i_{t+1}} \expect\left[\varepsilon_{i_{t+1}}\varepsilon_{i_{t+2}} \mid \mathbf{W}, \mathbf{U}\right] b_{i_{t+2}, i_{t+3}} \cdots b_{i_{2t+1}, i_{2t+2}} \\
		& = \frac{1}{n^{2t+2}p_n^{2t}} \sum_{i_1, ..., i_{2t+1}} b_{i_1, i_2} \cdots b_{i_{t}, i_{t+1}} b_{i_{t+1}, i_{t+2}} \cdots b_{i_{2t}, i_{2t+1}} \cdot \expect\left[\varepsilon_{i_{t+1}}^2 \mid \mathbf{W}, \mathbf{U}\right]\\
		& \leq  \frac{\bar{\sigma}^2}{n^{2t+2}p_n^{2t}} \iota_n' \tilde{B} \iota_n
\end{align*}}

\noindent The second line above follows since $\expect\left[\varepsilon_{i_{t+1}}\varepsilon_{i_{t+2}} \mid \mathbf{W}, \mathbf{U}\right] = 0$ unless $i_{t+1} = i_{t+2}$. The third line follows from Assumption \ref{assumption--varep_condmoments}. Here, $\tilde{B}$ is a mixed product of order $(2t, \tau)$, with $\tau \geq 2$. Applying Lemma \ref{lemma--order-B} gives us that 
\begin{equation*}
	\expect\left[\left(\frac{\iota_n'B_t\bm{\varepsilon}}{n^{t+1}p_n^t}\right)^2 \right]  = O\left( \frac{n^{2t}p_n^{2t-1}}{n^{2t+2}p_n^{2t}} \right) = O\left(\frac{1}{n^2p_n}\right) \quad \implies \quad \frac{\iota_n'B_t\bm{\varepsilon}}{n^{t+1}p_n^t} = o_p\left(\frac{1}{\sqrt{n}}\right)~.
\end{equation*}
Since \eqref{equation--Chat-C_varepsilon} involves only a finite number of $B_t$'s we conclude that it is $o_p(1/\sqrt{n})$. 

\subsubsection*{Asymptotic Distribution of Eigenvector}

Let $\hat{\pi}^{(\infty)} = (\mathbf{W}'\mathbf{W})^{-1}\mathbf{W}'\mathbf{\hat{C}}$. We first show that $\hat{\pi}^{(\infty)}  \overset{p}{\to} \pi^{(\infty)}$ when $np_n \gg \sqrt{\frac{\log n}{\log \log n}}$. Write:
\begin{equation*}
	\hat{\pi} = (\mathbf{W}'\mathbf{W})^{-1}\mathbf{W}'\mathbf{C} + (\mathbf{W}'\mathbf{W})^{-1}\mathbf{W}'(\hat{\mathbf{C}} - \mathbf{C})
\end{equation*}
We have already shown that first term has probability limit $\pi^{(\infty)}$. To analyze the second term, let $\mathbf{W}_{\cdot,j}$ denote the $j$-th column of $\mathbf{W}$. Then, 
\begin{equation*}
	\left\lVert \frac{1}{n} \mathbf{W}_{\cdot,j}'(\hat{\mathbf{C}} - \mathbf{C}) \right\rVert^2 \leq \left(\frac{1}{n}\sum_{i=1} W_{ij}^2\right)\left(\frac{1}{n}\sum_{i=1} (\hat{C}_{i} - C_i)^2\right)~,
\end{equation*}
where $\frac{1}{n}\sum_{i=1} W_{ij}^2 \pto \expect\left[W^2_{ij}\right] < \infty$. Moreover, 
\begin{align} \label{equation--consistency-eigen-noise-C-C}
	& \frac{1}{n}\sum_{i=1} (\hat{C}_{i} - C_i)^2 = \lVert v_1(\hat{A}) - v_1(A) \rVert^2 \nonumber \\
	& \leq \frac{8\lVert \hat{A} - A \rVert^2}{\lambda^2_1(A)} \quad \mbox{ by the Davis--Kahan inequality (Theorem 4.5.5 in \citealt{vershynin2018high})} \nonumber\\
	&= O_p\left(\frac{\left(np_n\right)^{(1+\nu)/2} \left(\frac{ \log n}{\log \log n}\right)^{(1-\nu)/4}}{np_n}\right) \quad \mbox{ by Lemma \ref{lemma--spectral_conc}}~.
\end{align}
As such, the second term is $o_p(1)$ and $\hat{\pi}^{(\infty)}$ is consistent.

As before, write:
\begin{align*}
	\hat{\beta}^{(\infty)} & = \beta^{(\infty)}\left(1 - \frac{\hat{\mathbf{C}}'\mathbf{M}_W(\mathbf{\hat{C}}-{\mathbf{C}})}{\hat{\mathbf{C}}'\mathbf{M}_W\hat{\mathbf{C}}}\right) + \frac{\hat{\mathbf{C}}'\mathbf{M}_W\bm{\varepsilon}}{\hat{\mathbf{C}}'\mathbf{M}_W\hat{\mathbf{C}}} 
\end{align*}
Now, 
\begin{equation*}
	\begin{aligned}
		\frac{1}{n}\hat{\mathbf{C}}'\mathbf{M}_W(\mathbf{\hat{C}}-{\mathbf{C}}) %& = \frac{1}{n} \sum_{i=1}^n (\hat{C}_i-C_i)(\hat{C}_i - W_i'\hat{\pi})\nonumber \\
		& = \frac{1}{n} \sum_{i=1}^n (\hat{C}_i-C_i)\hat{C}_i - \hat{\pi}'W_i(\hat{C}_i - C_i) \\
		& = O_p\left(\lVert v_1(\hat{A}) - v_1(A) \rVert\right) \quad \mbox{ by Cauchy-Schwarz} \\
		& = O_p\left(\frac{\lVert \boldsymbol{\xi} \rVert}{(\lambda_1 -\lambda_2)np_n}\right) \quad \mbox{ by Davis-Kahan} \\	
		& = O_p\left(\frac{1}{\sqrt{np_n}}\right) +O_p\left(\frac{1}{np_n}\sqrt{\frac{\log n}{\log (\log n/np_n)}}\right) 
	\end{aligned}
\end{equation*}
where the final rate estimate follows from Theorem 3.2 in \cite{benaych2020spectral}. By the same decomposition as in \eqref{equation--distribution--decomposition-CMC}, the above rate estimate also gives us that $\frac{1}{n} \hat{\mathbf{C}} \mathbf{M}_W\hat{\mathbf{C}} \pto \expect\left[(C^{*,(\infty)}-W_i\pi)^2\right] >0$.

It remains to show that $\mathbf{\hat{C}}\mathbf{M}_W\bm{\varepsilon}/n = \frac{C\mathbf{M}_W\varepsilon}{n} + o_p(1/\sqrt{n})$.  The decomposition in \eqref{equation--distribution--decomposition-CMvarepsilon} applies. Since $\hat{\pi}^{\infty} \pto \infty$ as with degree and diffusion, we only need to deal with $\frac{1}{n} \sum_{i=1}^{n} (\hat{C}_i -C_i)\varepsilon_i$. Write:
\begin{align*}
	\expect\left[\left(\frac{1}{n} \sum_{i=1}^{n} (\hat{C}_i -C_i)\varepsilon_i\right)^2 \; \bigg\lvert \; \mathbf{W}, 
	\mathbf{U}\right] \leq \frac{1}{n^2} \sum_{i=1}^{n} (\hat{C}_i -C_i)^2 \bar{\sigma}^2 = \frac{\bar{\sigma}^2}{n} \lVert v_1(\hat{A}) -v_1(A) \rVert~.
\end{align*}
Now, $\lVert v_1(\hat{A}) -v_1(A) \rVert \to 0$ with probability approaching 1 by \cite{benaych2020spectral}. It is also bounded in absolute value. Conclude that $\frac{1}{n} \sum_{i=1}^{n} (\hat{C}_i -C_i)\varepsilon_i= o_p(1/\sqrt{n})$.

\subsubsection*{Bias of Eigenvector under Assumption \ref{assumption--rank_r}}

For the rest of this proof, we will use the notation $\hat{v}_1 = v_1(\hat{A})$, $\hat{\lambda}_1 = \lambda_1(\hat{A})$, $\tilde{v}_1 = v_1({A})$, $\tilde{\lambda}_1 = \lambda_1({A})$.  When $np_n \geq \log n/ \log \log n$, Theorem 3.2 in \cite{benaych2020spectral} implies that $\Vert \boldsymbol{\xi} \rVert = O_p(\sqrt{np_n})$. By Weyl's inequality, $\hat{\lambda}_r - \tilde{\lambda}_r = O_p(\sqrt{np_n})$ for all $r \in [R]$. Since $\tilde{\lambda}_r/np_n \pto {\lambda}_r$ by Lemma \ref{lemma--eigenvector_of_A}, we also have that $\hat{\lambda}_r/np_n \pto \lambda_r$. In other words, $\lVert \boldsymbol{\xi}\rVert/\hat{\lambda}_r, \lVert \boldsymbol{\xi}\rVert/\tilde{\lambda}_r = O_p(1/\sqrt{np_n})$. By the Davis-Kahan inequality, we also have that $\lVert \hat{v}_r - \tilde{v}_r \rVert = O_p(1/\sqrt{np_n})$. 

As before, write:
\begin{equation*}
	\frac{1}{n}\hat{\mathbf{C}}'\mathbf{M}_W(\mathbf{\hat{C}}-{\mathbf{C}}) = \frac{1}{n} \sum_{i=1}^n (\hat{C}_i-C_i)\hat{C}_i - \hat{\pi}'W_i(\hat{C}_i - C_i)
\end{equation*}
Note that 
\begin{equation*}
	\frac{1}{n} \sum_{i=1}^n (\hat{C}_i-C_i)\hat{C}_i = 1- \tilde{v}_1'\hat{v}_1= O_p\left(\frac{1}{np_n}\right)
\end{equation*}
where the rate estimate follows since
\begin{align*}
	2\left(1 - \hat{v}_1'\tilde{v}_1\right) = \lVert \hat{v}_1 - \tilde{v}_1 \rVert^2 & = O_p\left(\frac{1}{np_n}\right) 
\end{align*}
by the Davis-Kahan inequality. It remains to show that $n^{-1}\sum_{i=1}^{n} W_i(\hat{C}_i - C_i) = O_p\left({1}/{np_n}\right)$. 

By definition, $\hat{\lambda}_1\hat{v}_1 = (A + \boldsymbol{\xi})\hat{v}_1$. Rearranging, we have
\begin{align}\label{equation--eigen_bias_neumann_expansion}
	\hat{v}_1 & = \left(\mathbf{I} - \frac{ \boldsymbol{\xi} }{\hat{\lambda}_1}\right)^{-1}\frac{1}{\hat{\lambda}_1}A\hat{v}_1 = \left(\mathbf{I} + \mathbf{K}_1\right)\frac{1}{\hat{\lambda}_1} A\hat{v}_1 = \left(\mathbf{I} + \mathbf{K}_1\right) \sum_{r=1}^R \frac{\tilde{\lambda}_r}{\hat{\lambda}_1} (\hat{v}_1'\tilde{v}_r)\tilde{v}_r
\end{align}
where $\mathbf{K}_1 := \sum_{k=1}^\infty \left( \boldsymbol{\xi}/\hat{\lambda}_1\right)^k$. The above Neumann expansion is valid w.p.a. 1 since $\Vert \boldsymbol{\xi} \rVert/\hat{\lambda}_1 = O_p\left(1/\sqrt{np_n}\right)$. Next, for $r \neq 1$, we can use the same Neumann expansion to write:
\begin{equation}\label{equation--eigen_bias_vk_v1_expansion}
	\begin{aligned}
		0 = \hat{v}_r'\hat{v}_1 & = \sum_{s=1}^R \frac{\tilde{\lambda}^2_s}{\hat{\lambda}_r\hat{\lambda}_1} (v_s'\hat{v}_r)(v_s'\hat{v}_1) \\
		& \qquad + 2 \sum_{t=1}^R \sum_{s=1}^R \frac{\tilde{\lambda}_s\tilde{\lambda}_t}{\hat{\lambda}_r\hat{\lambda}_1} (v_s'\hat{v}_r)(v_t'\hat{v}_1) v_s \mathbf{K}_1 v_t  \\
		& \qquad + \sum_{t=1}^R \sum_{s=1}^R \frac{\tilde{\lambda}_s\tilde{\lambda}_t}{\hat{\lambda}_r\hat{\lambda}_1} (v_s'\hat{v}_r)(v_t'\hat{v}_1) v_s \mathbf{K}^2_1 v_t  
	\end{aligned}
\end{equation}
We have already shown that the last two summands are $O_p(1/np_n)$. 

We now bound the terms in the first summand. Since $\{\hat{v}_r\}_{r=1}^R$ are orthonormal, 
\begin{equation}
	\begin{aligned}\label{equation--eigen_bias_lambda_v_joint_conv}
		1 = \hat{v}_1'\hat{v}_1 & = \left(\frac{\tilde{\lambda}_1}{\hat{\lambda}_1}\right)^2 (\tilde{v}_1'\hat{v}_1)^2 + \sum_{r =2}^R \left(\frac{\tilde{\lambda}_r}{\hat{\lambda}_1}\right)^2 (\tilde{v}_r'\hat{v}_1)^2 \\
		& \qquad + 2 \sum_{r=1}^R \sum_{s=1}^R \frac{\tilde{\lambda}_r\tilde{\lambda}_s}{\hat{\lambda}_1^2}(\hat{v}_1'\tilde{v}_r) (\hat{v}_1'\tilde{v}_s) \tilde{v}_s'\mathbf{K}_1\tilde{v}_r  \\
		& \qquad + \sum_{r=1}^R \sum_{s=1}^R \frac{\tilde{\lambda}_r\tilde{\lambda}_s}{\hat{\lambda}_1^2}(\hat{v}_1'\hat{v}_r) (\tilde{v}_1'\tilde{v}_s) \tilde{v}_s'\mathbf{K}_1^2\tilde{v}_r 
	\end{aligned}
\end{equation}
By the rate estimates in the preamble, $\tilde{\lambda}_r/\hat{\lambda}_1 = O_p(1)$ and
\begin{equation}\label{equation--eigen_bias_v_rv_1_rate}
	(\tilde{v}_r'\hat{v}_1)^2 = 	\left(\tilde{v}_r'(\hat{v}_1-\tilde{v}_1) \right)^2 \leq \lVert \hat{v}_1-\tilde{v}_1 \rVert^2 = O_p\left(\frac{1}{np_n}\right)~.
\end{equation}
Next, consider:
\begin{align}\label{equation--bias_eigen_W1}
	\tilde{v}_s'\mathbf{K}_1\tilde{v}_r = \frac{\tilde{v}_s'\boldsymbol{\xi}\tilde{v}_r}{\hat{\lambda}_1}  + \sum_{k=2}^\infty \frac{\tilde{v}_s'\boldsymbol{\xi}^k\tilde{v}_r}{\hat{\lambda}_1^k} 
\end{align}
For the first term above, write
\begin{equation*}
	\frac{\tilde{v}_s'\boldsymbol{\xi}\tilde{v}_r}{\hat{\lambda}_1}  = \sum_{q,t} \alpha_{r,q}\alpha_{s,t} \frac{\phi_q(\mathbf{U})'\boldsymbol{\xi}\phi_{t}(\mathbf{U})}{n^2p_n} \cdot \left(\frac{\hat{\lambda}_1}{np_n}\right)^{-1}
\end{equation*}
By Lemma \ref{lemma--debiased-B}, 
\begin{equation*}
 	\expect\left[\left(\phi_q(\mathbf{U})'\boldsymbol{\xi}\phi_t(\mathbf{U}) \right)^2\; \bigg\lvert \; \mathbf{W}, \mathbf{U}\right] \leq p_n \sum_{i_1, i_2} \phi^2_q(U_{i_1})\phi^2_t(U_{i_2})= n^2p_n \left(\frac{1}{n}\sum_{i_1=1}^n \phi^2_q(U_{i_1})  \right) \left(\frac{1}{n} \sum_{i_1=1}^n  \phi^2_t(U_{i_2}) \right)~.
\end{equation*}
Taking expectations again and using the fact that $2ab \leq (a^2+b^2)$, we have that
\begin{equation*}
	\expect\left[\left(\phi_q(\mathbf{U})'\boldsymbol{\xi}\phi_t(\mathbf{U}) \right)^2\right] \leq \frac{n^2p_n}{2}\expect\left[  \left(\frac{1}{n} \sum_{i_1=1}^n \phi^2_q(U_{i_1})  \right)^2 + \left(\frac{1}{n}\sum_{i_1=1}^n \phi^2_t(U_{i_1}) \right)^2\right] 
\end{equation*}
Now,
\begin{equation*}
	\expect\left[  \left(\frac{1}{n}\sum_{i_1=1}^n \phi^2_q(U_{i_1}) \right)^2\right] = \frac{1}{n^2} \sum_{i_1, i_2} \expect[\phi_q(U_{i_1})^2\phi^2_q(U_{i_2})] \leq \expect\left[\phi^4_q(U_{i_1})\right] < \infty
\end{equation*}
Similarly for $\phi_s$. As such, $\phi_q(\mathbf{U})'\boldsymbol{\xi}\phi_t(\mathbf{U}) = O_p(n\sqrt{p_n})$. By Lemma \ref{lemma--eigenvector_of_A}, $\alpha_{s,q}$ are $O_p(1)$. Moreover, $\hat{\lambda}_1/np_n \pto \lambda_1$. Putting all of these estimates together, we have that  
\begin{equation}\label{equation--eigen_bias_power_one_estimate}
	\frac{\tilde{v}_s'\boldsymbol{\xi}\tilde{v}_r}{\hat{\lambda}_1}  = O_p\left(\frac{1}{n\sqrt{p_n}}\right) = O_p\left(\frac{1}{np_n}\right)
\end{equation}
For the last term in \eqref{equation--bias_eigen_W1}, we have that:
\begin{equation}\label{equation--eigen_bias_quadratic_estimate}
	\sum_{k=2}^\infty \left\lvert  \frac{\tilde{v}_s'\boldsymbol{\xi}^k\tilde{v}_r}{\hat{\lambda}_1^k}  \right\rvert \leq \sum_{k=2}^\infty \frac{\lVert \boldsymbol{\xi} \rVert^k}{\hat{\lambda}_1^k} \leq \frac{\lVert \boldsymbol{\xi} \rVert^2}{\hat{\lambda}_1^2} \frac{1}{1 - {\lVert \boldsymbol{\xi} \rVert}/{\hat{\lambda}_1} } = O_p\left(\frac{1}{np_n}\right)~.
\end{equation}
Combining the above rate estimates gives that 
\begin{align}\label{equation--bias_eigen_W1_rate}
	\tilde{v}_s'\mathbf{K}_1\tilde{v}_r = O_p\left(\frac{1}{np_n}\right)
\end{align}

Finally, we bound the last term in \eqref{equation--eigen_bias_lambda_v_joint_conv}. Write: 
\begin{align}\label{equation--bias_eigen_K2_rate}
	\left\lvert \tilde{v}_s'\mathbf{K}_1^2\tilde{v}_r \right\rvert = \left\lvert  \sum_{j,k \geq 2}  \frac{\tilde{v}_s \boldsymbol{\xi}^{j+k}\tilde{v}_r}{\hat{\lambda}_1^{j+k}}  \right\rvert & \leq \sum_{j,k \geq 2}  \frac{\lVert \boldsymbol{\xi}^{j+k}\rVert}{\hat{\lambda}_1^{j+k}}  = \sum_{k=0}^\infty (k+1) \frac{\lVert \boldsymbol{\xi} \rVert^{2+k}}{\hat{\lambda}_1^{2+k}} \nonumber \\
	& = \frac{\lVert \boldsymbol{\xi} \rVert^{2}}{\hat{\lambda}_1^{2}} \left(1 -\frac{\lVert \boldsymbol{\xi} \rVert}{\hat{\lambda}_1} \right)^{-2} = O_p\left(\frac{1}{np_n}\right)~.
\end{align}
Conclude that
\begin{equation*}
 	1  = \left(\frac{\tilde{\lambda}_1}{\hat{\lambda}_1}\right)^2 (\tilde{v}_1'\hat{v}_1)^2  + O_p\left(\frac{1}{np_n}\right)
\end{equation*}
By the Delta method,
\begin{equation*}
	\frac{\tilde{\lambda}_1}{\hat{\lambda}_1} \tilde{v}_1'\hat{v}_1 = 1+ O_p\left(\frac{1}{np_n}\right)
\end{equation*}
since the above term is positive w.p.a. 1. Because the proof did not use any properties specific to the leading eigenvector, we in fact have for $r \in [R]$ that:
\begin{equation}
	\frac{\tilde{\lambda}_r}{\hat{\lambda}_r} \tilde{v}_r'\hat{v}_r = 1+ O_p\left(\frac{1}{np_n}\right)~. \label{equation--eigenvalue_eigenvector_product}
\end{equation}

When $s \notin \{1,r\}$, the same argument as in \eqref{equation--eigen_bias_v_rv_1_rate} gives us that
\begin{equation*}
	\lvert (v_s'\hat{v}_r)(v_s'\hat{v}_1) \rvert \leq \lvert (v_s'\hat{v}_r)\rvert^{1/2} \cdot \lvert (v_s'\hat{v}_1) \rvert^{1/2} =O_p\left(\frac{1}{np_n}\right)
\end{equation*}
Putting the above estimates into \eqref{equation--eigen_bias_vk_v1_expansion}, we have that
\begin{align*}
	0 & = \frac{\tilde{\lambda}^2_r}{\hat{\lambda}_r\hat{\lambda}_1} (v_r'\hat{v}_r)(v_r'\hat{v}_1) + \frac{\tilde{\lambda}^2_1}{\hat{\lambda}_r\hat{\lambda}_1} (v_1'\hat{v}_r)(v_1'\hat{v}_1) + O_p\left(\frac{1}{np_n}\right)\\
	& = \frac{\lambda_r}{\lambda_1} (v_r'\hat{v}_1) +  \frac{\lambda_1}{\lambda_r} (v_1'\hat{v}_r) +O_p\left(\frac{1}{np_n}\right) 
\end{align*}
where the final equation follows from \eqref{equation--eigenvalue_eigenvector_product} and the fact that $\hat{\lambda}_r/np_n, \tilde{\lambda}_r/np_n \pto  \lambda_r+ O_p(1/\sqrt{np_n})$. Since $\lambda_1 > \lambda_r$ and $v_1'\hat{v}_r, v_r'\hat{v}_1  = O_p(1/\sqrt{np_n})$, the above equation can be satisfied only if 
\begin{equation}\label{equation--eigen_bias_K_cross_estimate}
	v_1'\hat{v}_r = O_p\left(\frac{1}{np_n}\right) \quad \mbox{ and } \quad v_r'\hat{v}_1 = O_p\left(\frac{1}{np_n}\right) 
\end{equation}

We are now ready to evaluate $\mathbf{W}'(\hat{\mathbf{C}}-\mathbf{C})/n$. Write for the $j$-th component, 
\begin{align*}
	\frac{1}{a_*^{(\infty)}} \cdot \frac{\mathbf{W}_{\cdot,j}'(\hat{\mathbf{C}}-\mathbf{C})}{n} 
	& = \left(\frac{\tilde{\lambda}_1}{\hat{\lambda}_1} \tilde{v}_1'\hat{v}_1 - 1\right)\cdot \frac{\mathbf{W}_{\cdot, j}\sqrt{n}\tilde{v}_1}{n}\\
	& \quad + \sum_{r=2}^{R} \left(\frac{\tilde{\lambda}_r}{\hat{\lambda}_1} \tilde{v}_r'\hat{v}_1\right)\cdot \frac{\mathbf{W}_{\cdot, j}\sqrt{n}\tilde{v}_r}{n}  + \sum_{r=1}^R \left(\frac{\tilde{\lambda}_r}{\hat{\lambda}_1} \tilde{v}_r'\hat{v}_1\right)\cdot \frac{1}{n} \mathbf{W}_{\cdot,j}\mathbf{K}_1 \sqrt{n}\tilde{v}_{r} 
\end{align*}
We have already shown in the proof of Theorem \ref{theorem--benchmark} that
\begin{equation*}
	\frac{\mathbf{W}_{\cdot, j}\sqrt{n}\tilde{v}_1}{n} = \frac{\mathbf{W}_{\cdot, j}'\mathbf{C}^{(\infty)}}{n} \pto \expect\left[W_{i,j}\phi_1(U_i)\right]
\end{equation*}
The same argument gives us that $\frac{1}{n}\mathbf{W}_{\cdot, j}'\sqrt{n}\tilde{v}_r  \pto \expect\left[W_{i,j}\phi_r(U_i)\right]$. Meanwhile, an argument essentially identical to the proof of \eqref{equation--bias_eigen_W1_rate} gives us that $\frac{1}{n} \mathbf{W}_{\cdot,j}'\mathbf{K}_1 \sqrt{n}\tilde{v}_{r} = O_p\left(\frac{1}{np_n}\right)$. Combining this with \eqref{equation--eigenvalue_eigenvector_product} and \eqref{equation--eigen_bias_K_cross_estimate} then gives us that $n^{-1}\sum_{i=1}^{n} W_i(\hat{C}_i - C_i) = O_p\left({1}/{np_n}\right)$. 

Finally, to show that the rate is attained, note that in the absence of covariates, bias is
\begin{equation*}
	\frac{1}{n} \sum_{i=1}^n (\hat{C}_i - C_i) \hat{C}_i = (\hat{v}_1 - \tilde{v}_1)'\hat{v}_1 = \frac{1}{2}\lVert \hat{v}_1 - \tilde{v}_1 \rVert^2~.
\end{equation*}
Specializing \eqref{equation--eigen_bias_neumann_expansion} with $f = 1$ and $\tilde{v}_1 = \iota/\sqrt{n}$, and applying \eqref{equation--eigenvalue_eigenvector_product} yields:
\begin{equation*}
	\hat{v}_1 - \tilde{v}_1 = \left(\mathbf{I} + \mathbf{K}_1\right) \frac{\tilde{\lambda}_1}{\hat{\lambda}_1} (\hat{v}_1'\tilde{v}_1) \tilde{v}_1 -\tilde{v}_1 = O_p\left(\frac{1}{np_n}\right)\tilde{v}_1 + \left(1 +O_p\left(\frac{1}{np_n}\right)\right)  \mathbf{K}_1 \tilde{v}_1~.
\end{equation*}
Consequently, 
\begin{align*}
		\lVert \hat{v}_1 - \tilde{v}_1 \rVert^2 &= O_p\left(\frac{1}{n^2p^2_n}\right) + 2 \left(1 + O_p\left(\frac{1}{np_n}\right)\right) O_p\left(\frac{1}{np_n}\right) \tilde{v}_1'\mathbf{K}_1\tilde{v}_1 + \left(1 + O_p\left(\frac{1}{np_n}\right)\right)^2 \tilde{v}_1'\mathbf{K}^2_1\tilde{v}_1 \\
		& = \left(1 + O_p\left(\frac{1}{np_n}\right)\right)^2 \tilde{v}_1'\mathbf{K}^2_1\tilde{v}_1 + o_p\left(\frac{1}{np_n}\right) \quad \mbox{ by \eqref{equation--eigen_bias_power_one_estimate}~.}
\end{align*}
Moreover, 
\begin{align*}
	\tilde{v}_1'\mathbf{K}^2_1\tilde{v}_1 = \sum_{j,k=1}^\infty \frac{\iota' \boldsymbol{\xi}^{j+k} \iota}{n \cdot \hat{\lambda}_1^{j+k}} & = \frac{\iota' \boldsymbol{\xi}^{2} \iota}{n \cdot \hat{\lambda}_1^{2}} + \left(\sum_{j,k=2}^\infty \frac{\iota' \boldsymbol{\xi}^{j+k} \iota}{n \cdot \hat{\lambda}_1^{j+k}}\right) + 2\left(\sum_{k=2}^\infty \frac{\iota' \boldsymbol{\xi}^{k+1} \iota}{n \cdot \hat{\lambda}_1^{k+1}}\right) \\
	& = \frac{\iota' \boldsymbol{\xi}^{2} \iota}{n \cdot \hat{\lambda}_1^{2}}  + o_p\left(\frac{1}{np_n}\right) \quad \mbox{ by  \eqref{equation--eigen_bias_quadratic_estimate} and \eqref{equation--bias_eigen_K2_rate}~.}
\end{align*}
Meanwhile, by \eqref{equation--deg_bias_tight_estimate}, $\frac{\iota' \boldsymbol{\xi}^{2} \iota}{n \cdot \lambda_1^{2}(\hat{A})} = \Theta_p\left( \frac{1}{np_n} \right)$. Conclude that $\lVert v_1(\hat{A}) - v_1(A) \rVert_2^2$---and therefore $n^{-1} \sum_{i=1}^{n} (\hat{C}_i - C_i)C_i$---is $\Theta_p\left(1/np_n\right)$. 
\hfill \qedsymbol

\subsection{Proof of Theorem \ref{theorem--distribution_degree_diff_bias_corrected}}

We need to prove that $\check{B}^{(d)}_1 = B^{(d)}_1 + o_p\left(1/\sqrt{n}\right)$ and $\check{B}^{(d)}_2 = B^{(d)}_2 + o_p\left(1/\sqrt{n}\right)$. Given \eqref{equation--B_1d--wBv} and \eqref{equation--B_2d--wBv}, it suffices to show that for a mixed product of order $(t,\tau)$, $t \geq 1$, and $w, v \in \{\mathbf{W}_{\cdot, j}, \iota_n\}$,
\begin{equation}\label{bias_correction_target}
	\frac{\check{Q}(B,w,v)}{n^{t+1}p_n^{t}} = \frac{w'Bv}{n^{t+1}p_n^t} + o_p\left(\frac{1}{\sqrt{n}}\right)~.
\end{equation}
Let 
\begin{equation*}
	\tilde{Q}(B,w,v) = \sum_{g \in \mathcal{T}(B)} k_B(g) \cdot \sum_{i_1, ..., i_{n(g)}} w_{i_1} \left(\prod_{jk \in E(g)} p_nf(U_{i_j},U_{i_k})\right)v_{i_m}~.
\end{equation*}
where the summation over the indices $i_1,...,i_{n(g)}$ is such that no index is repeated. Because indices do not repeat, $\expect\left[\check{Q}(B,w,v) - \tilde{Q}(B, w, v) \mid \mathbf{W}, \mathbf{U}\right] = 0$. By the same argument as in the proof of Lemma \ref{lemma--debiased-B},  an index $\mathbf{i}$ makes a non-zero contributions to $$\expect\left[\left(\check{Q}(B,w,v) - \tilde{Q}(B, w, v)\right)^2 \mid \mathbf{W}, \mathbf{U}\right]$$ only when $\tilde{g}^{\text{simp}}(\mathbf{i})$ is a connected graph with a repeated edge. It can therefore have at most $2t-1$ edges and at most one more node than edges. Consequently, 
\begin{equation*}
	\expect\left[\left(\check{Q}(B,w,v) - \tilde{Q}(B, w, v)\right)^2 \mid \mathbf{W}, \mathbf{U}\right] \leq n^{2t-4}p_n^{2t-1} \cdot \sum_{i_1, i_2,i_3,i_4} |w_{i_1}v_{i_2}w_{i_3}v_{i_4}|~.
\end{equation*}
Since the fourth moments of $w$ and $v$ are finite, 
\begin{equation}\label{equation--bias_correction_Qcheck_Qtilde}
	\frac{\check{Q}(B,w,v)}{n^{t+1}p_n^{t}}  - \frac{\tilde{Q}(B,w,v)}{n^{t+1}p_n^{t}} = O_p\left(\sqrt{\frac{n^{2t}p_n^{2t-1}}{n^{2t+2}p_n^{2t}}}\right) 
	= o_p\left(\frac{1}{\sqrt{n}}\right)
\end{equation}
By Lemma \ref{lemma--order-B} and \ref{lemma--debiased-B}, we have that
\begin{equation}\label{equation--bias_correction_wBv_wBtildev}
	\frac{w'Bv}{n^{t+1}p_n^t} = \frac{w'\tilde{B}v}{n^{t+1}p_n^t} +o_p\left(\frac{1}{\sqrt{n}}\right)
\end{equation}
where $\tilde{B}$ is defined to be:
\begin{equation*}
	\tilde{B} =\prod_{s=1}^r  \Big(\mathbbm{1}\left\{q_s = 0\right\} \cdot A^{p_s} + \mathbbm{1}\left\{q_s = 1\right\} \cdot \tilde{\boldsymbol{\xi}}(p_s)  \Big)~.
\end{equation*}
By construction, 
\begin{equation*}
	\frac{w'\tilde{B}v}{n^{t+1}p_n^t} = \frac{\tilde{Q}(B,w,v)}{n^{t+1}p_n^{t}} +\frac{1}{n^{t+1}p_n^t}  \sum_{t \in \mathcal{T}(B)} k_B(t) \cdot \sum_{\mathbf{i}\in \mathcal{I}^{\text{repeat}}}  w_{i_1} \left(\prod_{jk \in E(t)} p_nf(U_{i_j},U_{i_k})\right)v_{i_m}
\end{equation*}
where $\mathcal{I}^{\text{repeat}}$ is the set of $\mathbf{i}$ in which some index $i_k$ is repeated. For a given tree $g \in \mathcal{T}(B)$, summand is of order $p_n^{n(t)}$. The number of indices of length $n(t)+1$ for which a node repeats must be no more than $n(t)$. As such, 
\begin{align*}
	&\left\lvert  \frac{1}{n^{t+1}p_n^t}  \sum_{g \in \mathcal{T}(B)} k_B(g) \cdot \sum_{\mathbf{i}\in \mathcal{I}^{\text{repeat}}}  w_{i_1} \left(\prod_{jk \in E(g)} p_nf(U_{i_j},U_{i_k})\right)v_{i_m}  \right\rvert \\
	& \qquad \qquad \leq \frac{1}{n^{t+1}p_n^t} \sum_{g \in \mathcal{T}(B)} k_B(g) \cdot n^{n(g)-2}p_n^{n(g)}\sum_{i_1, i_m} \lvert w_{i_1}v_{i_m} \rvert = O_p\left(\frac{1}{n}\right)~.
\end{align*}
where the final estimate uses the fact that $w_i$ and $v_i$ have finite second moments. Hence,
\begin{equation*}
	\frac{w'\tilde{B}v}{n^{t+1}p_n^t} = \frac{\tilde{Q}(B,w,v)}{n^{t+1}p_n^{t}} + o_p\left(\frac{1}{\sqrt{n}}\right)~,
\end{equation*}
which together with \eqref{equation--bias_correction_Qcheck_Qtilde} and \eqref{equation--bias_correction_wBv_wBtildev} yield \eqref{bias_correction_target}. \hfill \qedsymbol

\section{Proofs of Auxiliary Lemmas}\label{appendix--lemma_proofs}

\subsection{Proof of Lemma \ref{lemma--graphon-centrality}}

We start with degree and diffusion. By definition,
\begin{align*}
	C_i^{(1)} - C_i^{*,(1)} & = \frac{a_*^{(1)}}{n}\sum_{j=1}^{n} f(U_{i},U_{j}) - \mathbb{E}[f(U_{i}, U_{j}) \mid U_{i}] \\
	C_i^{(T)} - C_i^{*,(T)} & = a_*^{(T)} \sum_{t=1}^T \bigg( \frac{\left(\delta^*\right)^{t-1}}{n^t}  \sum_{j_1,...,j_t}^{n} f(U_{i},U_{j_1})f(U_{j_1},U_{j_2})\cdots f(U_{j_{t-1}}, U_{j_{t}})  \\
	& \qquad \qquad \qquad \qquad \qquad \qquad - \mathbb{E}[f(U_{i},U_{j_1})\cdots f(U_{j_{t-1}}, U_{j_{t}}) \mid U_{i}] \bigg)
\end{align*}
which are U-statistics (written above as V-statistics) of order $d$. Hoeffding's exponential tail bound  \citep[equation (5.7)]{hoeffding1963probability} gives us that
\begin{equation*}
	\mathbb{P}\left( \left\lvert C_i^{(d)} - C_i^{*,(d)}  \right\rvert \geq n^{-1/2+\eta} \right) \leq d\exp\left(\frac{-2n^{2\eta}}{d}\right)~.
\end{equation*}
Applying the union bound across $i$ and noting that $nd \exp\left(\frac{-2n^{2\eta}}{d}\right) \to 0$ yields the desired result.

For eigenvector, we will show that the term in the expectation converges in probability to $0$. Convergence in moment then follows from the fact that it is the squared $l^2$ distance of two unit eigenvectors and hence bounded above by 4. 

Observe that $T$ is a compact, self-adjoint operator. Let $\phi_r$ and $\lambda_r$ be its eigenvalues and associated eigenfunctions. Define:
\begin{align*}
	k(u,v) = \sum_{r=1}^\infty \lambda_r \phi_r(u)\phi_r(v) \quad , \quad	k_R(u,v) = \sum_{r=1}^R \lambda_r \phi_r(u)\phi_r(v)
\end{align*}
and let $K$ and $K_R$ be matrices such that $[K]_{ij} = k(U_i, U_j)$ and $K_{R,ij} = k_R(U_i, U_j)$. Then:
\begin{align*}
	\frac{1}{a^{(\infty)}_*}\sqrt{\frac{1}{n} \sum_{i=1}^n \left( C_i^{(\infty)} - C_i^{*,(\infty)} \right)^2} &= \left\lVert v_1(A) - \frac{\phi_1(\mathbf{U})}{\sqrt{n}} \right\rVert \\
	& \leq \lVert v_1(A) -v_1(K) \rVert + \left\lVert v_1(K) - \frac{\phi_1(\mathbf{U})}{\sqrt{n}}\right\rVert 
\end{align*}
where $\phi_1(\mathbf{U}) = [\phi_1(U_1) \cdots \phi_1(U_n)]'$. By Lemma 10.16 of \cite{lovasz2012large}, $A$ and $K$ are arbitrarily close to $f$ and $k$ in cut distance with probability approaching 1. Assumption \ref{assumption--eigenvector_identification} together with the operator version of the Davis-Kahan inequality \cite[Lemma 8]{avella2020centrality} gives us that $\lVert v_1(A) -v_1(K) \rVert \overset{p}{\to} 0$. 

To show that $\lVert v_1(K) - \phi_1(\mathbf{U})/\sqrt{n}\rVert  \overset{p}{\to} 0$, we need control the error from orthonormalization. Let $\eta > 0$ be given and write
\begin{equation*}
	\left\lVert v_1(K) -\frac{\phi_1(\mathbf{U})}{\sqrt{n}}\right\rVert  \leq \lVert v_1(K) - v_1(K_R) \rVert + \left\lVert v_1(K_R) - \frac{\phi_1(\mathbf{U})}{\sqrt{n}} \right\rVert~.
\end{equation*}
Choose $R$ such that $\sum_{r=1}^R \lambda_r \leq \eta/2$. Next, let $\Gamma_1$ be the event that 
\begin{align*}
	\frac{1}{n} \sum_{i=1}^n \phi^2_r(U_i) \in [1-\gamma_1, 1+\gamma_1] \quad \mbox{ for } \quad r \in [R] \\
	\frac{1}{n} \sum_{i=1}^n \phi_r(U_i)\phi_s(U_i)  \in [-\gamma_1, \gamma_1] \quad \mbox{ for } \quad r \neq s	
\end{align*}
Since $\mathbb{E}[\phi^2_r(U)] = 1$ and $\mathbb{E}[\phi_r(U)\phi_s(U)] = 0$ for $r\neq s$, $\mathbb{P}(\Gamma_1) \to 1$ as $n \to \infty$ for any fixed $\gamma_1 >0$. We wil choose its value below. Observe that $v_1(K_R)$ must be of the form $v_1(K_R) = \sum_{r=1}^R \alpha_r \phi_r(\mathbf{U})/\sqrt{n}$. This is because orthonormalization cannot change the eigenspace. Moreover,
\begin{align*}
	1 = v_1(K_R)'v_1(K_R) & = \frac{1}{n}\sum_{r=1}^R \alpha_r^2 \phi_r(\mathbf{U})'\phi_r(\mathbf{U}) + \frac{1}{n}\sum_{r \neq s} \alpha_r\alpha_s \phi_r(\mathbf{U})'\phi_s(\mathbf{U}) \\
	& \geq \frac{1}{n} \sum_{r=1}^R \alpha_r^2 \phi_r(\mathbf{U})'\phi_r(\mathbf{U})  - \frac{1}{n}\sum_{r\neq s} \frac{1}{2}(\alpha_r^2 + \alpha_s^2) \cdot |\phi_r(\mathbf{U})'\phi_s(\mathbf{U})| \\
	& \geq \sum_{r=1}^R \alpha_r^2 \left( 1 - R\gamma_1\right) \quad \mbox{ on } \quad \Gamma_1
\end{align*}
In particular, $\alpha_r^2 \leq (1-R\gamma_1)^{-1} \leq 2$ if $\gamma_1 \leq 1/2R$. Next, write the leading eigenvalue of $K_R$ as $v_1(K_R)'K_Rv_1(K_R) = \sum_{r=1}^R \lambda_r c_r$ where
\begin{multline*}
	c_r = \frac{\alpha_r^2}{n^2} (\phi_r(\mathbf{U})'\phi_r(\mathbf{U}))^2 + \frac{\alpha_r}{n^2} \sum_{q\in 1}^R  \alpha_q (\phi_r(\mathbf{U})'\phi_r(\mathbf{U})) \cdot (\phi_r(\mathbf{U})'\phi_q(\mathbf{U})) \\
	+ \frac{1}{n^2}\sum_{q=1}^R \sum_{s\neq r} \alpha_s\alpha_q (\phi_r(\mathbf{U})'\phi_q(\mathbf{U})) (\phi_r(\mathbf{U})'\phi_s(\mathbf{U}))~.
\end{multline*}
On $\Gamma_1$, $c_r \geq \alpha_r^2(1-\gamma_1^2) - 8R\gamma_1 - 4R^2\gamma_1^2$. As such, 
\begin{equation*}
	|\lambda_1 - \lambda_1(K_R)| =\left\lvert \sum_{r=2}^R c_r(\lambda_1 - \lambda_r) \right\rvert \geq \left\lvert \sum_{r=2}^R c_r \right\rvert \cdot \lvert \lambda_1 - \lambda_r \rvert~,
\end{equation*}
so that
\begin{equation*}
	\frac{|\lambda_1 - \lambda_1(K_R)| }{|\lambda_1 - \lambda_2|} + 8R^2\gamma_1 + 4R^3 \gamma_1^2 \geq \sum_{r=2}^R \alpha_r^2~.
\end{equation*}
Choose $\gamma_1$ such that $8R^2\gamma_1 + 4R^3 \gamma_1^2 \leq \eta^2/16$. Let $\Gamma_2$ be the event that $\frac{|\lambda_1 - \lambda_1(K_R)| }{|\lambda_1 - \lambda_2|} \leq \eta^2/16$. Another application of Lemma 10.16 in \cite{lovasz2012large}, together with Weyl's inequality gives us that $\mathbb{P}(\Gamma_2) \to 1$. Then on $\Gamma_1 \cap \Gamma_2$,
\begin{align*}
	\left\lVert v_1(K) - \frac{\phi_1(\mathbf{U})}{\sqrt{n}}\right\rVert^2 = \left\lVert \sum_{r=2}^R \alpha_r \cdot \left(\frac{\phi_r(\mathbf{U})}{\sqrt{n}} - \frac{\phi_1(\mathbf{U})}{\sqrt{n}}\right) \right\rVert^2 \leq 2\sum_{r=2}^R \alpha_r^2 \leq \left(\frac{\eta}{2}\right)^2 
\end{align*}
As such, $\mathbb{P}\left(\left\lVert v_1(K) - \frac{\phi_1(\mathbf{U})}{\sqrt{n}}\right\rVert  \leq \eta\right) \to 1$. \hfill \qed

\subsection{Proof of Lemma \ref{lemma--order-xit}}

Write:
\begin{align*}
	\expect\left[w'Bv \mid \mathbf{U}, \mathbf{W}\right]  & = \sum_{i_1, ..., i_{t+1}}^n  w_{i_1}\expect\left[b_{i_1, i_2} \cdots b_{i_{t}, i_{t+1}}\mid \mathbf{U}, \mathbf{W}\right] v_{i_{t+1}}
\end{align*}
Observe that for $ \expect\left[\xi_{i_1, i_2} \cdots \xi_{i_{t}, i_{t+1}}\mid \mathbf{U}, \mathbf{W}\right]$ to be non-zero, every edge in $g(\mathbf{i})$ must have multiplicity at least 2. Otherwise, there will be a lone $\xi_{i_r,i_{r+1}}$ such that
\begin{align*}
	& \expect\left[\xi_{i_1, i_2} \cdots \xi_{i_{t}, i_{t+1}}\mid \mathbf{U}, \mathbf{W}\right] \\
	& = \expect\left[\xi_{i_1, i_2} \cdots \xi_{i_{r-1}, i_r} \cdot \expect\left[\xi_{i_r,i_{r+1}}\mid \mathbf{U}, \mathbf{W}, \{\xi_{i_s}\}_{s=1}^t \setminus  \xi_{i_r,i_{r+1}}  \right] \cdot \xi_{i_{r+1}, i_{r+2}} \cdots \xi_{i_{t}, i_{t+1}} \mid \mathbf{U}, \mathbf{W} \right] = 0
\end{align*}

As such, the rest of the proof focuses on $\mathbf{i} \in \mathcal{T}(t) \cup \mathcal{T}^c(t)$. Denote an edge $(j,k) \in E(\mathbf{i})$ by $jk$ and let its multiplicity be $m(jk) \geq 2$. Then,
\begin{align}\label{equation--tree_approx}
	& \expect\left[\xi_{i_1, i_2} \cdots \xi_{i_{t}, i_{t+1}}\mid \mathbf{U}, \mathbf{W}\right]  \nonumber \\
	& = \prod_{jk \in E(\mathbf{i})} \left(1-p_nf(U_{j}, U_k)\right)^{m(jk)}p_nf(U_j,U_k) - (p_nf(U_j,U_k))^{m(jk)} \cdot \left(1-p_nf(U_j,U_k)\right) \nonumber \\
	& = \prod_{jk \in E(\mathbf{i})} \left(p_nf(U_j, U_k) + O(p^2_n)\right)
\end{align}
In other words, we can approximate $\expect[\xi_{ij}^t]$ with $\expect[\xi_{ij}^2]$ and incur only an error of $O(p_n)$. This error is deterministic since $|f| \leq 1$. As such, the expectation associated with $g(\mathbf{i})$ depends only on $g^{\text{simp}}(\mathbf{i})$. 

Suppose $\mathbf{i}$ is such that $i_1 = l$ and $i_{t+1} = m$. By definition, if $g^{\text{simp}}(\mathbf{i})$ is a tree, its to the exepctation is collected in $[\tilde{B}(t)]_{l,m}$. Otherwise, its contribution is collected in $[R(\boldsymbol{\xi}^t)]_{l,m}$. 

To bound $[\tilde{B}(t)]_{l,m}$, fix a tree $g$. Suppose $g^{\text{simp}}(\mathbf{i}) = g$ and that $i_1 = l$, $i_{t+1} = m$. Since $g$ is a tree, it has $|E(g)|$ edges and $|E(g)|+1$ nodes. Having fixed the start and end points, there remains $|E(g)|-1$ free nodes. As such, there are at most $n^{|E(g)|- 1}$ such terms. The contributions from graphs homomorphic to $g$ is therefore
\begin{equation*}
	(1+O(p_n))  \cdot n^{|E(g)|-1} p_n^{|E(g)|} \leq (1+O(p_n))  \cdot n^{\lfloor t/2 \rfloor - 1} p_n^{\lfloor t/2 \rfloor}~.
\end{equation*}
The inequality above follows because we require every edge to have multiplicity 2, so that $g^{\text{simp}}(\mathbf{i})$ can have at most $\lfloor t/2 \rfloor$ edges. We obtain the stated bound on $[\tilde{\boldsymbol{\xi}}(t)]_{l,m}$ by observing there must be fewer than $(t+1)^t$ such $g$'s, since this is the number of walks of length $t$ on $t+1$ nodes. 

Next, consider walks for which $g^{\text{simp}}(\mathbf{i})$ is not a tree. Since a cycle has at least 3 edges, for $g^{\text{simp}}(\mathbf{i})$ to contain a cycle, $g(\mathbf{i})$ must have at least 6 edges. Otherwise, this set is empty and we can set $R(B) = 0$. Let $t \geq 6$. Fix the start point $i_1$ and end point $i_{t+1}$ and suppose 
$g^{\text{simp}}(\mathbf{i}) = g$ is not a tree. For such an $\mathbf{i}$,
\begin{equation*}
	\expect\left[\xi_{i_1, i_2} \cdots \xi_{i_{t}, i_{t+1}}\mid \mathbf{U}, \mathbf{W}\right] \leq p_n^{E(g)}(1+O(p_n))
\end{equation*} 
Having fixed $i_1$ and $i_{t+1}$, there remains $|V(g)|-2$ free nodes. As such, there are at most $n^{|V(g)|- 2}$ such terms. The contribution from a given $g$ to the expectation is therefore 
\begin{equation*}
	(1+O(p_n))  \cdot n^{|V(g)|-2} p_n^{|E(g)|} \leq (1+O(p_n))   n^{|E(g)|-2}  p_n^{|E(g)|} \leq (1+O(p_n)) n^{{\lfloor t/2 \rfloor}}p_n^{{\lfloor t/2 \rfloor}}
\end{equation*}
 The first inequality above follows because when $g$ is not a tree, $|V(g)| \leq |E(g)|$. The last inequality follows since $|E(g)| \leq \lfloor t/2\rfloor$.

 As before, the number of walks $g(\mathbf{i})$ for which $g^{\text{simp}}(\mathbf{i})$ is not a tree is upper bounded by $(t+1)^t$. Let $R(\boldsymbol{\xi}^t)$ be the matrix for which the $(l,m)$-th entry is
 \begin{equation*}
 	\left[R(\boldsymbol{\xi}^t)\right]_{l,m} = \sum_{\substack{\mathbf{i} \in \mathcal{T}^c(t) \\ i_1 = l, i_{t+1} = m}} \prod_{jk \in E(\mathbf{i})}  \expect\left[\xi_{i_1, i_2} \cdots \xi_{i_{t}, i_{t+1}} \mid \mathbf{U}, \mathbf{W}\right]  
 \end{equation*}
 The above arguments then give us 
\begin{align*}
	\left\lvert   	\left[R(\boldsymbol{\xi}^t)\right]_{l,m}   \right\rvert \leq (1+O(p_n)) \cdot (t+1)^t \cdot n^{\lfloor t/2\rfloor -2}p_n^{\lfloor t/2 \rfloor} ~.
\end{align*}
Finally, write: {\small
\begin{align*}
	& \expect\left[w'Bv \mid \mathbf{U}, \mathbf{W}\right] \\
	& = \sum_{\mathbf{i} \in \mathcal{T}(t) \cup \mathcal{T}^c(t)}\prod_{jk \in E(\mathbf{i})}  \expect\left[w_{i_1}\xi_{i_1, i_2} \cdots \xi_{i_{t}, i_{t+1}}v_{i_{t+1}} \mid \mathbf{U}, \mathbf{W}\right]  \\
	& = \sum_{l, m} 	\sum_{\substack{\mathbf{i} \in \mathcal{T}(t) \cup \mathcal{T}^c(t) \\
	i_1 = l, i_{t+1} = m}} \prod_{jk \in E(\mathbf{i})}  \expect\left[w_{i_1}\xi_{i_1, i_2} \cdots \xi_{i_{t}, i_{t+1}}v_{i_{t+1}} \mid \mathbf{U}, \mathbf{W}\right]\\
	& = \sum_{l, m} w_l \cdot \left(	\sum_{\substack{\mathbf{i} \in \mathcal{T}(t) \\
	i_1 = l, i_{t+1} = m}} \prod_{jk \in E(\mathbf{i})}  \expect\left[\xi_{i_1, i_2} \cdots \xi_{i_{t}, i_{t+1}} \mid \mathbf{U}, \mathbf{W}\right] \right) \cdot v_m	+ \sum_{l, m} 	w_l 	\cdot \left[R(\boldsymbol{\xi}^t)\right]_{l,m}  \cdot v_m~.
\end{align*}}

\noindent In the last equality, we used the fact that $\mathcal{T}(t)$ and $\mathcal{T}^c(t)$ are disjoint. Applying \eqref{equation--tree_approx} to the bracketed term yields the desired formula. \hfill \qed

\subsection{Proof of Lemma \ref{lemma--order-B}}

Let $B$ be a mixed product of order $(t,\tau)$. Suppose $B$ has $r_\xi$ contiguous blocks of $\boldsymbol{\xi}^{p_s}$. Let $\mathbf{i}(s), s \in [r_\xi]$ be the subset of indices corresponding to the $s$-th contiguous block. 

\begin{example} Let $B = A^2\boldsymbol{\xi}^3 A \boldsymbol{\xi}^2 A$. In the above notation, $r = 5$, $r_\xi = 2$, $\mathbf{i}(1) = (i_3, i_4, i_5, i_6)$, $\mathbf{i}(2) = (i_7,i_8, i_9)$.
\end{example}

Let $\mathcal{I} = [n]^t$ and let $\tilde{\mathcal{T}}(t)$ be the set of indices with length $t$ for which every edge corresponding to $\boldsymbol{\xi}$ has multiplicity at least 2 and $g^\text{simp}(\mathbf{i}(s))$ is a tree for all $s$. $\tilde{B}$ contains the contributions from all $\mathbf{i} \in \tilde{\mathcal{T}}(t)$. Since $\tau$ is at least 2, the bound from Lemma \ref{lemma--order-xit} implies that
\begin{equation*}
	\left\lvert [\tilde{B}]_{l,m} \right\rvert \leq (t+1)^t \cdot n^{t-\lfloor \tau/2 \rfloor - 1}p_n^{t-\lfloor \tau/2 \rfloor } \leq (t+1)^t \cdot n^{t-2}p_n^{t-1}
\end{equation*}

The contributions of all other walks are captured in $R(B)$. 

Fix $i_1$ and $i_{t+1}$ and suppose $\mathbf{i} \in \mathcal{I}\setminus \tilde{\mathcal{T}}(t)$. Then there exists $s$ such that either (1) $g^\text{simp}(\mathbf{i}(s))$ has a cycle, or (2) $g(\mathbf{i}(s))$ has an edge with multiplicity one. 
Case (1) is possible only when $t \geq 6$. In this case, $g^{\text{simp}}(\mathbf{i})$ also has a cycle. By the same argument as in Lemma \ref{lemma--order-xit}, the contribution of such terms to the expectation is at most $(1+O(p_n))\cdot p_n^{t-1}$. Having fixed $i_1$ and $i_{t+1}$, the maximum number of such terms is $n^{t-3}$ since $g$ is not a tree and therefore has at most as many nodes as edges. In case (2), the edge with multiplicity 1 may or may not occur in another block of $\boldsymbol{\xi}^{p_s}$. If it is not repeated, the expectation of this term is $0$. If the edge is repeated, then there must be a cycle in the $g^{\text{simp}}(\mathbf{i})$ and the argument in case (1) applies. For an edge to occur in two separate blocks of $\boldsymbol{\xi}^{p_s}$, $t \geq 3$. 

In sum, for each $g$, the set of $\mathbf{i}$ for which $i_1 = l$, $i_{t+1} = m$, and $g(\mathbf{i})$ is homomorphic (under labelling) to $g$ contributes at most $n^{t-3}p_n^{t-1}$ to the expectation. As before, the number of such $g$'s is less than $(t+1)^t$. Factorizing out the $(1+O(p_n))$ term gives us:
\begin{align*}
	\left\lvert   	\left[R(B)\right]_{l,m}   \right\rvert \leq (t+1)^t \cdot n^{t-3}p_n^{t-1} ~.
\end{align*}
when $t \geq 3$. When $t < 3$, any $\mathbf{i} \in \mathcal{I}\setminus \tilde{\mathcal{T}}(t)$ is associated with an expectation of $0$. \hfill \qedsymbol

\subsection{Proof of Lemma \ref{lemma--debiased-B}}

Write:
\begin{align*}
	&\expect\left[\left(	w'Bv - \expect\left[w' B v  \mid \mathbf{U}, \mathbf{W}\right]  \right)^2  \mid \mathbf{U}, \mathbf{W} \right]  \\
	& = \sum_{i_1, ..., i_{2t+2}} w_{i_1}\left(b_{i_1, i_2} \cdots \xi_{i_{t}, i_{t+1}} -	\expect\left[b_{i_1, i_2} \cdots b_{i_{t}, i_{t+1}}\mid \mathbf{U}, \mathbf{W}\right] \right)v_{i_{t+1}} \\
	& \qquad \qquad \qquad \cdot w_{i_{t+2}}\left(b_{i_{t+2}, i_{t+3}} \cdots b_{i_{2t+1}, i_{2t+2}} -	\expect\left[b_{i_{t+2}, i_{t+3}} \cdots b_{i_{2t+1}, i_{2t+2}}\mid \mathbf{U}, \mathbf{W}\right] \right)v_{i_{2t+2}} 
\end{align*}
Let $\mathbf{i} = (i_1, ..., i_{2t+2})$. Let $\tilde{g}(\mathbf{i})$ be the graph that is obtained using all edges $i_{s,s+1}$ except for $s = t+1$. Let $\tilde{g}^{\text{simp}}(\mathbf{i})$ be the simple graph obtained from $\tilde{g}(\mathbf{i})$ by deleting repeated edges. 

The sub-indices $(i_1, ..., i_{t+1})$ and $(i_{t+2}, ..., i_{2t+2})$ define two walks. For the expectation associated with $\mathbf{i}$ to be non-zero, the two walks must have an overlapping edge so that $\tilde{g}^{\text{simp}}(\mathbf{i})$ is a connected graph. Otherwise, the expectation of their product is $0$ by construction. Suppose $g$ is a connected graph and let $t \geq 2$. For $\mathbf{i}$ such that $\tilde{g}^{\text{simp}}(\mathbf{i})$ is homomorphic under labelling to $g$, and for which $i_1, i_{t+1}, i_{t+2}$ and $i_{2t+2}$ take some fixed value, their contribution to the expectation is $(1+O(p_n))\cdot p_n^{|E(g)|}$. Since $g$ is connected, it can have at most $|E(g)|+1$ nodes. Having fixed 4 nodes, we have at most $|E(g)| -3$ degrees of freedom in choosing the remaining nodes. The contribution of such terms is therefore upper bounded by:
\begin{equation*}
 |w_{i_1}v_{i_2}w_{i_3} v_{i_4}| \cdot 	n^{|E(g)| -3}p_n^{|E|} \leq  |w_{i_1}v_{i_2}w_{i_3} v_{i_4}| \cdot n^{2t-4}p_n^{2t-1}~.
\end{equation*} 
The inequality above holds because we started with 2 walks that have at most $t$ edges each. Since $g$ is obtained by stitching at least one edge from each walk together, it can have at most $2t-1$ edges. Having fixed 4 nodes, the total number of such graphs must be fewer than $(2t+2)^{2t-2}$ since we have at most $(2t+2)$ choices at each of the $2t-2$ positions. 

When $t =1$, the expectation is non-zero only when $i_1 = i_3$ and $i_2 = i_4$. The expectation of such a term is therefore $p_n\cdot w_{i_1}^2 f(U_{i_1}, U_{i_2}) v_{i_2}^2$. \hfill \qedsymbol

\subsection{Proof of Lemma \ref{lemma--spectral_conc}}

Suppose $np_n \gg \log^2 n$. Theorem 1.1 in the Supplementary Material of \cite{lei2015consistency}, gives us that for any $r > 0$, we have
$
	\left\lVert A - \hat{A} \right\rVert \leq k(r)\sqrt{np_n}
$
with probability at least  $1-n^{-r}$. Here, $k(r)$ is a constant that depends only on $r$. 

Suppose instead that $\sqrt{\frac{\log n}{\log \log n}} \ll np_n \ll \log^2 n$. Our set up satisfies the requirements for Corollary 3.3 in \cite{benaych2020spectral}. Set their $\varepsilon^2$ to $\left(\sqrt{\frac{\log n}{\log \log n}}/(np_n)\right)^{1-\nu}$ and note that their $d$ is our $np_n$. Then, with probability at least {\small $$1-\exp\left(-\left(np_n\right)^{2+\nu} \left(\frac{\log n}{\log \log n}\right)^{(1-\nu)/2} k \right)~,$$}

\noindent we have that {\small
\begin{equation*}
	\left\lVert \hat{A}-A \right\rVert \leq k \left(np_n\right)^{(1+\nu)/2} \left(\frac{ \log n}{\log \log n}\right)^{(1-\nu)/4}~,
\end{equation*}}

\noindent where $k$ is a universal constant. Combine the above inequalities and we are done. \qedsymbol

\subsection{Proof of Lemma \ref{lemma--eigenvector_of_A}}

By definition,
\begin{equation*}
	{\lambda}_s(A)v_s(A) = \sum_{r=1}^{R} \left(v_s(A)' \frac{\phi_r(\mathbf{U})}{\sqrt{n}}\right) \cdot\frac{ \phi_r(\mathbf{U})}{\sqrt{n}} =: \sum_{r=1}^{R} \alpha_{s,r} \frac{\phi_r(\mathbf{U})}{\sqrt{n}}~.
\end{equation*}
Moreover,
\begin{align}\label{equation--low_rank_eigen_coef_bound}
	1 & = v_s(A)'v_s(A) = \sum_{r,q} \alpha_{s,r} \alpha_{s,q} \frac{\phi_r(\mathbf{U})'\phi_q(\mathbf{U})}{n}
\end{align}
Since $\phi_r(\mathbf{U})'\phi_q(\mathbf{U})/n \pto \expect[\phi_r(U)\phi_q(U)]$ and $\alpha_{s,r}\alpha_{s,q} \geq -(\alpha_{s,r}^2 + \alpha_{s,q}^2)/2$, we must have that for any $\varepsilon > 0$,
$
	1 \geq (1-\varepsilon) \sum_{r=1} \alpha_{s,r}^2 \mbox{ w.p.a. 1.}
$
Consequently, $|\alpha_{s,r}| \leq 1+\varepsilon$ w.p.a. 1 for all $\varepsilon > 0$. 

Let $\tilde{v} = \phi_1(\mathbf{U})/{\lVert \phi_1(\mathbf{U}) \rVert}$. By the Courant-Fischer principle, 
\begin{equation*}
	\frac{{\lambda}_1(A)}{np_n} \geq \frac{\tilde{v}'A\tilde{v}}{np_n} = \frac{n}{\lVert \phi_1(\mathbf{U})\rVert^2} \sum_{r=1}^{R} \frac{\lambda_r(f)}{n^2} \cdot (\phi_1(\mathbf{U})'\phi_r(\mathbf{U}))^2 \pto \lambda_1(f)~.
\end{equation*}
Next, let $0 < \varepsilon < 1$ and define $\Gamma$ to be the event on which
\begin{equation*}
	\left\lvert \frac{\phi_r(\mathbf{U})'\phi_q(\mathbf{U})}{n} - \expect\left[\phi_r(U)\phi_q(U)\right] \right\rvert \leq \varepsilon \quad \mbox{ for all } r,q \in [R]^2~,
\end{equation*}
${\lambda}_1(A)/np_n \geq \lambda_1(f) -\varepsilon$ and $|\alpha_{1,r}| \leq 2$ for all $r \in [R]$. By the previous analysis, $\prob(\Gamma) \to 1$. On $\Gamma$,
\begin{align*}
\lambda_1(f) - \varepsilon  \leq \frac{\lambda_1(A)}{np_n} & = \frac{v_1(A)'Av_1(A)}{np_n} \\
& = \sum_{q,r,s} \alpha_{1,q}\alpha_{1,r}\lambda_s(f) \cdot \frac{\phi_r(\mathbf{U})'\phi_s(\mathbf{U})}{n} \cdot \frac{\phi_q(\mathbf{U})'\phi_s(\mathbf{U})}{n} \\
& \leq 4R^3\lambda_1(f) \varepsilon + (1+\varepsilon)^2 \cdot \sum_{r=1}^{R} \alpha_{1,r}^2\lambda_r(f)
\end{align*}
where the final inequality follows since there are fewer than $R^3$ terms with at least one pair of mismatched index. Since $\varepsilon^2 \leq \varepsilon$ and $\lambda_r(f) < \lambda_1(f)$ for all $r \geq 2$, each mismatched term has magnitude smaller than $4\lambda_1(f)\varepsilon$ on $\Gamma$. Since $\lambda_r(f) < \lambda_1(f)$, the preceding inequality can hold for arbitrary $\varepsilon$ w.p.a. 1 only if $\alpha_{1,1} \pto 1$. By \eqref{equation--low_rank_eigen_coef_bound}, $\alpha_{1,r} \to 0$ for all $r \geq 2$. Finally, we also have that $\prob(\lambda_1(A)/np_n \leq \lambda_1(f) + \varepsilon) \to 1$. Conclude that $\lambda_1(A)/np_n \pto \lambda_1(f)$.

Given the above characterization of $v_1(A)$, now let ${v}_2$ to be $\phi_2(\mathbf{U}) - \left(\phi_2(\mathbf{U}), v_1(A)\right)v_1(A)$ and define $\tilde{v}_2 = v_2/\lVert v_2 \rVert$.  By the same argument as before, 
\begin{equation*}
	\frac{\lambda_2(A)}{np_n} \geq \frac{\tilde{v}'_2 A\tilde{v}_2}{np_n} \geq \lambda_2(f) - \varepsilon \quad \mbox{ w.p.a. 1}
\end{equation*}
Given our results on $\alpha_{1,r}$, we can write:
\begin{equation*}
	0 = \tilde{v}_1'\tilde{v}_2 =   (1-o_p(1)) \sum_{r=1}^R \alpha_{1,r}\alpha_{2,r} - o_p(1) = (1-o_p(1)) \alpha_{2,1} - o_p(1)
\end{equation*}
which implies that $\alpha_{2,1} \pto 0$. Again, to attain an eigenvalue that is at least $\lambda_2 - \varepsilon$ w.p.a 1, $|\alpha_{2,2}| \pto 1$, which implies that $\alpha_{2,r} \pto 0$ for all $r \geq 3$. Consequently, $\frac{{\lambda}_2(A)}{np_n} \leq \lambda_2(f) + \varepsilon$ w.p.a. 1 and have that ${\lambda}_2(A)/np_n \pto \lambda_2(f)$. The proof for $\tilde{\lambda}_r$ with $r \geq 3$ follows from iterating the same Gram-Schmidt type argument. \hfill \qed

% end of auxiliary Lemma
\end{document}